\documentclass[12pt,a4paper]{article}
\pdfoutput=1
\usepackage{color}
\usepackage{amssymb,amsmath,bm,bbm}
\usepackage{epsf}
\usepackage{epsfig}
\usepackage{afterpage}
\usepackage{longtable}
\usepackage[dvipsnames]{xcolor}
\usepackage[linktoc=page,bookmarks=false,colorlinks=false,linkbordercolor=RoyalBlue,citebordercolor=ForestGreen,urlbordercolor=CornflowerBlue]{hyperref}
\usepackage{latexsym,mathrsfs,dsfont}
\usepackage[normalem]{ulem} 
\usepackage[compress]{cite}
\usepackage{graphicx}
\usepackage{url}
\usepackage{paralist}
\usepackage{bbold}
\usepackage{slashed}
\usepackage{multirow}
\usepackage[hypcap]{caption, subcaption}

\usepackage{fancyhdr}
\advance \headheight by 3.0truept       

\newcommand{\beq}{\begin{equation}}
\newcommand{\eeq}{\end{equation}}
\newcommand{\be}{\begin{equation}}
\newcommand{\ee}{\end{equation}}
\newcommand{\bi}{\begin{itemize}}
\newcommand{\ei}{\end{itemize}}
\newcommand{\ba}{\begin{array}}
\newcommand{\ea}{\end{array}}
\newcommand{\beqa}{\begin{eqnarray}}
\newcommand{\eeqa}{\end{eqnarray}}
\newcommand{\bea}{\begin{eqnarray}}
\newcommand{\eea}{\end{eqnarray}}
\newcommand{\beqn}{\begin{eqnarray}}
\newcommand{\eeqn}{\end{eqnarray}}

\definecolor{red}{cmyk}{0,1,1,0.4}

\definecolor{green1}{rgb}{0.06,0.66,0.06}
\definecolor{orange1}{rgb}{0.98,0.60,0.07}
\definecolor{darkgreen}{rgb}{0.0,0.6,0.0}
\definecolor{darkblue}{RGB}{12,13,115}
\definecolor{darkred}{RGB}{204,6,0}

\newcommand{\green}{{\color{green1}$\bigstar$}}

\newcommand{\red}{{\color{red}\small \protect\raisebox{-0.05em}{$\blacksquare$}}}
\newcommand{\sgreen}{{\color{green1}$\star$}}
\newcommand{\sred}{{\color{red}\tiny \protect\raisebox{-0.05em}{$\blacksquare$}}}
\newcommand{\rSt}{{\color{darkblue}$\bigstar$}}
\newcommand{\bSt}{{\color{darkred}$\bigstar$}}
\newcommand{\rst}{{\color{darkblue}$\star$}}
\newcommand{\bst}{{\color{darkred}$\star$}}

\def \refeq#1{(\ref{#1})}
\def \refsec#1{Section~\ref{#1}}

\def \refapp#1{Appendix~\ref{#1}}
\def \reffig#1{Fig.~\ref{#1}}
\def \reftab#1{Table~\ref{#1}}

\newcommand{\tev}{\, {\rm TeV}}
\newcommand{\gev}{\, {\rm GeV}}
\newcommand{\mev}{\, {\rm MeV}}

\def\epe{\varepsilon'/\varepsilon}

\def\kpn{K^+\rightarrow\pi^+ \nu\bar\nu}
\def\klpn{K_{L}\rightarrow\pi^0 \nu\bar\nu}

\newcommand{\GSM}{{\mathrm{G_{SM}}}}
\newcommand{\GSMUpr}{{\mathrm{G^\prime_{SM}}}}
\newcommand{\GSMandUpr}{{\mathrm{G^{(\prime)}_{SM}}}}
\newcommand{\SUthreeC}{{\mathrm{SU(3)_c}}}
\newcommand{\SUtwoL}{{\mathrm{SU(2)_L}}}
\newcommand{\UoneY}{{\mathrm{U(1)_Y}}}
\newcommand{\UoneEM}{{\mathrm{U(1)_{\rm em}}}}
\newcommand{\UonePr}{{\mathrm{U(1)_{L_\mu-L_\tau}}}}

\newcommand{\wc}[3][{}]{[{\cal C}_{#2}^{#1}]_{#3}}
\newcommand{\dotwc}[3][{}]{[\dot{\cal C}_{#2}^{#1}]_{#3}}

\newcommand{\Wc}[2][{}]{{\cal C}_{#2}^{#1}}
\newcommand{\WcD}[2][{}]{{\cal C}_{#2}^{#1\dagger}}
\newcommand{\dotWc}[2][{}]{\dot{\cal C}_{#2}^{#1}}

\newcommand{\Yuk}[1]{Y_{#1}^{}}
\newcommand{\YukD}[1]{Y_{#1}^\dagger}

\newcommand{\muEW}{{\mu_{\rm EW}}}
\newcommand{\muLow}{{\mu_{\rm low}}}

\begin{document}


\vspace{-14mm}
\begin{flushright}
  FLAVOUR(267104)-ERC-120 \\ 
  LMU-ASC 01/16 \\
  TUM-HEP-1060/16
\end{flushright}

\vspace{8mm}

\begin{center}

{\Large\bf
  \boldmath{Patterns of Flavour Violation in Models \\[0.2cm] 
            with Vector-Like Quarks}
} \\[8mm]

{\bf 
  Christoph Bobeth,${}^{1,2,4}$
  Andrzej~J.~Buras,${}^{1,2}$ 
  Alejandro Celis, ${}^3$ 
  Martin Jung ${}^{1,4}$ 
}\\[1cm]

{\small
${}^1$TUM Institute for Advanced Study, 
  Lichtenbergstr.~2a, D-85748 Garching, Germany \\[2mm]

${}^2$Physik Department, TU M\"unchen, 
  James-Franck-Stra{\ss}e, D-85748 Garching, Germany \\[2mm]

${}^3$Ludwig-Maximilians-Universit\"at M\"unchen, Fakult\"at f\"ur Physik,\\
  Arnold Sommerfeld Center for Theoretical Physics, 80333 M\"unchen, Germany \\[2mm]

${}^4$Excellence Cluster Universe, Technische Universit\"at M\"unchen,
  Boltzmannstr. 2, D-85748 Garching, Germany}

\end{center}

\vspace{3mm}

\begin{abstract}
\noindent

We study the patterns of flavour violation in renormalisable extensions of the
Standard Model (SM) that contain vector-like quarks (VLQs) in a single complex
representation of either the SM gauge group $\GSM$ or $\GSMUpr \equiv \GSM \otimes
\UonePr$. We first decouple VLQs in the $M=(1 - 10)$~TeV range and then at the 
electroweak scale also $Z, Z'$ gauge bosons and additional scalars to study the
phenomenology. The results depend on the relative size of $Z$- and $Z'$-induced
flavour-changing neutral currents, as well as the size of $|\Delta F|=2$
contributions including the effects of renormalisation group Yukawa evolution 
from $M$ to the electroweak scale that turn out to be very important for models
with right-handed currents through the generation of left-right operators.
In addition to rare decays like $P\to \ell\bar\ell$, $P\to P' \ell\bar\ell$,
$P\to P'\nu\bar\nu$ with $P=K, B_s, B_d$ and $|\Delta F|=2$ observables we
analyze  the ratio $\epe$ which appears in the SM to be significantly below
the data. We study patterns and correlations between these
observables which taken together should in the future allow
for differentiating between VLQ models. In particular the patterns in models with
left-handed and right-handed currents are markedly different from each other.
Among the highlights are large $Z$-mediated new physics effects in Kaon
observables  in some of the models and significant
effects in $B_{s,d}$-observables. $\epe$ can easily be made consistent
with the data, implying then uniquely the suppression of $\klpn$. Significant
enhancements of $Br(\kpn)$ are still possible.
We point out that the combination of NP effects to $|\Delta F|=2$ and $|\Delta F|=1$
observables in a given meson system generally allows to determine the masses
of VLQs in a given representation independently of the size of VLQ couplings.
\end{abstract}

\setcounter{page}{0}
\thispagestyle{empty}
\newpage
\enlargethispage{3ex}
\setcounter{tocdepth}{2}
\tableofcontents

\newpage

%
%
%

\section{Introduction}

Among the simplest renormalisable extensions of the Standard Model (SM) that do
not introduce any additional fine tunings of parameters are models in which the
only new particles are vector-like fermions. Such fermions can be much heavier
than the SM ones as they can acquire masses in the absence of electroweak
symmetry breaking.  If in the process of this breaking mixing with the SM
fermions occurs, the generation of flavour-changing neutral currents (FCNC)
mediated by the SM $Z$ boson is a generic implication.  If in addition the gauge
group is extended by a second $\mathrm{U(1)}$ factor, a new heavy gauge boson $Z^\prime$
is present and additional heavy scalars are necessary to provide mass for the
$Z^\prime$ and to break the extended gauge-symmetry group down to the SM gauge
group.  There is a rich literature on FCNCs implied by the presence of
vector-like quarks (VLQs), see in particular \cite{Nir:1990yq, Branco:1992wr, 
delAguila:2000rc, Barenboim:2001fd, Buras:2009ka, Botella:2012ju, Fajfer:2013wca,
Buras:2013td, Altmannshofer:2014cfa, Alok:2015iha, Ishiwata:2015cga, Arnan:2016cpy}.

The goal of the present paper is an extensive study of patterns of flavour
violation in models with VLQs that are based on the following gauge groups:
\begin{align}
  \GSM & \equiv \SUthreeC \otimes \SUtwoL \otimes \UoneY ,
\\
  \GSMUpr & \equiv \GSM \otimes \UonePr .
\end{align} 
The choice of the particular symmetry group $\UonePr$ \cite{He:1990pn,
  He:1991qd} is phenomenologically motivated by the fact that it allows in a
simple manner to address successfully the LHCb anomalies
\cite{Altmannshofer:2014cfa, Altmannshofer:2015mqa}, while being anomaly-free
and containing less parameters than general $Z^\prime$ models
\cite{Buras:2013qja}.

In our paper we will be guided by the analyses 
in Refs.~\cite{delAguila:2000rc, delAguila:2008pw, Ishiwata:2015cga} which
identified all renormalisable models with additional fermions residing in a
single vector-like complex representation of the SM gauge group with a mass
$M$. It turns out that there are 11 models where new fermions have the proper
quantum numbers so that they can couple in a renormalisable manner to the SM
Higgs and SM fermions, thereby implying new sources of flavour violation. Our
analysis will concentrate on FCNCs in the $K$, $B_d$ and $B_s$ systems,
therefore only the five models with couplings to down quarks are relevant for
us, as specified in Section~\ref{sec:models}. We call this class of models
$\GSM$-models.

Consequently the models based on the gauge group $\GSMUpr$ are called
$\GSMUpr$-models.  The VLQs in these models belong to the same representations
under $\GSM$ as in $\GSM$-models, but are additionally charged under
$\UonePr$. These models also contain new heavy scalars.

As we will discuss in detail in \refsec{sec:models} and \refsec{sec:Comparison},
the patterns of flavour violation in $\GSM$-models and $\GSMUpr$-models differ
significantly from each other:
\begin{itemize}
\item In $\GSM$-models Yukawa interactions of the SM scalar doublet $H$
  involving ordinary quarks and VLQs imply flavour-violating $Z$ couplings to
  ordinary quarks, which then dominate $|\Delta F|=1$ FCNC transitions. However,
  the situation in $|\Delta F|=2$ transitions is much more involved and 
  depends on whether right-handed (RH) or left-handed (LH) flavour-violating quark 
  couplings to the $Z$ are present. If they are RH the effects of renormalisation group (RG) 
  evolution from $M$ (the common VLQ mass) down to the electroweak scale, $\muEW$, 
  generate left-right operators \cite{Bobeth:2017xry} via top-Yukawa induced mixing.
  These operators are
  strongly enhanced through QCD RG effects below the electroweak scale and in the
  case of the $K$ system through chirally enhanced hadronic matrix elements. 
  They dominate then new physics (NP) contributions to $\varepsilon_K$, but in the
  $B_{s,d}$ meson systems for  VLQ-masses above $5\tev$ they have to compete with 
  contributions from box diagrams with VLQs ~\cite{Ishiwata:2015cga}. If they are
  LH the Yukawa enhancement is less important, because left-right operators are not
  present and box diagrams play an important role both in the $B_{s,d}$ and $K$ systems.
\item In $\GSMUpr$-models the pattern of flavour violation depends on the scalar
  sector involved. We consider only models in which at least one of the
  additional scalars is charged under $\UonePr$ in such a way that Yukawa
  couplings between the given VLQ and ordinary quarks are allowed. If this is
  the case for a new scalar which is just a singlet $S$ under the SM group, the
  latter imply flavour-violating $Z^\prime$ couplings to ordinary quarks without
  any FCNCs mediated by the $Z$. In the following we refer to these models as
  $\GSMUpr(S)$-models.  If, on the other hand, such a Yukawa coupling requires
  the scalar to be a doublet $\Phi$, both tree-level $Z^\prime$ and $Z$
  contributions to flavour observables will be present. Their relative size
  depends on the model parameters, specifically the $Z^\prime$ mass. In these
  cases we introduce again an additional scalar singlet, but without Yukawa
  couplings, since otherwise the $Z^\prime$ mass would have to be of the order
  of the electroweak scale, which is phenomenologically very difficult to
  achieve.  In the following we refer to these models as $\GSMUpr(\Phi)$-models.
\end{itemize}

In this manner we will consider three classes of VLQ models with rather
different patterns of flavour violation:
\begin{align}
   \GSM\,, &&  \GSMUpr(S)\,, && \GSMUpr(\Phi)\,,
\end{align}
in which $|\Delta F|=1$ FCNCs are mediated by the $Z$, $Z^\prime$ and both,
respectively. In $\GSMUpr(\Phi)$ models $|\Delta F|=2$ transitions
are dominated for $M\ge 5\tev$ by box diagrams with VLQs and scalar exchanges,
while in the $\GSMUpr(S)$ models also tree-level $Z^\prime$ exchanges can play
an important, sometimes dominant, role. A particular feature of
$\GSM$ models are the top-Yukawa induced RG effects to $|\Delta F|=2$ transitions
that are largest for RH scenarios and are absent in $\GSMUpr$ models. 

In \cite{Ishiwata:2015cga} an extensive analysis of the $\GSM$-models has been
performed and a subset of $\GSMUpr$-models has been analyzed in
\cite{Altmannshofer:2014cfa, Altmannshofer:2015mqa}. Therefore it is mandatory
for us to state what is new in our article regarding these models:
\begin{itemize}
\item The authors of \cite{Ishiwata:2015cga} concentrated on the derivation of
  bounds on the Yukawa couplings as functions of $M$ but did not study the
  correlations between various flavour observables which is the prime target of
  our paper. Similar comments apply to \cite{Altmannshofer:2014cfa}.
\item NP contributions to flavour observables depend in each model on the
  products of complex Yukawa couplings $\lambda^*_s\lambda_d$,
  $\lambda^*_b\lambda_d$ and $\lambda^*_b\lambda_s$ for $s\to d$, $b\to d$ and
  $b\to s$ transitions, respectively, as well as the VLQ mass $M$. This
  structure allows to set one of the $\lambda_q$-phases to zero, such that each
  model depends on only five Yukawa parameters and $M$, implying a number of
  correlations between flavour observables. The strongest correlations are,
  however, still found between observables corresponding to the same
  flavour-changing transition, and we concentrate our analysis on them. The
  correlations between observables with different transitions are weaker, but
  could turn out to be useful in the future when the data and theory improve, in
  particular in the context of models for Yukawa couplings.
\item An important novelty of our paper, relative to \cite{Ishiwata:2015cga,
    Altmannshofer:2014cfa, Altmannshofer:2015mqa}, is the inclusion of the ratio
  $\epe$ in our study. Recent analyses indicate that the measurement of $\epe$
  is significantly above its SM prediction~\cite{Bai:2015nea, Buras:2015yba,
    Buras:2015xba,Kitahara:2016nld}; it is hence of interest to see which of the
  models analyzed by us, if any, are capable of addressing this tension and what
  the consequences for other observables are.
\item
  Another important novelty in the context of VLQ models and $|\Delta F|=2$
  transitions in general is the inclusion of the effects of RG top-Yukawa evolution 
  from $M$ to the electroweak scale that turn out to be very important for models
  with RH currents through the generation of left-right
  operators contributing to these transitions as mentioned above. This changes
  markedly the pattern of flavour violation in such models relative to models
  with LH  currents where no left-right operators are generated.
\end{itemize}

Our paper is organized as follows. In \refsec{sec:models} we present the
particle content of the considered VLQ models, together with the gauge
interactions, Yukawa interactions and the scalar sector.  In
\refsec{sec:VLQ-decoupling} we perform the decoupling of the VLQs and construct
the effective field theory ($\GSMandUpr$-EFT) for each model for scales
$\mu_{\rm EW}< \mu< M$. \refsec{sec:EFT-down-pheno} is devoted to the matching
of these EFTs to phenomenological ones describing $|\Delta F| = 1,2$ processes
below the scale $\mu_{\rm EW}$. This results in explicit flavour-violating
couplings of the $Z$ and $Z^\prime$ to the SM quarks.  These enter the effective
Lagrangians for the various flavour-changing processes, from which we derive the
explicit formulae for the considered observables.  In \refsec{sec:Comparison} we
describe the patterns of flavour violation expected in different models,
summarizing them with the help of two DNA tables. In \refsec{sec:numerics},
after formulating our strategy for the phenomenology, we present numerical
results of our study. We conclude in \refsec{sec:summary}.  Several appendices
collect additional information on the models, the decoupling of 
VLQs, RG equations in the $\GSM$-EFT, the considered decays, some 
technical details and the input and statistical procedure used 
in the numerical analysis.

%
%
%

\section{The VLQ Models \label{sec:models}}

Throughout the article we focus on models with vector-like fermions residing in
complex representations, either of the the SM gauge group $\GSM$ or its
extension by an additional gauged $(L_\mu-L_\tau)$ symmetry, $\UonePr$. For both
models we adapt the usual SM fermion content of the three generations
($i = 1,2,3$) of quarks ($q_L^i =(u_L^i, d_L^i)^T, u_R^i, d_R^i$) and leptons
($L_L^i = (\nu_i, \ell_L^i)^T, \ell_R^i$), which acquire masses via spontaneous
symmetry breaking from the standard scalar $\SUtwoL$ doublet~$H$.

The gauged $(L_{\mu} - L_{\tau})$ symmetry is anomaly-free in the
SM~\cite{He:1990pn, He:1991qd}. The only non-vanishing $(L_{\mu} - L_{\tau})$
charges of the SM fermions are introduced as
\begin{align}
  \label{eq:UonePr-lep-charges}
  Q'(L_L^2) & = Q'(\mu_R)  =  Q'_{\ell}, &
  Q'(L_L^3) & = Q'(\tau_R) =  - Q'_{\ell}.
\end{align}
Here $L_L^2 = (\nu_{\mu}, \mu_L)$ and $L_L^3 = (\nu_{\tau}, \tau_L)$ are
left-handed $\mathrm{SU(2)_L}$ doublets and $\mu_R$ and $\tau_R$ right-handed
singlets. We normalize the $(L_{\mu} - L_{\tau})$ charges of the leptons without
loss of generality by setting $Q'_{\ell} = 1$.  The SM quarks do not couple
directly to the $\UonePr$ gauge boson $Z^\prime$. However, such couplings are
generated in $\GSMUpr$ models through Yukawa interactions of SM quarks with VLQs
that couple directly to $Z^\prime$.

%
%

\subsection{VLQ Representations}

As we are mainly interested in the phenomenology of down-quark physics, we will
restrict our analysis to $\SUthreeC$ triplets and consider the following five
models with $\SUtwoL$ singlets, doublets and triplets:
\begin{equation}
\begin{aligned}
  \mbox{singlets} & : & D(1, -1/3, -X), & &  & &  {\rm (V)}
\\
  \mbox{doublets} & : & Q_V(2, +1/6, +X), & & Q_d(2, -5/6, -X), & & {\rm(IX,XI)}
\\
  \mbox{triplets} & : & T_d(3, -1/3, -X), & & T_u(3, +2/3, +X), & & {\rm (VII,VIII)}
\end{aligned}
\end{equation}
where the transformation properties are indicated as
$(\SUtwoL, \UoneY, \UonePr)$, \emph{i.e.} $X$ denotes the charge under
$\UonePr$.  It is implied that in $\GSM$-models the $\UonePr$ charge should be
omitted. The representations $D$, $Q_V$, $Q_d$, $T_d$, $T_u$ correspond to the
models V, IX, XI, VII, VIII introduced in Ref.~\cite{Ishiwata:2015cga}, where a
complete list of renormalisable models with vector-like fermions under $\GSM$
can be found, see also \cite{delAguila:2000rc, delAguila:2008pw}. 
Concerning $\GSMUpr$, the combination of representations $D$,
$Q_V$ and additionally $U(1, +2/3, -X)$ has been studied first in
\cite{Altmannshofer:2014cfa}.

The kinetic and gauge interactions of the new VLQs are given by
\begin{align}
  \label{eq:Lag:kin:VLQ}
  {\cal L}_{\rm kin} & 
  = \overline{D} (i {\cal D}\!\!\!\!/ - M_D) D
  + \sum_{a=V,d} \overline{Q}_a (i {\cal D}\!\!\!\!/ - M_{Q_a}) \, Q_a 
  + \sum_{a=d,u} \mbox{Tr} \left[ \overline{T}_{a} (i {\cal D}\!\!\!\!/ - M_{T_a})\,  T_{a} \right] , 
\end{align}
with appropriate covariant derivatives ${\cal D}_\mu$ and we follow
\cite{Ishiwata:2015cga} for the triplet representations as given in (2.13) and
(2.14) of that paper. The masses~$M$ of the VLQs introduce a new scale, which we
will assume to be significantly larger than all other scales. The covariant
derivative is, omitting the $\SUthreeC$ part,
\begin{align}
  \label{eq:def-cov-derivative}
  {\cal D}_\mu & =
  \partial_\mu - i g_1 \frac{\sigma^a}{2} W^a_\mu - i g_2 Y B_\mu - i g' Q' \hat{Z}'_\mu
\end{align}
with the gauge couplings $g_{2,1}$ and $g'$ of $\SUtwoL$, $\UoneY$ and
$\UonePr$, respectively, and charges $Y$ and $Q'$ of $\UoneY$ and $\UonePr$. The
Pauli-matrices are denoted by $\sigma^a$. The ``hat'' on $\hat{Z}'_\mu$
indicates that we deal here with the gauge eigenstate and not mass eigenstate,
see \refeq{eq:Zmixing}.

%
%

\subsection{Yukawa interactions of VLQs}
\label{sec:VLQ:Yuk}


\subsubsection[$\GSM$]{\boldmath $\GSM$}

The scalar sector consists of the SM scalar doublet $H$ with its usual scalar
potential. The VLQs interact with SM quarks ($q_L,\, u_R,\, d_R$) via Yukawa
interactions
\begin{equation}
  \label{eq:Yuk:H}
\begin{aligned}
  - {\cal L}_{\rm Yuk}(H) & =  
  \left( \lambda_i^D \, H^\dagger \overline{D}_R 
       + \lambda_i^{T_d} \, H^\dagger \overline{T}_{dR}
       + \lambda_i^{T_u} \, \widetilde{H}^\dagger \overline{T}_{uR} \right) q_L^i
\\ 
  &  
  + \lambda_i^{V_u} \, \bar{u}_R^i \widetilde{H}^\dagger Q_{VL} 
  + \bar{d}_R^i \left( \lambda_i^{V_d} \,  H^\dagger Q_{VL}
          + \lambda_i^{Q_d} \, \widetilde{H}^\dagger Q_{dL} \right) 
  + \mbox{h.c.}\,,
\end{aligned}
\end{equation}
where $\widetilde H \equiv i\sigma_2 H^*$.  The complex-valued Yukawa couplings
$\lambda_i^{\rm VLQ}$ give rise to mixing with the SM quarks and
flavour-changing $Z$-couplings, which have been worked out in detail
\cite{delAguila:2000rc, Ishiwata:2015cga} and are discussed in \refsec{sec:tree-decoupl}.


\subsubsection[$\GSMUpr(S)$]{\boldmath $\GSMUpr(S)$}

In models with an additional $\UonePr$ the scalar sector has to be extended in
order to generate the mass of the corresponding gauge boson $Z'$.  A complex
scalar $S(1,0,X)$ ($\SUthreeC$ singlet) is added in the minimal version.  As
VLQs are charged under $\UonePr$, their Yukawa couplings with the SM doublet $H$
are forbidden, but the ones involving $S$ are allowed for
$Q'_S=\pm Q'_{\rm VLQ}$ and given by \cite{Altmannshofer:2014cfa}
\begin{align}
  \label{eq:Yuk:S}
  - {\cal L}_{\rm Yuk}(S)&  
  = \left(\lambda_i^{D} \, \bar{d}_R^i \, D_L 
        + \lambda_i^{V} \, \overline{Q}_{VR} \, q_{L}^i \right) S
  + \mbox{h.c.} \,.
\end{align}
In fact this scalar system is sufficient for models with VLQs having $\UoneY$
charges $Y = -1/3$ and $+1/6$ of the SM fermions $d_R$ and $q_L$, respectively.
In the following we refer to these models as $\GSMUpr(S)$-models. The special
feature of these models is that because of the absence of tree-level 
$Z$ contributions tree-level $Z^\prime$ exchanges dominate
$\Delta F=1$ transitions and in some part of the parameter space can also
compete with contributions from box diagrams with VLQs and scalars in the case
of $\Delta F=2$ transitions. 


\subsubsection[$\GSMUpr(\Phi)$]{\boldmath $\GSMUpr(\Phi)$}

For VLQs with $\GSM$ quantum numbers different from one of the SM quark fields,
the simple extension by a scalar singlet is not possible. In a next-to-minimal
version we therefore add to the scalar sector an additional scalar $\SUtwoL$
doublet $\Phi(2,+1/2,X)$, besides the SM-like $H(2,+1/2,0)$. We require
$|X| \neq 1, 2$ in order to avoid lepton-flavour violating (LFV) Yukawa
couplings --- see for example \cite{Crivellin:2015mga} --- and in consequence
there are no LFV $Z'$ couplings, which are subject to strong constraints at low
energies. The vacuum expectation value (VEV) of $\Phi$ gives an unavoidable
contribution to the $Z'$~mass of the order of the electroweak scale, contributes
to the mass of $H$ and generates potentially large $Z-Z'$ mass mixing
effects. The latter would be strongly constrained by electroweak precision tests
\cite{ALEPH:2005ab}, in particular there would be sizeable corrections to the $Z$
couplings to muons. In order to avoid these difficulties, $\Phi$ is accompanied
by an additional complex scalar singlet $S(1,0,Y)$, which breaks the $\UonePr$
symmetry at the TeV scale. The $L_{\mu}-L_{\tau}$ charge of $S$ is chosen to be
$Y = X/2$ in order to avoid the appearance of a Goldstone boson in the scalar
sector and to forbid Yukawa couplings of $S$ with SM fermions and VLQs.

The Yukawa interactions of the VLQs with $\Phi$ are
\begin{align} 
  \label{eq:Yuk:Phi}
  -{\cal L}_{\rm Yuk}(\Phi) & =
  \left( \lambda_i^D \, \Phi^\dagger \overline{D}_R 
       + \lambda_i^{T_d} \, \Phi^\dagger \overline{T}_{dR}
       + \lambda_i^{T_u} \, \widetilde{\Phi}^\dagger \overline{T}_{uR} \right) q_L^i
  + \lambda_i^{Q_d} \, \widetilde{\Phi}^\dagger \bar{d}_R^i \, Q_{dL} 
  + \mbox{h.c.} ,
\end{align}
with $\widetilde\Phi \equiv i\sigma_2 \Phi^*$ and we will refer to these models
as $\GSMUpr(\Phi)$-models. We note that the structure of couplings equals the
one of $\GSM$ models given in Eq.~\refeq{eq:Yuk:H} upon $H \leftrightarrow \Phi$.
For the VLQ $D(1,-1/3,X)$ we consider thus two versions, one in $\GSMUpr(S)$ and
one in the $\GSMUpr(\Phi)$-model. We refrain from the same procedure for
$Q_V(2,+1/6,X)$.  In $\GSMUpr(\Phi)$ models FCNCs are mediated by both $Z$ and
$Z^\prime$ but in the case of $\Delta F=2$ transitions box diagrams with VLQs
and scalars play the dominant role for sufficiently large $M$. 

For ease of notation, we will sometimes refrain below from explicitly labelling
the $\lambda_i$ by the VLQ representation, as should be done if several of them
are considered simultaneously.


\subsubsection{Yukawa couplings of several representations}

In our numerics we will consider one VLQ representation at a time as this
simplifies the analysis significantly. In particular the number of parameters is
quite limited. Still it is useful to make a few comments on the structure of
flavour-violating interactions and at various places in our paper to state how
our formulae would be modified through the presence of several VLQ
representations in a given model. We plan to return to the phenomenology of such
models in the future.

When admitting several VLQ representations $F^m$ and $F^n$ simultaneously,
potentially additional locally gauge-invariant Yukawa couplings
$\sim \widetilde{\lambda}_{mn} \overline{F}^m_L \varphi_{mn} F^n_R$ with
$\varphi_{mn}=H$ have to be included in the case of
$\GSM$-models~\cite{delAguila:2000rc}. They give rise to flavour-changing
neutral Higgs currents at tree level. In the $\GSMUpr$-models the
$\UonePr$-charges of the additional $\varphi_{mn} = S, \Phi$ have been chosen
following the criteria explained above, which fixes in turn the
$\UonePr$-charges of the VLQs.  In consequence such couplings to
$\varphi_{mn} = S, \Phi$ are not permitted, however they are still allowed for
$\varphi_{mn} = H$, which has zero $\UonePr$-charge. In $\GSMUpr(S)$ models,
only the particular choice of the $\UonePr$ charges $Q'_{Q_V} = -Q'_{D}$
\cite{Altmannshofer:2014cfa} forbids these couplings to $H$, whereas the choice
$Q'_{Q_V} = Q'_{D}$ would allow them, due to the possibility to replace
$\overline{Q}_{VR} \, q_{L}^i \to \bar{q}_{L}^i \, Q_{VR}$ in Eq.~\refeq{eq:Yuk:S},
which maintains gauge invariance since $S$ is a singlet. On the other hand, in
$\GSMUpr(\Phi)$ models such couplings arise for $Q_{d}$ with $D$ and $T_d$.

Another important consequence of the presence of several representations is 
the generation of left-right $|\Delta F|=2$ operators in models with both 
LH and RH currents via box diagrams discussed in \refsec{sec:loop-decoupl}, 
which is the case when singlets or triplets together with doublets are present.
In the case of a single representation such operators can also be generated in
models with doublets through the top-Yukawa RG evolution from $M$ to the 
electroweak scale, see \refsec{sec:GSM-RGE}.

%
%

\subsection{Scalar sectors}
\label{sec:scalar-sector}

In the $\GSM$-models, the scalar sector contains only the standard doublet
$H(2,+1/2,0)$, which provides masses to gauge bosons and standard fermions in
the course of spontaneous symmetry breaking of
$\SUtwoL \otimes \UoneY \to \UoneEM$ via the VEV $v\simeq 246$~GeV, where
\begin{align}
  \langle H \rangle & = (0, v/\sqrt{2})^T .
\end{align}

In $\GSMUpr(S)$-models the doublet $H(2,+1/2,0)$ fulfils again the same role,
whereas the singlet $S(1,0,X)$ provides via its VEV
$\langle S\rangle = v_S / \sqrt{2}$ a mass for the additional $\UonePr$
$Z'$-gauge boson
\begin{align}
  M_{Z'}^2 & = g'^2 v_S^2 X^2 .
\end{align}

In $\GSMUpr(\Phi)$-models the doublet $\Phi_2 \equiv H(2,+1/2,0)$ gives masses
to the chiral fermions, whereas $\Phi_1 \equiv \Phi(2,+1/2,X)$ contributes to
the masses of the $Z$ and $Z'$ gauge bosons in combination with
$S(1, 0, X/2)$.\footnote{This convention corresponds to that of the Type I
  2HDM.} The neutral components of the doublets acquire VEV's
\begin{align}
  \label{eq:GSMpr-Phi-vev}
  \langle \Phi_a^0 \rangle & 
  = \frac{v_a}{\sqrt{2}} \,, &
  \tan\beta &
  \equiv \frac{v_2}{v_1} \,, &
  v &
  = \sqrt{v_1^2 + v_2^2} \simeq 246\gev\,, 
\end{align}
with $0 \leq \beta \leq \pi/2$. In this case, neutral gauge boson mixing occurs
with details given in \refapp{app:scalar:S+H+Phi}.

Further details on the scalar sectors of the $\GSMUpr(S)$ and $\GSMUpr(\Phi)$
models are collected in \refapp{app:scalar:S+H} and~\ref{app:scalar:S+H+Phi},
respectively.  In \reftab{tab:Models} we summarize all $\GSMUpr$-models and
indicate which diagrams dominate NP contributions to $|\Delta F|=1$ and
$|\Delta F|=2$ transitions in a given model.

\begin{table}[!htb]
\renewcommand{\arraystretch}{1.1}
\centering{
\resizebox{\textwidth}{!}{
\begin{tabular}{|r|c|c|c|c|}
\hline
  VLQ Representation & Scalar Singlet & Scalar Doublets & $|\Delta F|=1$ & $|\Delta F|=2$
\\
\hline \hline
 $D_a(3,1,-1/3,-X)$  & $S(1,1,0,X)$   & $H(1,2,1/2,0)$ 
& $Z^\prime$ & $Z^\prime$, Box
\\
$D_b(3,1,-1/3,-X)$   & $S(1,1,0,X/2)$ & $\Phi_1(1,2,1/2,X)$, $\Phi_2(1,2,1/2,0)$
& $Z^\prime,~Z$ & Box
\\
 $Q_V(3,2,+1/6,+X)$   & $S(1,1,0,X)$   & $H(1,2,1/2,0)$
& $Z^\prime$  & $Z^\prime$, Box
\\
$Q_d(3,2,-5/6,-X)$  & $S(1,1,0,X/2)$  & $\Phi_1(1,2,1/2,X)$, $\Phi_2(1,2,1/2,0)$
& $Z^\prime,~Z$ & Box
\\
 $T_d(3,3,-1/3,-X)$ & $S(1,1,0,X/2)$  & $\Phi_1(1,2,1/2,X)$, $\Phi_2(1,2,1/2,0)$
& $Z^\prime,~Z$ & Box
\\
$T_u(3,3,+2/3,+X)$    & $S(1,1,0,X/2)$  & $\Phi_1(1,2,1/2,X)$, $\Phi_2(1,2,1/2,0)$
& $Z^\prime,~Z$ & Box
\\
\hline
\end{tabular}
}
}
\renewcommand{\arraystretch}{1.0}
\caption{\small
  Fermion and scalar representations under $\SUthreeC \otimes \SUtwoL \otimes 
  \UoneY \otimes \UonePr$ in $\GSMUpr$-models. In the last two columns we show 
  which diagrams dominate NP contributions to $|\Delta F|=1$ and $|\Delta F|=2$ 
  transitions  for $M\ge 5\tev$.
}
\label{tab:Models}
\end{table}

%
%
%

\section{Decoupling of VLQs}
\label{sec:VLQ-decoupling}

The VLQ models are characterised by the masses $M$ of the VLQs, the various
Yukawa couplings $\lambda_i^{\rm VLQ}$ ($i=1,2,3$) of \refsec{sec:VLQ:Yuk} and
the VEVs of the respective scalar sectors, see \refsec{sec:scalar-sector}.  The
present lower bound on $M$ from the LHC is in the ballpark of $1\tev$, while the
lower bounds on $M_{Z^\prime}$ are typically close to $3\tev$ if $Z^\prime$ has
a direct coupling to light quarks. But as emphasized in
\cite{Altmannshofer:2014cfa, delAguila:2014soa, Altmannshofer:2015mqa}, $Z^\prime$ 
of $\UonePr$ does not have such couplings, implying a much weaker lower bound on its mass,
which could in fact be as low as the electroweak scale and even lower.  While it
could also be as heavy as the VLQ mass, we will assume the hierarchy
\begin{align}
  \label{eq:scale-hierarchy-1}
  M_{Z} \lesssim M_{Z^\prime} \ll M, 
  \quad\mbox{or equivalently}\quad
  v  \lesssim  v_S  \ll M\,,
\end{align}
in order to simplify the analysis.  It is then natural to decouple first the
VLQs and to consider EFTs for $\GSM$ and $\GSMUpr$ valid between the scales
$\mu_M \sim M$ and $\mu_{\rm EW} \sim v \simeq v_S$. These are subsequently
matched in one step onto $\SUthreeC \otimes \UoneEM$-invariant phenomenological
EFTs of $|\Delta F|=1,2$ decays, which are valid between $\mu_{\rm EW}$ and
$\mu_b \sim m_b$, where $m_b$ denotes the bottom mass. The coefficients
determined in the process will indicate which operators are the most important.
In principle one could consider an intermediate EFT which is constructed by
integrating out $Z^\prime$ and the new scalars before integrating out top quark,
$W$ and $Z$, but from the point of view of renormalisation group effects,
integrating out all these heavy fields simultaneously appears to be an adequate
approximation.

In this section we present the results from the decoupling of the VLQs that are
important for our phenomenological applications within the framework of the
$\GSMandUpr$-EFTs. The matching step of the $\GSMandUpr$-EFTs to
phenomenological EFT's of $|\Delta F| = 1,2$ processes at the scale
$\mu_{\rm EW}$ is given in \refsec{sec:EFT-down-pheno}. The Lagrangian of the
$\GSMandUpr$-EFT consists of the dimension-four interactions of the light fields
and dimension six interactions generated by the decoupling of VLQs
\begin{align}
  \label{eq:GSM:EFT}
  {\cal L}_{\GSMandUpr-{\rm EFT}} & 
  = {\cal L}_{{\rm dim}-4} + \sum_a {\cal C}_a {\cal O}_a, 
\end{align}
which are invariant under either $\GSM$ or $\GSMUpr$, depending on the model.
Thus in $\GSM$-models ${\cal L}_{{\rm dim}-4}$ coincides with the SM Lagrangian
and the corresponding non-redundant set of operators of dimension six has been
classified in Ref.~\cite{Grzadkowski:2010es}. In $\GSMUpr$-models operators that are
invariant under $\GSMUpr$ must be added, which involve the $Z'$-boson and the
additional scalar singlets and/or doublets. The Wilson coefficients
${\cal C}_a$\footnote{The Wilson coefficients of $\GSMandUpr$-EFTs are denoted
  with calligraphic ${\cal C}_i$, whereas the ones of phenomenological EFTs with
  $C_i$. \label{foot:1}} are effective couplings, which are suppressed by
$1/M^2$ and their effects on observables by $v_i^2/M^2$ compared to the SM,
with $v_i = (v, v_1, v_S)$ depending on the model. They are
determined at the scale $\mu_M$ when decoupling VLQs. The decoupling proceeds
either by explicit matching calculations starting at tree-level and including
subsequently higher orders or by integrating them out in the path integral
method \cite{delAguila:2000rc}. The tree-level decoupling has been known for a
long time for $\GSM$ models \cite{delAguila:2000rc} and is given for
$\GSMUpr(S)$ models in Ref.~\cite{Altmannshofer:2014cfa}.

Within the EFT, RG equations allow to evolve the Wilson
coefficients from $\mu_M$ down to $\mu_{\rm EW}$.In leading logarithmic
approximation and retaining only the first logarithm~(1stLLA)
it has the approximate solution
\begin{align}
  \label{eq:SMEFT-RGE}
  {\cal C}_a (\mu_{\rm EW}) &
  =  \left[ \delta_{ab} 
     - \frac{\gamma_{ab}}{(4 \pi)^2} \ln \frac{\mu_M}{\mu_{\rm EW}} \right]
     {\cal C}_b (\mu_M)\,,
\end{align}  
which holds as long as the second term remains small compared to the first.
The anomalous dimension matrices (ADM) $\gamma_{ab}$
depend in general on couplings of the gauge, Yukawa and scalar sectors and are
known for the $\GSM$-EFT \cite{Jenkins:2013zja, Jenkins:2013wua,
  Alonso:2013hga}.  Largest contributions might be expected for the case of
$\gamma_{ab} \propto Y_u^\dagger Y_u^{} \sim y_t^2$ mixing due to the top-quark
Yukawa coupling $y_t \sim 1$ of the order of a few percent in the case of
self-mixing ($a = b$) and from the mixing due to QCD under $\alpha_s$. On the
other hand, for $a\neq b$ non-zero Wilson coefficients can be generated at 
1stLLA order.\footnote{\label{footnoteX} Note that the 1stLLA neglects 
``secondary mixing'' effects that are present in LLA, i.e. summing all large
logarithms, because although operator ${\cal O}_A$ might not have ADM entry 
with operator ${\cal O}_B$ (no ``direct mixing''), it can still contribute
to the Wilson coefficient $C_{B}(\mu_{\rm EW}$), if it mixes directly with 
some operator ${\cal O}_C$ that in turn mixes directly into ${\cal O}_B$.}
In particular, as we will see below, in the case of models with
right-handed neutral currents left-right operators can be generated in this
manner with profound direct impact on $|\Delta F|=2$ transitions, 
thereby affecting the predictions for $|\Delta F|=1$ observables.

The VLQs have a very limited set of couplings to light fields, which are either
via gauge interactions \refeq{eq:Lag:kin:VLQ} to the gauge bosons or via Yukawa
interactions \refeq{eq:Yuk:H}--\refeq{eq:Yuk:Phi} to light --- w.r.t. to VLQ
mass $M$ --- SM quarks and scalars $\varphi = H, S$ or $\Phi$, depending on the
model. At tree-level, this particular structure of interactions can give rise
only to flavour-changing $Z$ and $Z'$ couplings, whereas all other decoupling
effects are loop-suppressed \cite{Arzt:1994gp}.

The decoupling of the VLQs proceeds in the unbroken phase of
$\SUtwoL \otimes \UoneY$, hence quark fields are flavour-eigenstates and neutral
components of scalar fields are without VEV at this stage. After the
RG evolution from $\mu_M$ to $\mu_{\rm EW}$, spontaneous symmetry breaking
will take place within the $\GSMandUpr$-EFTs and the transformation from flavour- 
to mass-eigenstates for fermions and gauge bosons can be performed, accounting for
the dimension six part in Eq.~\refeq{eq:GSM:EFT}.

%
%

\subsection[Tree-level decoupling and $Z$ and $Z'$ effects]{\boldmath
  Tree-level decoupling and $Z$ and $Z'$ effects}
\label{sec:tree-decoupl}

\begin{figure}
  \begin{subfigure}[t]{0.32\textwidth}
    \centering
    \includegraphics[width=0.8\textwidth]{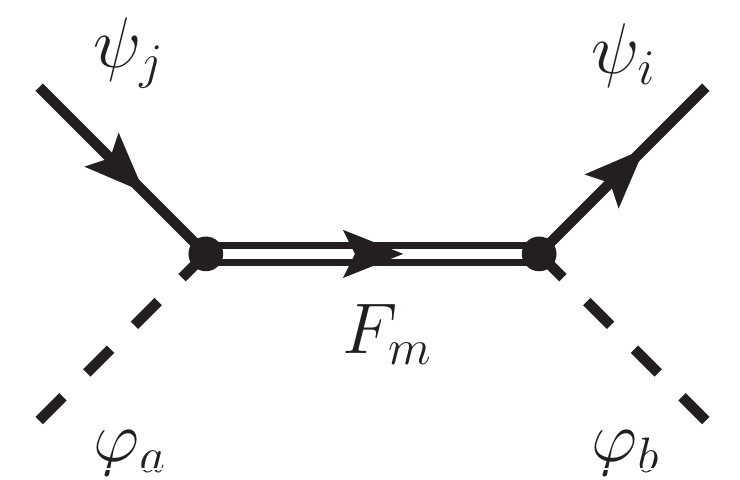}
    \caption{}
    \label{fig:tree-decoupl-1}
  \end{subfigure}
  \begin{subfigure}[t]{0.32\textwidth}
    \centering
    \includegraphics[width=0.8\textwidth]{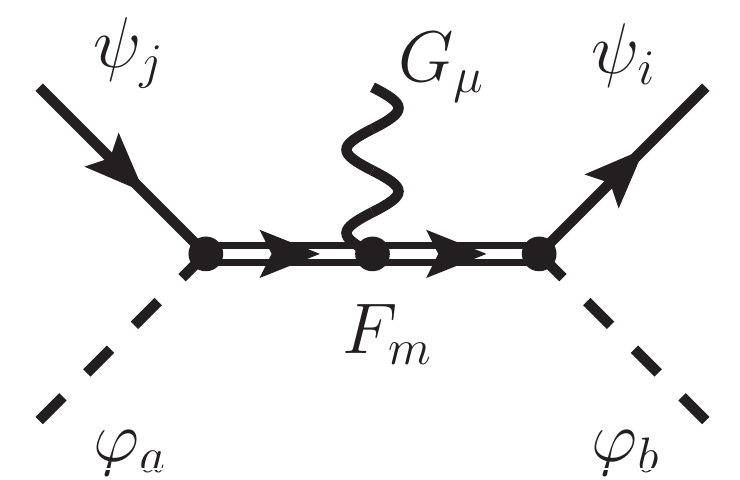}
    \caption{}
    \label{fig:tree-decoupl-2}
  \end{subfigure}
  \begin{subfigure}[t]{0.32\textwidth}
    \centering
    \includegraphics[width=0.8\textwidth]{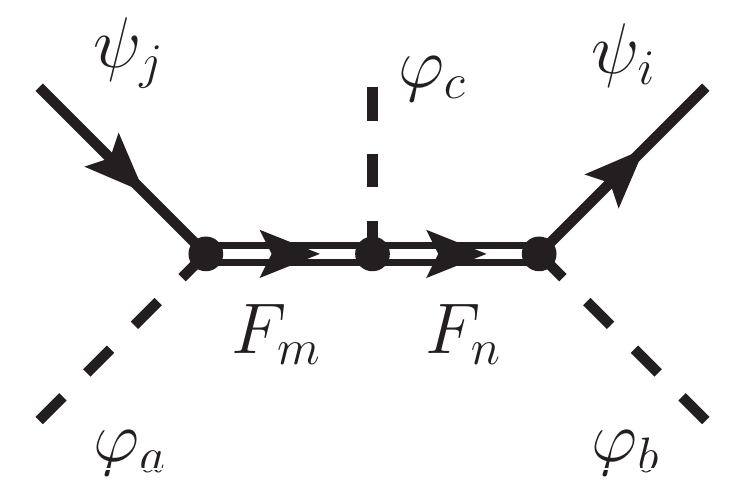}
    \caption{}
    \label{fig:tree-decoupl-3}
  \end{subfigure}
\caption{\small 
  Tree-level graphs (a) and (b) of the decoupling of a VLQ $F_m$
  that give rise to $\psi^2 \varphi^2 D$ operators. They proceed via their
  Yukawa interactions with scalars $\varphi = (H, S, \Phi)$ and SM quarks
  $\psi = (q_L, u_R, d_R)$. The gauge boson $G_\mu$ depends on the
  representation.  Tree-level graph~(c) requires two representations $F_{m,n}$
  with a Yukawa coupling via $\varphi_c$ and give rise to $\psi^2 \varphi^3$
  operators.  
}
\end{figure}

\begin{table}
\centering
\renewcommand{\arraystretch}{1.5}
\resizebox{\textwidth}{!}{
\begin{tabular}{|l|c|l|c|l|c|}
\hline
  \multicolumn{2}{|c|}{$\GSM$}
& \multicolumn{2}{c|}{$\GSMUpr(S)$}
& \multicolumn{2}{c|}{$\GSMUpr(\Phi)$}
\\
\hline \hline
  \multicolumn{6}{|c|}{$\psi^2 \varphi^2 D$}
\\
\hline
  ${\cal O}_{Hq}^{(1)}$
& $(H^\dagger i \overleftrightarrow{\cal D}_{\!\!\!\mu} H) [\bar{q}_L^i \gamma^\mu q_L^j]$
& ${\cal O}_{Sq}$
& $(S^\ast i \overleftrightarrow{\cal D}_{\!\!\!\mu} S) [\bar{q}_L^i \gamma^\mu q_L^j]$
& ${\cal O}_{\Phi q}^{(1)}$
& $(\Phi^\dagger i \overleftrightarrow{\cal D}_{\!\!\!\mu} \Phi) [\bar{q}_L^i \gamma^\mu q_L^j]$
\\
  ${\cal O}_{Hq}^{(3)}$
& $(H^\dagger i \overleftrightarrow{\cal D}^a_{\!\!\!\mu} H) [\bar{q}_L^i \sigma^a \gamma^\mu q_L^j]$
& ---
& ---
& ${\cal O}_{\Phi q}^{(3)}$
& $(\Phi^\dagger i \overleftrightarrow{\cal D}^a_{\!\!\!\mu} \Phi) [\bar{q}_L^i \sigma^a \gamma^\mu q_L^j]$
\\
  ${\cal O}_{Hu}$
& $(H^\dagger i \overleftrightarrow{\cal D}_{\!\!\!\mu} H) [\bar{u}_R^i \gamma^\mu u_R^j]$
& ${\cal O}_{Su}$
& $(S^\ast i \overleftrightarrow{\cal D}_{\!\!\!\mu} S) [\bar{u}_R^i \gamma^\mu u_R^j]$
& ${\cal O}_{\Phi u}$
& $(\Phi^\dagger i \overleftrightarrow{\cal D}_{\!\!\!\mu} \Phi) [\bar{u}_R^i \gamma^\mu u_R^j]$
\\
  ${\cal O}_{Hd}$
& $(H^\dagger i \overleftrightarrow{\cal D}_{\!\!\!\mu} H) [\bar{d}_R^i \gamma^\mu d_R^j]$
& ${\cal O}_{Sd}$
& $(S^\ast i \overleftrightarrow{\cal D}_{\!\!\!\mu} S) [\bar{d}_R^i \gamma^\mu d_R^j]$
& ${\cal O}_{\Phi d}$
& $(\Phi^\dagger i \overleftrightarrow{\cal D}_{\!\!\!\mu} \Phi) [\bar{d}_R^i \gamma^\mu d_R^j]$
\\
  ${\cal O}_{Hud}$
& $(\widetilde{H}^\dagger i {\cal D}_{\!\mu} H) [\bar{u}_R^i \gamma^\mu d_R^j]$
& ---
& ---
& ${\cal O}_{\Phi ud}$
& $(\widetilde{\Phi}^\dagger i {\cal D}_{\!\mu} \Phi) 
   [\bar{u}_R^i \gamma^\mu d_R^j]$
\\
\hline
  \multicolumn{6}{|c|}{$\psi^2 \varphi^3$}
\\
\hline
  ${\cal O}_{uH}$
& $(H^\dagger H) [\bar{q}_L^i u_R^j \widetilde{H}]$
& ${\cal O}_{uS}$
& $(S^\ast S) [\bar{q}_L^i u_R^j \widetilde{H}]$
& ${\cal O}_{u\Phi}$
& $(\Phi^\dagger \Phi) [\bar{q}_L^i u_R^j \widetilde{H}]$
\\
  ${\cal O}_{dH}$
& $(H^\dagger H) [\bar{q}_L^i d_R^j H]$
& ${\cal O}_{dS}$
& $(S^\ast S) [\bar{q}_L^i d_R^j H]$
& ${\cal O}_{d\Phi}$
& $(\Phi^\dagger \Phi) [\bar{q}_L^i d_R^j H]$
\\
\hline
\end{tabular}
}
\renewcommand{\arraystretch}{1.0}
\caption{
  We follow the definitions of \cite{Grzadkowski:2010es} for $\psi^2 \varphi^2 D$
  operators, except for the signs of gauge couplings in the covariant derivatives,
  and $(\psi^2 \varphi^3 + \mbox{h.c.})$ operators  in the case of $\GSM$ models
  and extend them to $\GSMUpr(S)$ and $\GSMUpr(\Phi)$-models ($\varphi = H, S,\Phi$).
  Superindices $i,j = 1, 2, 3$ on quark fields denote the generations. These are all
  operators that could arise from tree-level decoupling of VLQs, depending on the
  model.
}
\label{tab:psi2H2D-ops}
\end{table}

The couplings of the VLQs permit at tree level only a dimension six contribution
from the generic 4-point diagram in \reffig{fig:tree-decoupl-1}. Since its
dimension-five contribution vanishes~\cite{delAguila:2000rc}, it is equivalent
to consider the 5-point diagram \reffig{fig:tree-decoupl-2}, where either
$\SUtwoL$ or $\UoneY$ gauge bosons in $\GSM$-models or in addition a~$\hat{Z}'$
in $\GSMUpr$-models is radiated off the VLQ~\cite{delAguila:2000rc,
  Altmannshofer:2014cfa}.  As a consequence, in $\GSM$- and $\GSMUpr$-models
only operators of the type $\psi^2 \varphi^2 D \propto (\varphi^\dagger i
\overleftrightarrow{\cal D}_{\!\!\!\mu}\, \varphi) [\overline\psi_i \gamma^\mu \psi_j]$
($\varphi = H, S, \Phi$) receive non-vanishing contributions at tree-level,
which are projected in part onto $\psi^2 \varphi^3$-type operators via equation
of motions (EOM) \cite{Buchmuller:1985jz, Grzadkowski:2010es}.
We list the corresponding definitions of the operators in \reftab{tab:psi2H2D-ops},
following the notation of \cite{Grzadkowski:2010es} in the case of the $\GSM$-EFT 
and extending it to $\GSMUpr$-EFTs.

After spontaneous symmetry breaking the $\psi^2\varphi^3$ operators contribute
to the quark masses $m_\psi$ ($\psi = u, d$) at the scale $\mu_{\rm EW}$ via
\begin{align}
  \label{eq:quark-mass-dim-6}
  m_\psi^{ij} & 
  = \frac{v_2}{\sqrt{2}} \left( Y_\psi^{ij} 
     - \frac{v_2^2}{2} {\cal C}_{\psi H}^{ij}
     - \frac{v_S^2}{2} {\cal C}_{\psi S}^{ij}
     - \frac{v_1^2}{2} {\cal C}_{\psi \Phi}^{ij}\right) ,
\end{align}
which allows to substitute Yukawa couplings $Y_\psi$ in terms of measured
$m_\psi$ and new physics parameters ${\cal C}_{\psi^2 \varphi^3} \propto
Y_\psi\, {\cal C}_{\psi^2 \varphi^2 D}$, see \refapp{app:psi2H3-operators}. 
If several representations of VLQs are present in a given model and two of them
$F_{m,n}$ couple to a scalar $\varphi_c$\footnote{As discussed above
  $\varphi_c = H$ in $\GSM$ and $\GSMUpr$-models.} via Yukawa couplings
$\widetilde{\lambda}_{mn}$, a third possibility is allowed at tree-level
depicted in~\reffig{fig:tree-decoupl-3}, which contributes directly to
$\psi^2\varphi^3$ operators and gives rise to flavour-changing neutral
$H \bar{\psi}_i \psi_j$ interactions at tree-level~\cite{delAguila:2000rc}. The
various possibilities for $\GSM$ models, where $\varphi_c = H$, can be found in
\cite{delAguila:2000rc}.

The relation of quark masses to the Yukawa interactions
\refeq{eq:quark-mass-dim-6} includes now also $1/M^2$ contributions. Their
diagonalisation proceeds as usual for the quark fields with the help of
$3\times 3$ unitary rotations in flavour space:
\begin{align}
  \label{eq:trafo-weak-mass}
  \psi_L & \to V^\psi_L \psi_L\,, & \psi_R & \to V^\psi_R \psi_R\,,
\end{align}
implying
\begin{align}
  V^{\psi\dagger}_L m_\psi V^\psi_R & = m_\psi^{\rm diag}, &
  V & = (V^{u}_L)^\dagger V^d_L\,,
\end{align}
with diagonal up- and down-quark masses $m_\psi^{\rm diag}$ and the
unitary quark-mixing matrix $V$. In the limit of vanishing dimension-six contributions,
$V$ will become the Cabibbo-Kobayashi-Maskawa (CKM) matrix of the SM. Throughout
we will assume for down quarks the weak basis in which the mass term $m_d$
is already diagonal, implying $q_L = (V^\dagger u_L, d_L)^T$. This fixes also 
the definition of the Wilson coefficients $\Wc{\psi^2 \varphi^2 D}$ (for more details
see \cite{Aebischer:2015fzz}) and the basis for the VLQ Yukawa couplings 
$\lambda_i^{\rm VLQ}$.

After spontaneous symmetry breaking the $\psi^2\varphi^2 D$ operators give rise
to flavour-changing $Z$ and $Z'$ interactions for fermions $(f = \ell, u, d)$,
which we parametrise as follows:
\begin{align}  
  \label{eq:Zcouplings}
  \mathcal{L}_{\rm VLQ}^{(Z)} & =  
  \bar f^{i} \left[ \Delta_L^{ij}(Z) \, \gamma^{\mu} P_L 
       \,+\, \Delta_R^{ij}(Z)  \, \gamma^{\mu} P_R \right] f^j   Z_{\mu} \,,
\\
  \label{eq:Zprimecouplings}
  \mathcal{L}^{(Z^{\prime})} & =  
  \bar f^{i} \left[ \Delta_L^{ij}(Z^{\prime}) \, \gamma^{\mu} P_L   
       + \Delta_R^{ij}(Z^{\prime}) \, \gamma^{\mu} P_R \right] f^j  \, Z^{\prime}_{\mu} \,.  
\end{align}
For completeness, we provide the matching conditions for the Wilson 
coefficients in \refapp{app:VLQ-decoupl}. We note that RG effects have been neglected
in \refeq{eq:Zcouplings} and \refeq{eq:Zprimecouplings} since they are only due to
self-mixing of $\psi^2 \varphi^2 D$ operators as listed in \refapp{app:SMEFT-ADMs}.

The flavour-diagonal ($i =j$) couplings of leptons to the $Z$ will be set to the
ones of the SM as corrections from NP to them are in $\mathrm{G}_{\rm{SM}}$-models
one-loop suppressed.
This is also the case of $\GSMUpr(S)$ models where $Z$ does not play any role in FCNCs. In
$\GSMUpr(\Phi)$ models modifications of the $Zf\bar{f}$ couplings come from
$Z-Z^\prime$ mixing. These shifts are relevant for leptons in partial widths of
$Z\to \ell\bar\ell$ (see \refapp{app:scalar:S+H+Phi}) and could be of relevance
in electroweak precision tests. In the semi-leptonic $|\Delta F|=1$ FCNCs we
will include them for consistency in $\GSMUpr(\Phi)$ models, although they are
negligible in comparison to other effects.


\subsubsection{\boldmath $\GSM$-models}
\label{sec:Zcoupl:GSM}

In the case of $\GSM$-models, the decoupling of VLQs gives the results for 
$\Delta_{L,R}(Z)$ couplings collected for down-quarks in \reftab{tab:GSM}, 
where
\begin{align}
  \label{eq:Dij}
  \Delta^{ij} &
  \equiv  \frac{\lambda_i^* \lambda_j}{g_Z} \frac{M_Z^2}{M^2} , &
  g_Z &
  \equiv \sqrt{g_1^2 + g_2^2} .  
\end{align}
Except for the sign in the case of $T_u$, our results agree with those in 
\cite{Ishiwata:2015cga}. Furthermore, also non-zero couplings to up-type quarks
arise \cite{Ishiwata:2015cga} but they will not play any role in our paper.

\begin{table}[!htb]
\renewcommand{\arraystretch}{1.3}
\centering{
\begin{tabular}{|c|c||c|c|c|c|c|}
\hline
 Coupling & $q$ 
& $D$ & $Q_V$ & $Q_d$ & $T_d$ & $T_u$ 
\\
\hline \hline
  \multirow{2}{*}{$\Delta^{q_i q_j}_L(Z)$} & $d$ 
& $\Delta^{ij} $ & $0$ & $0$ & $ \Delta^{ij}/2$ & $- \Delta^{ij} $ 
\\
& $u$ 
& $0$ & $0$ & $0$ & $V_{im} \Delta^{mn} (V^\dagger)_{nj}$ & $-V_{im} \Delta^{mn} (V^\dagger)_{nj}/2$   
\\
\hline
 \multirow{2}{*}{$\Delta^{q_i q_j}_R(Z)$} & $d$ 
& $0$ & $-(\Delta^{ij})^*$ & $(\Delta^{ij})^*$ & $0$ & $0$
\\
& $u$ 
& $0$ & $(\Delta^{ij}_u)^*$ & $0$ & $0$ & $0$    
\\
 \hline
\end{tabular}
}
\renewcommand{\arraystretch}{1.0}
\caption{\small
  $\Delta^{q_i q_j}_{L,R}(Z)$ for down- and up-type-quark couplings ($i,j=1,2,3$)
  to the $Z$ boson in $\GSM$-models. Here $V_{ij}$ is the CKM matrix and $\Delta_u =
  \Delta (\lambda^{V_d}_i \to \lambda^{V_u}_i)$, see \refeq{eq:Yuk:H}.
}
\label{tab:GSM}
\end{table}


\subsubsection{\boldmath $\GSMUpr$-models}
\label{sec:Zcoupl:GSMUpr}
 
In the $\GSMUpr$-models, the $(L_{\mu}- L_{\tau})$ symmetry fixes the $Z^{\prime}$
coupling to leptons to be
\begin{align}
  \label{eq:Zpr-lepton-coupl}
  \Delta^{\ell\bar\ell}_{L}(Z^{\prime}) &
  = \Delta^{\ell\bar\ell}_{R}(Z^{\prime}) 
  = \Delta^{\nu_\ell\bar\nu_\ell}_{L}(Z^{\prime}) 
  = g^{\prime} Q'_\ell ,
\end{align}
with $Q'_\ell = \{0, +1, -1\}$ for $\ell = \{e,\mu,\tau\}$. 
Here we have neglected $Z-Z'$ mixing effects existing in
$\GSMUpr(\Phi)$-models. However, for consistency we have to include these
effects in the couplings of the $Z$ to leptons
\begin{align}
  \label{eq:Z-lepton-coupl}
  \Delta^{\ell\bar\ell}_{L}(Z) &
  = - g_Z \left(\frac{1}{2} - s_W^2 \right) + g' Q'_\ell \xi_{ZZ'} , &
  \Delta^{\ell\bar\ell}_{R}(Z) 
  = g_Z s_W^2 + g' Q'_\ell \xi_{ZZ'} ,
\end{align}
to first order in the small mixing angle $\xi_{ZZ'}$ (see
\refapp{app:scalar:S+H+Phi} for details). On the other hand, the gauge couplings
to quarks are model dependent.

In $\GSMUpr(S)$-models the scalar sector of $S$ and $H$ generates only non-zero
quark couplings to $Z'$, whereas in $\GSMUpr(\Phi)$-models the scalar sector of
$S$, $H$ and $\Phi$ gives rise to non-zero couplings of SM quarks to both $Z'$
and $Z$. We define
\begin{align}
  \label{eq:def:Gij-Kij}
  G^{ij} & \equiv
  - \frac{\lambda_i^* \lambda_j}{2 X g^{\prime}}  \frac{M_{Z^{\prime}}^2}{M^2} \,, &
  K^{ij} & \equiv 
  c_{\beta}^2 \frac{\lambda_i^* \lambda_j}{g_Z} \frac{M_Z^2}{M^2} 
  = c_{\beta}^2 \Delta^{ij}\,,
\end{align}
with $ \Delta^{ij}$ defined in Eq.~\refeq{eq:Dij} and the $Z-Z'$ mixing angle 
[see \refeq{eq:ZZp-mixing-angle}]
\begin{align}
  \label{eq:def-rprime}
  \xi_{ZZ'} & \simeq 
    r' \, c_{\beta}^2 \frac{M_Z^2}{M_{Z^{\prime}}^2}\,, &
  r' & \equiv \frac{2 X g^{\prime}}{g_Z} \,.
\end{align}
Here $c_{\beta} \equiv \cos \beta$ is a parameter associated with the scalar
sector (see \refeq{eq:GSMpr-Phi-vev}) of $\GSMUpr(\Phi)$-models, i.e.
$v_1 = v \cos\beta$.  The $\xi_{ZZ'}$ describes $Z-Z^\prime$ mixing, which is
phenomenologically constrained to be small, $\xi_{ZZ'} < 0.1$, due to
constraints from the $Z$-boson mass, $M_Z$, and partial widths
$Z\to \ell\bar\ell$ measured at LEP, as described in more detail in
\refapp{app:scalar:S+H+Phi}. The down- and up-quark couplings to $Z^\prime$ and
$Z$ are collected for these models in \reftab{tab:GSMP}.  We confirm previous
findings \cite{Altmannshofer:2014cfa} for the $\GSMUpr(S)$-models.

We note that the $Z'$ couplings are suppressed/enhanced by the ratio $r'$
w.r.t. the $Z$-couplings. Enhancement takes place for
$2\, g' X > g_Z \approx 0.75$, such that for example $r' \approx 3$ can be
reached with $g' X \approx 1.1$, still within the perturbative regime. The
couplings of $T_d$ and $T_u$ differ just by a sign and factors 1/2. In
distinction to $Z$-contributions in $\GSM$-models, both $Z$- and
$Z'$-contributions in $\GSMUpr(\Phi)$ models decouple with large $\tan\beta$,
see $K^{ij}$ in Eq.~\refeq{eq:def:Gij-Kij}.

\begin{table}[!htb]
\renewcommand{\arraystretch}{1.3}
\centering{
\resizebox{\textwidth}{!}{
\begin{tabular}{|c|c||c|c|c|c|}
\hline
 Model & $q$ & $\Delta^{q_i q_j}_L(Z^\prime)$  &  $\Delta^{q_i q_j}_R(Z^\prime)$  & $\Delta^{q_i q_j}_L(Z)$   &
 $\Delta^{q_i q_j}_R(Z)$  
\\
\hline \hline
  \multicolumn{6}{|c|}{$\GSMUpr(S)$}
\\
\hline
  $D$  &  $d$  &  $0$  &  $(G^{ij})^*$  &  $0$  &  $0$
\\
\cline{2-6}
  \multirow{2}{*}{$Q_{V}$}  &  $d$  &  $G^{ij}$  &  $0$  &  $0$  &  $0$ 
\\
  &  $u$  &  $V_{im}\, G^{mn}\, (V^{\dag})_{nj}$  &  $0$  &  $0$  &  $0$ 
\\
\hline \hline
  \multicolumn{6}{|c|}{$\GSMUpr(\Phi)$} 
\\
\hline
  $D$  &  $d$  &  $-r' K^{ij}$  &  $0$  &  $\left[1 - r' \xi_{ZZ'} \right] K^{ij}$  &  $0$ 
\\
\cline{2-6}
  $Q_d$  &  $d$  &  $0$  &  $-r' (K^{ij})^*$  &  $0$  &  $\left[1 - r' \xi_{ZZ'} \right](K^{ij})^*$
\\
\cline{2-6}
  \multirow{2}{*}{$T_d$} & $d$ & $-r' {K^{ij}}/{2}$ & $0$ & $ \left[1 - r' \xi_{ZZ'} \right] {K^{ij}}/{2}$ & $0$ 
\\
& $u$ & $-r' \, V_{im} K^{mn} (V^{\dag})_{nj}$ & $0$ & $ \left[1 - r' \xi_{ZZ'} \right] V_{im} K^{mn} (V^{\dag})_{nj}$ & $0$ 
\\
\cline{2-6}
  \multirow{2}{*}{$T_u$} & $d$ & $r' K^{ij}$ & $0$ & $-\left[1 - r' \xi_{ZZ'} \right] K^{ij} $ & $0$
\\
& $u$ & $r' V_{im} K^{mn}(V^{\dag})_{nj}/2$ & $0$ & $-\left[1 - r' \xi_{ZZ'} \right] V_{im} K^{mn}(V^{\dag})_{nj}/2$ & $0$
\\
 \hline
\end{tabular}
}
}
\renewcommand{\arraystretch}{1.0}
\caption{\small
  $\Delta^{q_i q_j}_{L,R}(Z^\prime)$ and $\Delta^{q_i q_j}_{L,R}(Z)$ for 
  down- and up-type quark couplings ($i,j=1,2,3$) in $\GSMUpr$-models.
  Here $V_{ij}$ is the CKM matrix.
}
\label{tab:GSMP} 
\end{table}

%
%

\subsection{Decoupling at one-loop level}
\label{sec:loop-decoupl}

\begin{figure}
  \centering
  \begin{subfigure}[t]{0.32\textwidth}
    \centering
    \includegraphics[width=0.8\textwidth]{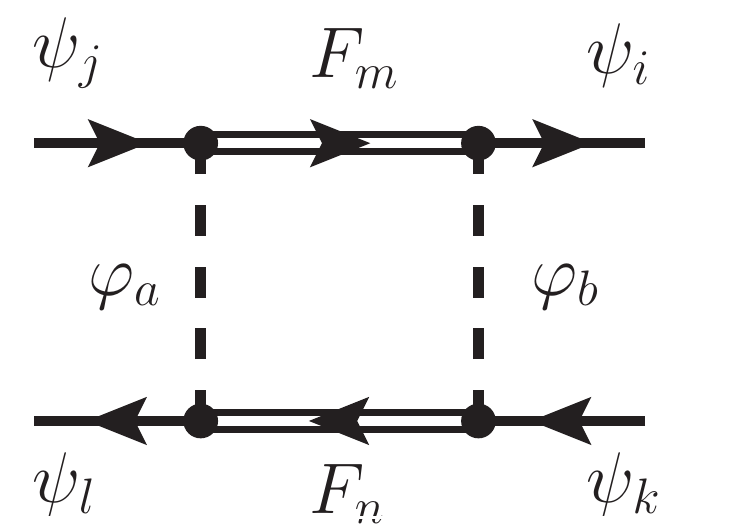}
    \caption{}
    \label{fig:box-decoupl1}
  \end{subfigure}
  \begin{subfigure}[t]{0.32\textwidth}
    \centering
    \includegraphics[width=0.8\textwidth]{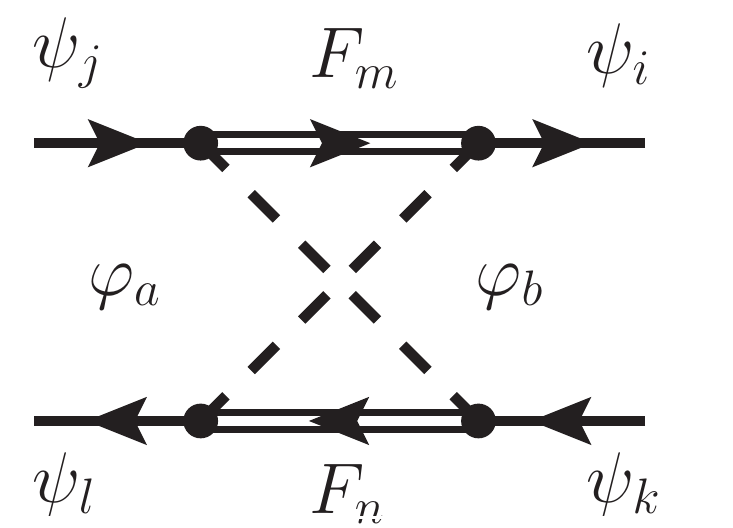}
    \caption{}
    \label{fig:box-decoupl2}
  \end{subfigure}
\caption{\small 
  Box graphs for the decoupling of VLQs in representations
  $F_{m,n}$ due to their Yukawa interactions with scalars
  $\varphi = H, S, \Phi$ and SM quarks $\psi = (q_L, d_R, u_R)$. The crossed
  graph appears for certain representations $F_m \neq F_n$. The
  $|\Delta F| = 2$ graphs are found for $k=j$ and $l=i$.
}
\end{figure}

All other decoupling processes proceed via loops. Those that would lead to
non-canonical kinetic terms in the $\GSMandUpr$-EFTs can be absorbed by a
suitable choice of wave-function renormalisation constants in the full theory
above the scale $\mu_M$, resulting in non-minimal renormalisation of
interactions and giving rise to finite threshold effects of coupling
constants. In $\GSMUpr$-models this is the case for kinetic mixing of $B_\mu$
and $\hat{Z}'_\mu$, which enters our analysis only as a higher order effect.

All other effects enter as dimension six operators.  The ones with four quarks
are most important for quark-flavour phenomenology. They involve only VLQ-Yukawa
interactions, as depicted in \reffig{fig:box-decoupl1} and
\reffig{fig:box-decoupl2}, and give rise to $\psi^4$-type operators, among which
are also $|\Delta F| = 2$ operators.  Here we match directly to the operators
present in the phenomenological EFT of $|\Delta F| = 2$ decays, using the
conventions in \refapp{app:DF2:EFT}, avoiding thereby the intermediate matching
to the $\GSM$-invariant form.\footnote{Note that the set of $\psi^4$-type
  operators is the same in all $\GSMandUpr$ models and a non-redundant set can
  be found in Ref.~\cite{Grzadkowski:2010es}.} Still, we outline this
step for completeness here. In the VLQ models considered, there are four 
relevant $\psi^4$ operators in $\GSMandUpr$-EFTs at the VLQ scale $\mu_M$ and 
a fifth operator is generated due to QCD mixing via RG evolution from $\mu_M$ to 
$\mu_{\rm EW}$. These are the $(\overline{L}L)(\overline{L}L)$ operators
\begin{align}
  \label{eq:GSM-EFT:DF2-op-LLLL}
  [{\cal O}_{qq}^{(1)}]_{ijkl} & 
  = [\bar{q}_L^i \gamma_\mu q_L^j] [\bar{q}_L^k \gamma^\mu q_L^l] , &
  [{\cal O}_{qq}^{(3)}]_{ijkl} & 
  = [\bar{q}_L^i \gamma_\mu \sigma^a q_L^j] [\bar{q}_L^k \gamma^\mu \sigma^a q_L^l] ,
\end{align}
the $(\overline{L}L)(\overline{R}R)$ operators
\begin{align}
  \label{eq:GSM-EFT:DF2-op-LLRR}
  [{\cal O}_{qd}^{(1)}]_{ijkl} & 
  = [\bar{q}_L^i \gamma_\mu q_L^j] [\bar{d}_R^k \gamma^\mu d_R^l] , &
  [{\cal O}_{qd}^{(8)}]_{ijkl} & 
  = [\bar{q}_L^i \gamma_\mu T^A q_L^j] [\bar{d}_R^k \gamma^\mu T^A d_R^l] ,
\end{align}
and the $(\overline{R}R)(\overline{R}R)$ operator
\begin{align}
  \label{eq:GSM-EFT:DF2-op-RRRR}
  [{\cal O}_{dd}]_{ijkl} & 
  = [\bar{d}_R^i \gamma_\mu d_R^j] [\bar{d}_R^k \gamma^\mu d_R^l] ,
\end{align}
with $kl = ij$ for $|\Delta F| = 2$ processes and the $T^A$ denoting $\SUthreeC$ colour
generators. Their Wilson coefficients are matched to the ones of the 
$|\Delta F| = 2$ phenomenological EFT at the electroweak scale $\mu_{\rm EW}$
\cite{Aebischer:2015fzz} as
\begin{equation}
  \label{eq:GSM-DF2-matching}
\begin{aligned}
  C_{\rm VLL}^{ij} & =
     - {\cal N}_{ij}^{-1} \left( \wc[(1)]{qq}{ijij} + \wc[(3)]{qq}{ijij} \right) , &
  C_{\rm VRR}^{ij} & = - {\cal N}_{ij}^{-1} \wc{dd}{ijij} ,
\\
  C_{{\rm LR}, 1}^{ij} & =
   -{\cal N}_{ij}^{-1} \left( \wc[(1)]{qd}{ijij} 
            - \frac{\wc[(8)]{qd}{ijij}}{2 N_c} \right) , &
  C_{{\rm LR}, 2}^{ij} & = {\cal N}_{ij}^{-1} \wc[(8)]{qd}{ijij} ,
\end{aligned}
\end{equation}
where ${\cal N}_{ij}$ is given in \refeq{eq:DF2:norm-factor}.
Here we anticipate this matching to the VLQ scale $\mu_M$ as there are no RG effects 
of phenomenological importance for the discussion of $B$-meson and Kaon sectors. 
For more details see \refsec{sec:GSM-RGE}, where also QCD mixing is given for these
operators. Since the Wilson coefficients of these operators are generated at $\mu_M$
at one-loop, their interplay with other sectors in quark-flavour physics due to
RG mixing are considered higher order and hence beyond the scope of our work.

In $\GSM$-models VLQs contribute to $|\Delta F| = 2$ operators $O_a^{ij}$ for
$a = {\rm VLL, VRR, LR1}$ via box diagrams (see Figs.~\ref{fig:box-decoupl1} and
\ref{fig:box-decoupl2}), which contain two heavy VLQ propagators with
representations $F_m$ and $F_n$ and massless components of the standard doublet
$H = (H^+, H^0)^T$. These box diagrams yield the general structure of the Wilson
coefficients
\begin{align}
  \label{eq:DF2-GSM-wilson}
  {\cal C}_a^{ij} (\mu_M) &
  = \frac{\eta_{mn} }{(4 \pi)^2} 
    \frac{\Lambda_{ij}^m \, \Lambda_{ij}^n}{{\cal N}_{ij}} f_1(M_m, M_n)
\end{align}
at the scale $\mu_M$.  Here the prefactor  corresponds to the SM
normalisation of the $|\Delta F| = 2$ EFT, see \refeq{eq:DF2:norm-factor}. The
function
\begin{align}
  f_1(M_m, M_n) & =
  \frac{\ln (M_m^2/M_n^2)}{M_m^2 - M_n^2}\,, &
  \mbox{with} && 
  f_1(M_m, M_m) & = \frac{1}{M_m^2}\,,
\end{align}
depends on the VLQ masses of representations $F_{m,n}$. The couplings $\Lambda_{ij}^m$ 
are
\begin{align}
  \Lambda_{ij}^m & = (\lambda^m_i)^{\ast} \lambda_j^m &
  \mbox{for} & & 
  F_m & = D, T_d, T_u ,\nonumber
\\
  \Lambda_{ij}^m & = \lambda_i^m (\lambda^m_j)^{\ast}
  & \mbox{for} & &
  F_m & = Q_d, Q_V .\label{eq:Lambda}
\end{align}
The index $a$ of the operator and the numerical factors $\eta_{mn}$ are
collected in \reftab{tab:DF2-GSM-matching}. Note that $a = {\rm VLL}$ for
$F_{m,n} = D,\, T_d,\, T_u$, and $a = {\rm VRR}$ for $F_{m,n} = Q_d,\, Q_V$,
whereas $a = {\rm LR1}$ for $F_m = D,\, T_d,\, T_u$ and $F_n = Q_d,\, Q_V$. The
factors $\eta_{mn}$ are positive except for interference of
$F_m = D,\, Q_d,\, T_d$ with $F_n = Q_V,\, T_u$, because in this case the scalar
propagators are crossed, which gives rise to an additional sign w.r.t. the
diagram with non-crossed scalar propagators. For $F_m = F_n$, these results
agree with \cite{Ishiwata:2015cga} for $D$, $T_u$, $T_d$, but for $Q_d$ (model
XI) we find an additional factor of $2$. Concerning $Q_V$ (model IX) we find a
contribution to $\sim O_{\rm VRR}$ instead of $\sim O_{\rm VLL}$ and also
opposite sign. For completeness we provide also the results for $F_m \neq F_n$.

\begin{table}[!htb]
\renewcommand{\arraystretch}{1.3}
\centering{
\begin{tabular}{|c||c|c|c|c|c|}
\hline
 $(F_m,\, F_n)$
&  $D$  &  $Q_d$  &  $Q_V$  &  $T_d$  &  $T_u$  
\\
\hline \hline
  $D$
& $\rm VLL$, $+1/8$
& $\rm LR1$, $+1/4$
& $\rm LR1$, $-1/4$
& $\rm VLL$, $+1/16$
& $\rm VLL$, $-1/8$
\\
  $Q_d$ & 
& $\rm VRR$, $+1/4$
& $\rm VRR$, $-1/4$
& $\rm LR1$, $+3/8$ 
& $\rm LR1$, $-3/8$
\\
  $Q_V$ & & 
& $\rm VRR$, $+1/4$ 
& $\rm LR1$, $-3/8$  
& $\rm LR1$, $+3/8$ 
\\
  $T_d$ & & & 
& $\rm VLL$, $+5/32$ 
& $\rm VLL$, $-1/8$
\\
  $T_u$ & & & & 
& $\rm VLL$, $+5/32$
\\
\hline
\end{tabular}
}
\renewcommand{\arraystretch}{1.0}
\caption{\small
  The index $a = {\rm VLL, VRR, LR1}$ appearing in Eq.~\refeq{eq:DF2-GSM-wilson} 
  for representations $(F_m,\, F_n)$, followed by corresponding $\eta_{mn}$.
}
\label{tab:DF2-GSM-matching}
\end{table}

In $\GSMUpr(S)$ models we consider only VLQs $D$ and $Q_V$ and their interference
\begin{equation}
  \label{eq:DF2-GSMprS-wilson}
\begin{aligned}
  D & : &
  {\cal C}_{\rm VRR} & = \frac{1}{(4 \pi)^2} 
  \frac{(\lambda^D_i \lambda^{D\ast}_j)^2}{{\cal N}_{ij}} \frac{1}{8 M_D^2} ,
\\
  Q_V & : &
  {\cal C}_{\rm VLL} & = \frac{1}{(4 \pi)^2} 
  \frac{(\lambda^{V\ast}_i \lambda^V_j)^2}{{\cal N}_{ij}} \frac{1}{8 M_V^2} ,
\\ 
  D \times Q_V & : & 
  {\cal C}_{\rm LR1} & = -\frac{1}{(4 \pi)^2} 
  \frac{(\lambda^D_i \lambda^{D\ast}_j)(\lambda^{V\ast}_i \lambda^V_j)}{{\cal N}_{ij}}
  \frac{f_1(M_D, M_V)}{4} ,
\end{aligned}
\end{equation}
which agrees with \cite{Altmannshofer:2014cfa} except for a minus sign from
crossed scalar propagators in the interference term $D \times Q_V$.

The results for $\GSMUpr(\Phi)$ models can be found straight-forwardly from the
ones of the $\GSM$ models, bearing in mind that \refeq{eq:Yuk:H} and
\refeq{eq:Yuk:Phi} are equivalent up to the replacement $H \to \Phi$.

%
%

\subsection{Renormalisation group evolution
  \label{sec:GSM-RGE}
}

The VLQ tree-level exchange in the considered VLQ scenarios generates only 
$\psi^2\varphi^2 D$- and $\psi^2\varphi^3$-type operators at the scale $\mu_M$ 
with nonvanishing Wilson coefficients (see \refapp{app:VLQ-decoupl})
\begin{align}
  \label{eq:GSM-wc-list}
  \GSM : & &  
  \Wc{Hd}, \; \Wc[(1)]{Hq}, \; \Wc[(3)]{Hq}, & & 
  \Wc{uH}, \; \Wc{dH}\,,
\\
  \GSMUpr(S) : &  & 
  \Wc{Sd}, \; \Wc{Sq}, \; & &
  \Wc{uS}, \; \Wc{dS} \,,
\\
  \GSMUpr(\Phi) : & & 
  \Wc{\Phi d}, \; \Wc[(1)]{\Phi q}, \; \Wc[(3)]{\Phi q}, & &
  \Wc{u\Phi}, \; \Wc{d\Phi}\,,
\end{align}
depending on the VLQ scenario.\footnote{\label{footnote:2} We assume
that in the VLQ scenario $Q_V$ the VLQ Yukawa couplings $\lambda^{V_u}_i =0$,
otherwise in this scenario also $\Wc{Hu}$ and $\Wc{Hud}$ must be considered.} 
The RG evolution from $\mu_M$ down to $\muEW$ can induce via operator mixing leading
logarithmic contributions also to other classes of operators in $\GSMandUpr$ EFTs
at the scale $\muEW$. These operators are possibly related to a variety of processes 
and thus imply additional potential constraints. 

The largest enhancements can appear if the ADM $\gamma_{ab}$ in \refeq{eq:SMEFT-RGE} 
is proportional
to the strong coupling $4\pi \alpha_s \sim 1.4$ or the top-Yukawa coupling
$y_t \sim 1$. Note that QCD mixing is flavour-diagonal and hence can not give
rise to new genuine phenomenological effects, i.e. one can not expect qualitative
changes. On the other hand, Yukawa couplings are the main source of flavour-off-diagonal
interactions and we will focus on these here. The $\SUtwoL$ gauge interactions
induce via ADMs $\gamma_{ab} \propto g_2^2$ \cite{Alonso:2013hga} only 
intra-generational mixing between $u_L^i \leftrightarrow d_L^i$ and are parametrically
smaller than $y_t$-induced effects, such that we do not consider them here. The
$\UoneY$ gauge interactions are only flavour-diagonal and numerically even more
suppressed. 

Concerning $\GSMUpr$ models, RG effects due to top-Yukawa couplings are absent
for $\psi^2 \varphi^2 D$ and $\psi^2 \varphi^3$ operators, because $\varphi = S, \Phi$ 
do not have Yukawa couplings to $q_L, u_R, d_R$, which are forbidden by their
additional $\UonePr$ charge. Hence RG effects as discussed below are not present 
in these scenarios.

The ADMs due to Yukawa interactions can be found in \cite{Jenkins:2013wua} for
the $\GSM$-EFT ($\varphi = H$) and we collect the ones involving the Wilson coefficients 
\refeq{eq:GSM-wc-list} in \refapp{app:SMEFT-ADMs}. The RG equations of these Wilson
coefficients are also coupled with those of SM couplings, such as the
quartic Higgs coupling and quark-Yukawa couplings \cite{Jenkins:2013zja},
but in 1stLLA they decouple. The modification of SM couplings due to dim-6 effects
can be neglected when discussing the RG evolution of dim-6 effects themselves 
in first approximation. Moreover, the quartic Higgs coupling is irrelevant for the
processes discussed here and the quark masses are determined from low-energy 
experiments, i.e. much below $\muEW$. Hence phenomenologically most interesting
are RG effects of mixing of $\psi^2 H^2 D$ and $\psi^2 H^3$ operators into
other operator classes that do not receive tree-level matching contributions
at $\mu_M$. Those classes are 
\begin{align}
  H^6 \; (1)\,, \qquad 
  H^4 D^2 \; (2)\,, \qquad 
  \psi^4 \; (5) \,,
\end{align}
where we list in parentheses the number of operators.\footnote{Implying footnote
\ref{footnote:2}.} We focus on the $\psi^4$ operators, which all turn out to
be four-quark operators, because they are most relevant for processes of down-type 
quarks considered here. We comment shortly on the $H^6$ and $H^4 D^2$ classes in 
\refapp{app:SMEFT-ADMs}.

The RG equation \refeq{eq:SMEFT-RGE} implies for a specific $a \in \psi^4$,
see also \cite{Bobeth:2017xry},
\begin{align}
  \label{eq:GSM-RGE-DF2}
  {\cal C}_a (\muEW) &
  = -  \frac{1}{(4 \pi)^2} \ln \frac{\mu_M}{\muEW}
     \sum_{b\, \in\, \psi^2 H^2 D} \gamma_{ab}\; {\cal C}_{b} (\mu_M) ,
\end{align}
where $a\neq b$, such that 1stLLA contributions are \emph{one-loop suppressed} w.r.t
tree-level generated $\psi^2 H^2 D$ contributions. Three of the $\psi^4$ 
operators (${\cal O}_{qq}^{(1,3)}$ and ${\cal O}_{qd}^{(1)}$) can mediate 
down-type quark $|\Delta F|=2$ processes and all five $|\Delta F|=1$ processes,
see again \refapp{app:SMEFT-ADMs}.

The $|\Delta F|=1$ four-quark operators modify directly hadronic $|\Delta F|=1$ 
processes, whereas they enter semileptonic $|\Delta F|=1$ processes only
via additional operator mixing in both SMEFT and phenomenological EFTs, therefore
receiving another suppression in semileptonic processes.
The 1stLLA contribution is a novel effect for $|\Delta F|=2$ processes, where 
it competes with the direct one-loop box contribution in VLQ models discussed
in \refsec{sec:loop-decoupl}. On the other hand, semileptonic and hadronic 
$|\Delta F|=1$ processes are generated directly by $\psi^2 H^2 D$ operators in
the next matching step of $\GSM$ to phenomenological EFTs at $\muEW$ (see
\refsec{sec:EFT-down-pheno} and \reffig{fig:DF1-tree-Z}), which are therefore
enhanced in these processes compared to the 1stLLA contributions 
discussed here. Consequently, the 1stLLA is one-loop suppressed in
VLQ models in hadronic $|\Delta F|=1$ processes, unless the potentially novel
chiral structure of the $\psi^4$ operators enhances a specific hadronic 
observable. We will return to this point in \refsec{sec:matching:ddqq:DF1}.

Under the transformation from weak to mass eigenstates for up-type quarks
\refeq{eq:trafo-weak-mass}
\begin{align}
  \Yuk{u} &
  \; \stackrel{\rm dim-4}{\approx} \;
   \frac{\sqrt{2}}{v} V_L^u m_U^{\rm diag} V_R^{u\dagger}
  \; = \; \frac{\sqrt{2}}{v} V_{\rm CKM}^\dagger m_U^{\rm diag} ,
\end{align}
the corresponding ADMs of $\psi^4$ operators in \refapp{app:SMEFT-ADMs}
transform as
\begin{align}
  [\YukD{u}\Yuk{u}]_{ij} &
  \; = \;\frac{2}{v^2} \sum_k m_k^2 \delta_{ki} \delta_{kj}
  \; \approx \; \frac{2}{v^2} m_t^2 \delta_{3i} \delta_{3j}\,,
\\
  \label{eq:ADM-mass-basis:DF2}
  [\Yuk{u}\YukD{u}]_{ij} & 
  \; = \;\frac{2}{v^2} \sum_{k=u,c,t} m_k^2 V_{ki}^* V_{kj}^{}
  \; \approx \; \frac{2}{v^2} m_t^2 \lambda^{(t)}_{ij}\,,
\end{align}
with up-type quark mass $m_k$ and the definition of CKM-products $\lambda^{(t)}_{ij}$
given in \refeq{eq:def-CKM-product}. Since the ADMs are needed here for the 
evolution of dim-6 Wilson coefficients themselves, we have used tree-level
relations derived from the dim-4 part of the Lagrangian only, thereby 
neglecting dim-6 contributions, which would constitute a dim-8 corrections
in this context. In the sum over $k$ only the top-quark contribution
is relevant ($m_{u,c} \ll m_t$), if one assumes that the unitary 
matrix $V$ is equal to the CKM matrix up to dim-6 corrections.\footnote{
We expect only tiny contributions from $k=c$ in case that $ij = sd$,
for $ij = bd, bs$ such contributions are entirely negligible.} 

The $|\Delta F|=2$ mediating $\psi^4$ operators involve the combination 
\refeq{eq:ADM-mass-basis:DF2}. We obtain via \refeq{eq:GSM-RGE-DF2}
and explicit matching conditions \refeq{eq:GSM:matching:SMEFT:psi2H2D}
\begin{align}
  \label{eq:DF2-1stLLA}
  {\cal C}^{ij}_a (\muEW) &
  = \frac{\kappa_m}{(4\pi)^2} \frac{\Lambda_{ij}^m \lambda^{(t)}_{ij}}{{\cal N}_{ij}} 
    \frac{1}{M^2} \frac{2 m_t^2}{v^2}\ln\frac{\mu_M}{\muEW}\,, 
\end{align}
with $\Lambda_{ij}^m$ from \refeq{eq:Lambda}, the chirality of the $|\Delta F|=2$ 
operator
\begin{equation}
\begin{aligned}
  a & = {\rm VLL}  \qquad\, \mbox{for} \qquad F_m = D, T_d, T_u \,,
\\
  a & = {\rm LR,1} \qquad   \mbox{for} \qquad F_m = Q_d, Q_V\,,
\end{aligned}  
\end{equation}
and the VLQ-model-dependent factor
\begin{align}
  \label{eq:kappam}
  \kappa_m &
  = \left(0,\; -\frac{1}{2},\; +\frac{1}{2},\; -\frac{1}{2},\; +\frac{1}{4} \right)
&
  \mbox{for}&
&
  F_m = (D,\, Q_d,\, Q_V,\, T_d,\, T_u).
\end{align}

We note the relations
\begin{align}
  \kappa_m \frac{\Lambda_{ij}^m}{M^2} &
  = [\Wc[(1)]{Hq} - \Wc[(3)]{Hq}]_{ij} \qquad (F_m = D,\, T_d,\, T_u),
\end{align}
where the relative sign comes from relative signs in \refeq{eq:ADM:qq1} and
\refeq{eq:ADM:qq3} when inserted in \refeq{eq:GSM-DF2-matching} and
\begin{align}
  \kappa_m \frac{\Lambda_{ij}^m}{M^2} &
  =  \wc{Hd}{ij}, \qquad (F_m = Q_d,\, Q_V)\,.
\end{align}

We point out the different flavour structure of the 1stLLA contribution
\refeq{eq:DF2-1stLLA} compared to the one of the direct box-contribution 
\refeq{eq:DF2-GSM-wilson} discussed in the previous section \refsec{sec:loop-decoupl}:
\begin{align}
  {\cal C}^{ij}_a |_{\rm 1stLLA} &
  \; \sim \; \Lambda_{ij} \times \lambda^{(t)}_{ij} , &
  {\cal C}^{ij}_b |_{\rm Box} &
  \; \sim \; (\Lambda_{ij})^2  ,
\end{align}
showing linear versus quadratic dependence on the product of VLQ Yukawa 
couplings~$\Lambda_{ij}$. A detailed comparison of both contributions is given 
in \refsec{sec:Comparison}.

The LLA RG equations of $|\Delta F| = 2$ Wilson coefficients from 
QCD, only \cite{Ciuchini:1997bw, Buras:2000if}, are given as
\begin{equation}
  \label{eq:DF2-RGE:Nf6}
\begin{aligned}
  {\cal C}_{{\rm VLL (VRR)}}(\mu_{\rm EW}) & 
  = \eta_6^{2/7} {\cal C}_{{\rm VLL (VRR)}}(\mu_M) \,,
\\
  {\cal C}_{{\rm LR},1}(\mu_{\rm EW}) & 
  = \eta_6^{1/7} {\cal C}_{{\rm LR},1}(\mu_M)\,,
\\
  {\cal C}_{{\rm LR},2}(\mu_{\rm EW}) & 
  = \frac{2}{3} \left(\eta_6^{1/7} - \eta_6^{-8/7}\right) {\cal C}_{{\rm LR},1}(\mu_M) 
  + \eta_6^{-8/7} {\cal C}_{{\rm LR},2}(\mu_M)\,,  
\end{aligned}
\end{equation}
with $N_f = 6$ denoting the number of active quark flavours and
$\eta_6 = \alpha_s^{(6)}(\mu_M)/ \alpha_s^{(6)}(\mu_{\rm EW})$.  The initial
conditions of ${\cal C}_a^{ij}(\mu_M)$ from box-diagrams are collected
in \refeq{eq:DF2-GSM-wilson} and~\refeq{eq:DF2-GSMprS-wilson}. Note that
${\cal C}_{{\rm LR},2}(\mu_M) = 0$, and ${\cal C}_{{\rm LR},1}(\mu_M) \neq 0$
only in the presence of several VLQ representations.

%
%
%

\section{Implications for the down-quark sector}
\label{sec:EFT-down-pheno}

In the previous section the decoupling of the VLQs at tree-level and for
$|\Delta F| = 2$ at one-loop level at the scale $\mu_M$ has been presented,
including the most important effects from the RG evolution down to the
electroweak scale $\mu_{\rm EW}$. In this section we discuss the decoupling of
degrees of freedom of the order of $\mu_{\rm EW}$ by matching onto
phenomenological $|\Delta F| = 1, 2$ EFTs. In the $\GSM$-models these are the
$W$ and $Z$ bosons, the top-quark and the standard Higgs $h^0$ that are all in
the mass range $\mu_{\rm EW} \in [80,\, 180]$~GeV.  In $\GSMUpr$ models
the $Z'$ and additional scalars are present, which we allow to be 
heavier, up to the $\sim 1$~TeV range. For the purpose of the decoupling, however,
we ignore this hierarchy with the heavy standard sector $\sim 100$~GeV.

In our analysis we will frequently use general formulae for flavour observables
in models with tree-level neutral gauge boson exchanges that are collected in
\cite{Buras:2012jb}. These formulae were given in terms of the so-called master
one-loop functions which have been already used before in many concrete
extensions of the SM, see \cite{Buras:2010wr} for a review. Therefore our task
is to calculate NP contributions to these functions in the VLQ models, using the
results obtained in the previous section. To this end it will be useful to adopt
the notations of \cite{Buras:2012jb, Buras:2010wr}.

We define the relevant CKM factors by\footnote{This notation differs
  sufficiently from the one for Yukawa couplings $\lambda_i$ so that there
  should not be any problem in distinguishing them.}
\begin{align}
  \label{eq:def-CKM-product}
  \lambda_{ij}^{(U)} & = V_{Ui}^* V_{Uj}^{} &
  \mbox{with} &  &  U & \in \{u,c,t\}\quad\mbox{and}\quad i,j\in\{d,s,b\}\,.
\end{align}
We introduce further
\begin{align}
  \label{gsm}
  g_{\text{SM}}^2 & =
  4 \frac{G_F^2 M_W^2}{2 \pi^2} = 1.78137\times 10^{-7} \gev^{-2}\,.
\end{align}
The relevant master functions in the SM are
\begin{align}
  S_0(x_t),\qquad X_0(x_t), \qquad Y_0(x_t), \qquad  Z_0 (x_t) \,.
\end{align}
They are {\it flavour universal} and {\it real valued}. For completeness their
explicit expressions can be found in the appendices. In the considered VLQ
models new contributions not only break flavour universality, but also bring in
new CP-violating phases, so that minimal flavour violation (MFV) is violated.

%
%

\subsection[$|\Delta F| = 2$]{\boldmath $|\Delta F| = 2$}

\begin{figure}
  \centering
  \begin{subfigure}[t]{0.32\textwidth}
    \centering
    \includegraphics[width=0.8\textwidth]{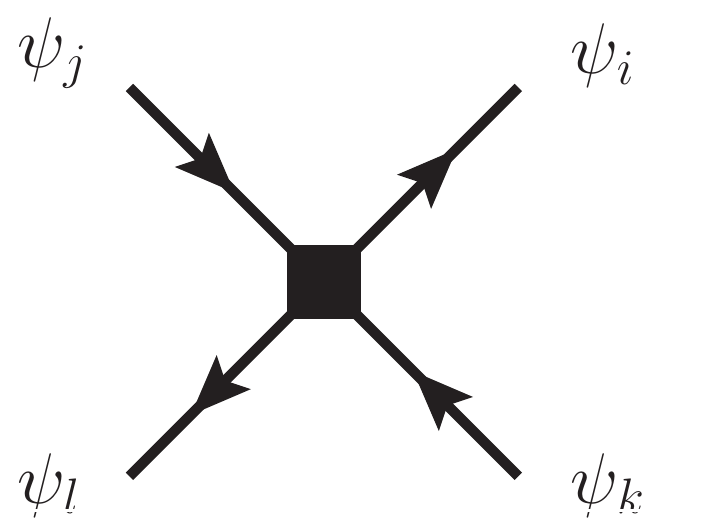}
    \caption{}
    \label{fig:psi4-local}
  \end{subfigure}
  \begin{subfigure}[t]{0.64\textwidth}
    \centering
    \includegraphics[width=0.4\textwidth]{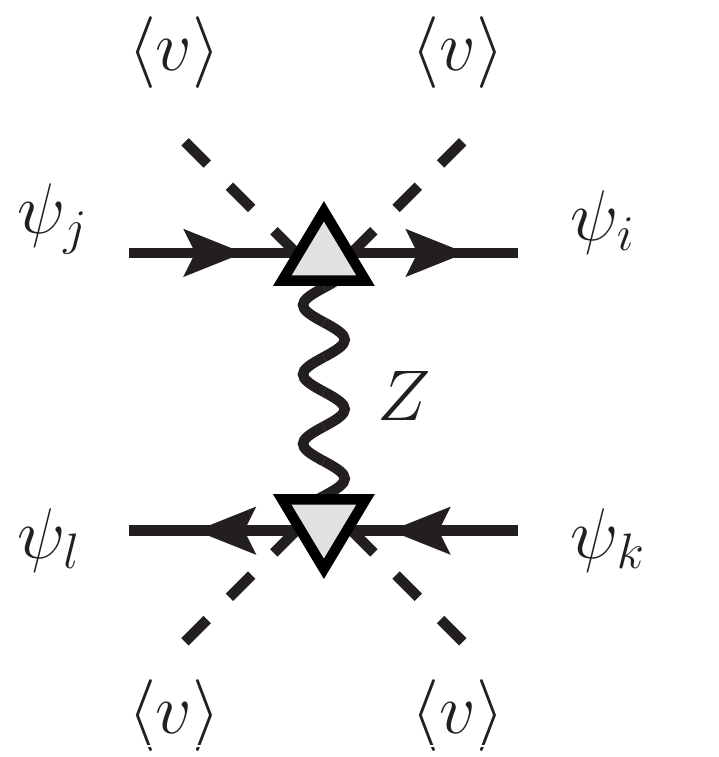}
    \hskip 0.05\textwidth
    \includegraphics[width=0.4\textwidth]{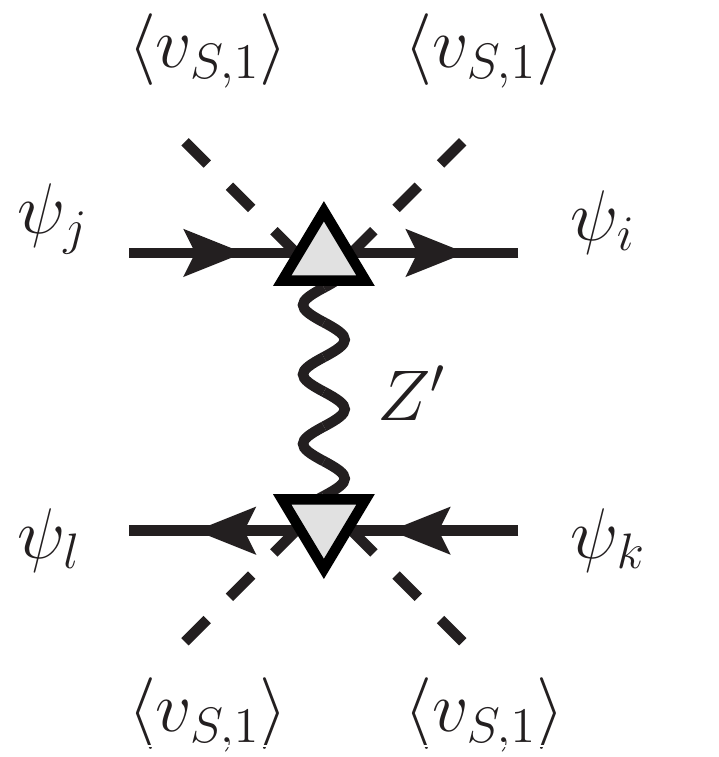}
    \caption{}
    \label{fig:psi4-bilocal}
  \end{subfigure}
\caption{\small 
  \reffig{fig:psi4-local} shows flavour-changing four-quark
  transitions in the $\GSMandUpr$-EFT that are mediated by local
  $\psi^4$-operators, generated at the scale $\mu_M$ at one-loop level
  (indicated by the filled square).  \reffig{fig:psi4-bilocal} shows
  contributions from double insertions of $\psi^2 \varphi^2 D$-operators via
  intermediate $Z$ or $Z'$ exchange, which are formally of higher power, but
  are generated by tree-level VLQ exchange (indicated by the triangles).
}
\label{fig:psi4}
\end{figure}

The Wilson coefficients\footnote{See footnote \ref{foot:1}.} of $|\Delta F| = 2$
operators governing neutral kaon and $B_q$-meson mixing ($q=d,s$), defined in
\refapp{app:DF2:EFT}, can receive at the scale $\mu_{\rm EW}$ several
contributions depicted in \reffig{fig:psi4}, depending on the model. Firstly,
there are the local contributions, \reffig{fig:psi4-local}, from the one-loop
decoupling presented in \refsec{sec:loop-decoupl}, which are formally of order
$v^2/M^2$, but one-loop suppressed. Secondly, there are also local
1stLLA contributions in $\GSM$ models due to top-Yukawa RG effects from 
$\psi^2 H^2 D$ operators presented in \refsec{sec:GSM-RGE}, which are formally of 
order $v^2/M^2 \ln(v/M)$ and also one-loop suppressed. Thirdly, there are 
double-insertions of flavour-changing $Z^{(\prime)}$ couplings, \reffig{fig:psi4-bilocal},
that count due to the double insertion formally as $v^4/M^4$, but are generated
already at tree-level. Fourthly, when considering several VLQ 
representations also double-insertions of $\psi^2 \varphi^3$-type operators
\cite{delAguila:2000rc}, generating flavour-changing neutral Higgs exchange,
can contribute in analogy to \reffig{fig:psi4-bilocal} when replacing the
$Z^{(\prime)}$ by $h^0$. As a consequence in this case also non-vanishing 
contributions can arise to the operators $O_{{\rm S\chi\chi},1}$ with $\chi =L,R$ 
and $O_{{\rm LR},2}$ \cite{Aebischer:2015fzz}.

Unless we consider several VLQ representations simultaneously, new physics
contributions from box diagrams, the top-Yukawa generated 1stLLA 
contributions in LH $\GSM$ models and the double-insertions of flavour-changing
$Z^{(\prime)}$-couplings involve only the operators $O_{{\rm VLL}}^{ij}$ and
$O_{{\rm VRR}}^{ij}$.  Below $\muEW$, they obey the same RG evolution 
\refeq{eq:DF2-RGE:Nf6}
--- with appropriate change of number of active quark flavours $N_f = 6 \to 5$
--- and enter the $M_{12}$ element of the mass-mixing matrix as the linear
combination
\begin{align}
  \label{eq:Seff}
  \left[C_{\rm VLL}^{ij} + C_{\rm VRR}^{ij}\right](\mu_{\rm EW}) &
  \equiv S_{ij} = S_0(x_t) + \Delta S_{ij} 
\end{align}
with $\Delta S_{ij}$ denoting VLQ contributions. The SM contribution is given 
at LO by $S_0(x_t)$, see \refeq{eq:DF2:S0}. We have
\begin{align}
  \label{eq:DF2:S-sum}
  \Delta S_{ij} & 
  = [\Delta S_{ij}]_{\rm VLL} + [\Delta S_{ij}]_{\rm VRR}\,, 
\end{align}
although in a given model only one of these contributions is present. If two
different models containing LH and RH couplings are combined, the most
important transitions in $|\Delta F|=2$ are not these two operators, but
$O_{{\rm LR},1}^{ij}$ and $O_{{\rm LR},2}^{ij}$.

The $[\Delta S_{ij}]_{\rm V\chi\chi}$ with $\chi = L,R$ include quite generally
box diagrams with VLQs  and scalar exchanges, the top-Yukawa generated
1stLLA contributions in LH $\GSM$ models as well as tree-level $Z$ and $Z^\prime$ 
contributions. We can therefore write
\begin{align}
  \label{eq:Zprime1}
  [\Delta S_{ij}]_{\rm V\chi\chi} & 
  = {\cal C}_{\rm V\chi\chi}^{ij}(\mu_{\rm EW})
  + \frac{4 r_Z}{g_{\rm SM}^2 M_Z^2} \left[\frac{\Delta_\chi^{ij}(Z)}{\lambda_{ij}^{(t)} }\right]^2
  + \frac{4 r_{Z^\prime}}{g_{\rm SM}^2 M_{Z'}^2} \left[\frac{\Delta_\chi^{ij}(Z^\prime)}{\lambda_{ij}^{(t)}}\right]^2 ,
\end{align}
where ${\cal C}_{\rm V\chi\chi}^{ij}(\mu_{\rm EW})$ are given by
\refeq{eq:DF2-RGE:Nf6} for $\chi = R$ or the sum of \refeq{eq:DF2-RGE:Nf6} and 
\refeq{eq:DF2-1stLLA} for $\chi = L$.  
The $r_V$ for $V = Z, Z^\prime$ are NLO QCD corrections\footnote{Since we decouple
$Z$ and $Z^\prime$ simultaneously at $\mu_{\rm EW} \sim M_Z$, we do not resum
logarithms between scales $M_{Z^\prime}$ and $\mu_{\rm EW}$ as for example in
Ref.~\cite{Buras:2012dp}.} to \reffig{fig:psi4-bilocal} from decoupling of the
$V$ boson at the scale $\mu = \mu_{\rm EW}$ \cite{Buras:2012fs}, Note the 
model-dependence of the factors $\Delta_\chi^{ij}(Z)$ and $\Delta_\chi^{ij}(Z^\prime)$,
given in \reftab{tab:GSM} and \reftab{tab:GSMP}, and the different dependence on
the VLQ mass of these factors and ${\cal C}_{\rm V\chi\chi}^{ij}(\mu_{\rm EW})$.

The top-Yukawa operator mixing generates in RH $\GSM$ models also LR
operators for a single VLQ representation. When two or more representations are
considered, also LR and SLL (SRR) operators contribute in principle. The Wilson
coefficients of LR operators can receive contributions from box diagrams, 
top-Yukawa generated RG effects and tree-level $Z^{(\prime)}$ exchanges, whereas SLL
(SRR) and LR,2 from tree-level $h^0$ exchange. The results for all box contributions
${\cal C}^{ij}_{{\rm LR},1}$ are given in formulae \refeq{eq:DF2-GSM-wilson} and
\refeq{eq:DF2-GSMprS-wilson} and the RG evolution in \refeq{eq:DF2-RGE:Nf6},
to which the top-Yukawa generated 1stLLA contributions \refeq{eq:DF2-1stLLA}
have to be added in RH $\GSM$ models. Adding the $Z$- and $Z^\prime$-contributions,
one arrives at
\begin{equation}
\begin{aligned}
  C_{{\rm LR},1}^{ij}(\mu_{\rm EW}) &
  = {\cal C}^{ij}_{{\rm LR},1}(\mu_{\rm EW})
  + \frac{1}{{\cal N}_{ij}} \left[
    \frac{\Delta^{ij}_L(Z) \Delta^{ij}_R(Z)}{M_{Z}^2}  
  + \frac{\Delta^{ij}_L(Z^\prime) \Delta^{ij}_R(Z^\prime)}{M_{Z^\prime}^2} \right],
\\
  C_{{\rm LR},2}^{ij}(\mu_{\rm EW}) &
  = {\cal C}^{ij}_{{\rm LR},2}(\mu_{\rm EW})\,,
\end{aligned}
\end{equation}
with the couplings $\Delta^{ij}_{\chi}(Z^{(\prime)})$ ($ \chi=L,R$) collected in
\reftab{tab:GSM} and \reftab{tab:GSMP}. ${\cal N}_{ij}$ is defined in
\refeq{eq:DF2:norm-factor}.

The RG evolution from $\mu_{\rm EW}$ to $m_b$ is done at NLLA accuracy 
for the SM contribution and LLA accuracy for the VLQ contribution.

%
%

\subsection[$|\Delta F| = 1$: Semi-leptonic  $d_j \to d_i + (\ell\bar\ell, \nu\bar\nu)$]
{ \boldmath 
  $|\Delta F| = 1$: Semi-leptonic  $d_j \to d_i + (\ell\bar\ell, \nu\bar\nu)$
  \label{sec:MF1}
}

\begin{figure}
  \centering
  \begin{subfigure}[t]{0.64\textwidth}
    \centering
    \includegraphics[width=0.4\textwidth]{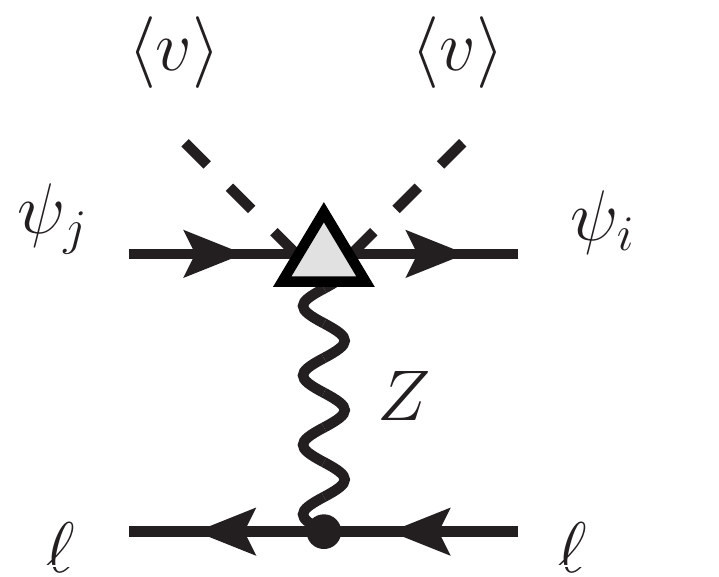} 
    \hskip 0.05\textwidth
    \includegraphics[width=0.4\textwidth]{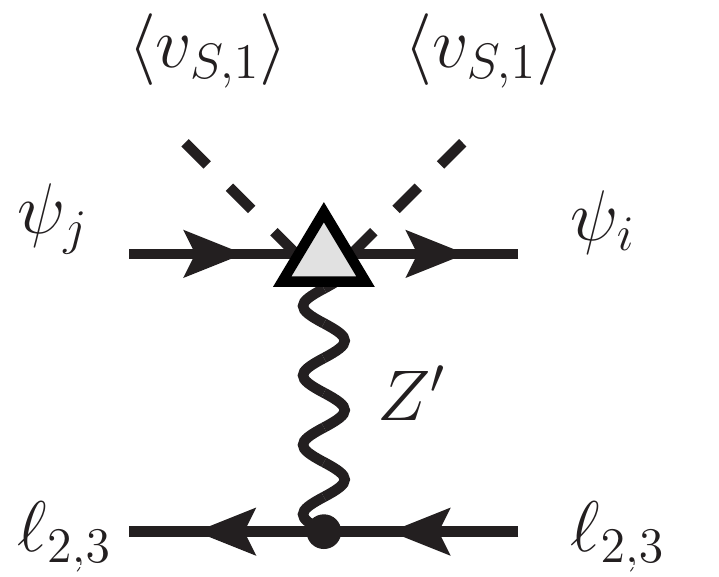}    
    \caption{}
    \label{fig:DF1-tree-Z-semilep}
  \end{subfigure}
  \begin{subfigure}[t]{0.32\textwidth}
    \centering
    \includegraphics[width=0.8\textwidth]{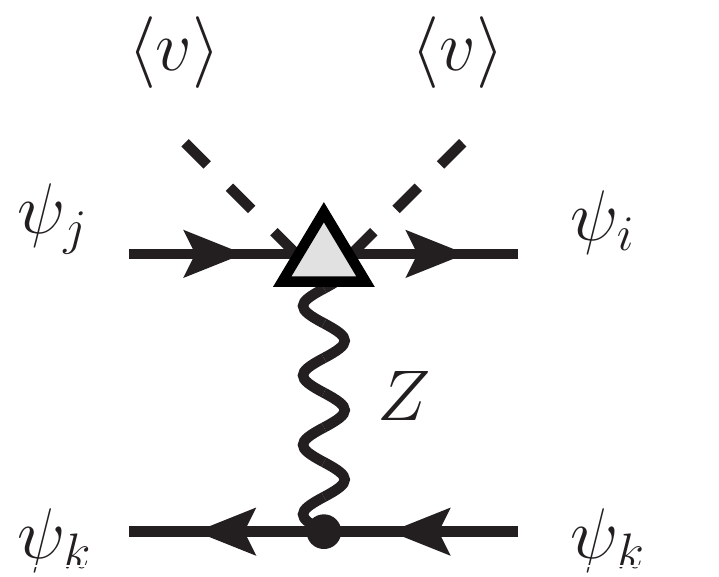}
    \caption{}
    \label{fig:DF1-tree-Z-hadr}
  \end{subfigure}
\caption{\small 
  Flavour-changing $\Delta F = 1$ processes that are mediated in
  the $G_{\rm SM}^{(\prime)}$-EFTs by dimension six
  $\psi^2 \varphi^2 D$-operators (indicated by the triangle), which are in
  turn generated at the scale $\mu_M$ at tree level. Semileptonic transitions
  $\psi_j \to \psi_i\, \ell\bar\ell$ in \reffig{fig:DF1-tree-Z-semilep} can be
  mediated by both $Z$ and $Z'$ exchange, depending on the model. Not shown
  are analogous transitions $\psi_j \to \psi_i\, \nu\bar\nu$. Note that the
  $Z'$ couples only to the second and third generations of leptons and
  neutrinos. Hadronic transitions
  $\psi_j \to \psi_i \, \psi_k \overline{\psi}_k$ in
  \reffig{fig:DF1-tree-Z-hadr} are mediated only by $Z$ exchange, with
  $\psi = (q_L, u_R, d_R)$, depending on the operator.
  \label{fig:DF1-tree-Z}
}
\end{figure}

Semileptonic decays in the down-quark sector receive in VLQ models contributions
via the $Z$ and $Z'$ tree-level exchanges depicted in
\reffig{fig:DF1-tree-Z-semilep}.  They lead to modifications of the Wilson
coefficients of the corresponding phenomenological EFTs of
$d_j\to d_i \nu\bar\nu$ and $d_j\to d_i \ell\bar\ell$ decays given in
\refapp{app:d->dvv} and \refapp{app:d->dll}, respectively. All Wilson
coefficients in this section are formally at $\mu_{\rm EW}$, but since the
corresponding operators are conserved currents under QCD, the RG evolution to
the scale $\mu_b$ is trivial in all cases.\footnote{The usual mixing of $Q_9$
  operators with current-current operators $Q_{1,2}$ present in the SM and
  affecting $C_9$ coefficient is fully negligible here because NP contributions
  to $C_{1,2}$ are tiny in all models.}

The $V=Z,Z'$ contributions modify the Wilson coefficients and one-loop functions  
\begin{align}
  \label{eq:X_LR}
  C_{L\,(R)}^{ij,\nu} &
  = - \sum_{V} \frac{X_{L\,(R)}^{ij,\nu}(V)}{s_W^2}, &
  X_{L\,(R)}^{ij,\nu} (V) &
  = \frac{\Delta_L^{\nu\bar\nu}(V)}{g^2_{\rm SM} M_{V}^2}
    \frac{\Delta_{L\,(R)}^{ij}(V)}{\lambda_{ij}^{(t)}},
\end{align}
which enter the expressions for $d_j\to d_i \bar\nu\nu$ decays like $\kpn$,
$\klpn$ and also $B\to K^{(*)}\nu\bar\nu$ with more details in
\refapp{app:d->dvv}.

The Wilson coefficients of the operators entering the $d_j\to d_i \ell\bar\ell$
transitions receive the following contributions
\begin{align}
  \label{eq:DF1:bsll-C9}
  C_{9\,(9')}^{ij,\ell} &
  = - \sum_V \frac{\big[\Delta_R^{\ell\bar\ell}(V) + \Delta_L^{\ell\bar\ell}(V)\big]}
     {s_W^2 g_{\text{SM}}^2 M_{V}^2} \frac{\Delta_{L\,(R)}^{ij}(V)} {\lambda_{ij}^{(t)}} ,
\\
  \label{eq:DF1:bsll-C10}
  C_{10\,(10')}^{ij,\ell} & 
  = - \sum_V \frac{\big[\Delta_R^{\ell\bar\ell}(V) - \Delta_L^{\ell\bar\ell}(V)\big]}
     {s_W^2 g_{\text{SM}}^2 M_{V}^2} \frac{\Delta_{L\,(R)}^{ij}(V)}{\lambda_{ij}^{(t)}},
\end{align}
where the leptonic $Z$ couplings are taken to be the ones of the SM except for
$\GSMUpr(\Phi)$ models, where $Z-Z'$ mixing is included following
\refeq{eq:Z-lepton-coupl}. There are no $Z'$ contributions to $C_{10\,(10')}$,
as the lepton couplings are vectorial, see \refeq{eq:Zpr-lepton-coupl}.

The purely leptonic decay $K_L \to \mu\bar\mu$ is described by ($\bar{s} \to \bar{d}$) 
\begin{align}
  \label{eq:YAK}
  Y_{\rm A}(K) & = Y_L^{\rm SM}
  + \frac{\left[\Delta_R^{\mu\bar\mu}(Z) - \Delta_L^{\mu\bar\mu}(Z) \right]}
         {g_\text{SM}^2 M_{Z}^2}
  \left[\frac{\Delta_L^{sd}(Z) - \Delta_R^{sd}(Z)}{\lambda_{sd}^{(t)}}\right] ,
\end{align}
with $Y_L^{\rm SM} = 0.942$ \cite{Bobeth:2013tba}.

%
%

\subsection[$|\Delta F| = 1$: Hadronic $d_j \to d_i q\bar{q}$ and $\epe$]
{\boldmath $|\Delta F| = 1$: Hadronic $d_j \to d_i q\bar{q}$ and $\epe$
  \label{sec:matching:ddqq:DF1}
}

Purely hadronic flavour-changing decays $d_j \to d_i q\bar{q}$ receive in the
considered VLQ models predominantly contributions from $Z$ exchange depicted in
\reffig{fig:DF1-tree-Z-hadr}. Other contributions from scalar boxes,
\reffig{fig:psi4-local}, or double-insertions of $Z$ or $Z'$ exchange in
\reffig{fig:psi4-bilocal} are either loop- or power-suppressed.  The
phenomenological EFT of these transitions is given in \refapp{app:d->dqq}. Since
the flavour-diagonal $Z$ couplings are given by the SM ones to the order 
we are working in, no dependence on $q$ arises.  The non-vanishing contributions
to the $|\Delta F| = 1$ Wilson coefficients are conveniently rewritten as NP
contributions to the Inami-Lim $Z$-penguin function $C$ (see
Appendices~\ref{app:d->dvv} and \ref{app:d->dll})\footnote{Note that 
whereas the SM contribution to the function $C$ is gauge dependent this shift 
is gauge independent.}
\begin{align}
  C_{L(R)}^{ij} &
  = - \frac{g_Z}{2 g_{\rm SM}^2 M_Z^2} \frac{\Delta_{L(R)}^{ij}(Z)}{\lambda_{ij}^{(u)}}.
\end{align}
It contributes at the scale $\mu_{\rm EW}$ to the Wilson coefficients of the
QCD- and EW-penguin operators \cite{Buras:2015jaq},
\begin{align}
  \label{eq:ddqq-Z-contr}
  C_{3(5')}^{ij} &
  = \frac{\alpha}{6\pi}\frac{C_{L(R)}^{ij}}{s_W^2}, &
  C_{7(9')}^{ij} & 
  = \frac{\alpha}{6\pi} 4 C_{L(R)}^{ij}, &
  C_{9(7')}^{ij} & 
  = - \frac{\alpha}{6\pi} \frac{c_W^2}{s_W^2} 4 C_{L(R)}^{ij}\,. &  
\end{align}
The RG evolution induces also non-vanishing contributions for the remaining QCD-
and EW-penguin operators at lower scales relevant for Kaon and $B$-meson
decays. Here we are mainly interested in CP violation in the Kaon sector,
especially $\epe$.

It is known from various analyses of $\epe$, see \cite{Buras:2015jaq} and
references therein, that NP has to generate contributions to the Wilson
coefficients of $O_8 \propto (V-A) \otimes (V+A)$ or
$O'_{8} \propto (V+A) \otimes (V-A)$ operators at the low energy scale in order
to be able to modify significantly the SM predictions. This requires the
presence of both LH flavour-violating couplings and RH flavour-diagonal
couplings of $Z$ or $Z^\prime$ in the case of $O_8$, or RH flavour-violating
couplings and LH flavour-diagonal couplings in the case of $O_{8'}$. But in the
models considered quark couplings of the $Z^\prime$ are either LH \emph{or} RH,
hence such contributions can only be generated as a higher-order effect. Given
that $(V-A)$ and $(V+A)$ flavour-diagonal $Z$ couplings to SM quarks are always
present, tree-level $Z$ exchanges fully dominate.  NP contributions to
$O_{9,10} \propto (V-A) \otimes (V-A)$ or $O'_{9,10} \propto(V+A) \otimes (V+A)$
operators are negligible due to their suppressed hadronic matrix elements
relative to the ones of $O_8$ and $O'_8$. This can be clearly seen in the
semi-numeric expression \refeq{eq:epe-seminum} for $\epe$, where the
coefficients of $C^{(\prime)}_7$, which mixes into $C^{(\prime)}_8$, is largely enhanced
w.r.t.  all others. Whether $O_8$ or $O'_8$ is generated depends on whether a
given model has $(V-A)$ or $(V+A)$ flavour-violating couplings:
\begin{itemize}
\item Within the $\GSM$- and $\GSMUpr(\Phi)$-models, the pattern of NP
  contributions to $\epe$ is as follows
\begin{equation}
\begin{aligned}
  \mbox{singlets} & : & D & &  & \to &  (O_8)\,,
\\
\label{QVQd}
  \mbox{doublets} & : & Q_V, & & Q_d & \to & (O'_8)\,,
\\
  \mbox{triplets} & : & T_d, & & T_u & \to & (O_8)\,.
\end{aligned}
\end{equation}
\item In $\GSMUpr(S)$-models $\epe$ remains SM-like, which could become
  problematic as we discuss briefly below.
\end{itemize}

Tree-level $Z$ contributions to $\epe$ have been recently considered in detail
in Ref.~\cite{Buras:2015jaq}, where explicit expressions for the relevant hadronic
matrix elements $\langle Q_8(m_c)\rangle_2$ and
$\langle Q^\prime_8(m_c)\rangle_2$ can be found. Whereas these matrix elements
differ only by sign from each other, their Wilson coefficients differ also in
magnitude, the one of $Q^\prime_8$ being larger by a factor of
$c_W^2/s_W^2=3.33$.  This can also be seen in Eq.~\refeq{eq:ddqq-Z-contr},
remembering that the Wilson coefficients of $Q_8$ and $Q^\prime_8$ at $\mu=m_c$
are directly related to the Wilson coefficients of $Q_7$ and $Q^\prime_7$ at
$\mu_{\rm EW}$, respectively.

Finally let us mention that the top-Yukawa generated 1stLLA contributions to
$|\Delta F|=1$ operators in $\GSM$ models discussed in \refsec{sec:GSM-RGE} 
induce operators with the same chiral structure as already present from the
$Z$-exchange due to $\psi^2 H^2 D$ operators. In particular the $\psi^2 H^2 D$
Wilson coefficients generate $\psi^4$ Wilson coefficients
via the mixing given in \refeq{eq:ADM:qq1} -- \refeq{eq:RGE:ud1}
\begin{align}
  \Wc[(1,3)]{Hq} & \;\; \to \;\; \Wc[(1,3)]{qq},\; \Wc[(1)]{qu}
\\
  \Wc{Hd}        & \;\; \to \;\; \Wc[(1)]{qd},\; \Wc[(1)]{ud}
\end{align}
where ${\cal O}^{(1,3)}_{qq} \sim (V-A) \otimes (V-A)$, ${\cal O}^{(1)}_{qu,qd} 
\sim (V-A) \otimes (V+A)$ and ${\cal O}^{(1)}_{ud} \sim (V+A) \otimes (V+A)$.
Given their additional suppression w.r.t. existing contributions we do not 
consider these contributions further.

The status of $\epe$ in the SM can be summarized as follows. The RBC-UKQCD
lattice collaboration calculating hadronic matrix elements of all operators, but
not including isospin-breaking effects, finds \cite{Blum:2015ywa, Bai:2015nea}
\begin{align}
  \label{eq:epe:LATTICE}
  (\epe)_\text{SM} & = (1.38 \pm 6.90) \times 10^{-4}\qquad {\rm (RBC-UKQCD)}.
\end{align}
Using the hadronic matrix elements of QCD- and EW-penguin $(V-A)\otimes (V+A)$
operators from RBC-UKQCD lattice collaboration \cite{Blum:2015ywa, Bai:2015nea}
but extracting the matrix elements of $(V-A)\otimes (V-A)$-penguin operators
from the CP-conserving $K\to\pi\pi$ amplitudes and including isospin breaking
effects, one finds \cite{Buras:2015yba}
\begin{align}
  \label{eq:epe:LBGJJ}
  (\epe)_\text{SM} & = (1.9 \pm 4.5) \times 10^{-4}\qquad {\rm (BGJJ)}\,.
\end{align}
This result differs by $2.9\,\sigma$ from the experimental world average from
the NA48 \cite{Batley:2002gn} and KTeV \cite{AlaviHarati:2002ye, Abouzaid:2010ny}
collaborations,
\begin{align}
  \label{eq:epe:EXP}
  (\epe)_\text{exp} & = (16.6 \pm 2.3) \times 10^{-4} ,
\end{align}
suggesting that models providing enhancement of $\epe$ are favoured. A new
analysis in Ref.~\cite{Kitahara:2016nld} confirms these findings
\begin{align}
  \label{KNT}
  (\epe)_\text{SM} & = (1.1 \pm 5.1) \times 10^{-4}\qquad {\rm (KNT)}\,.
\end{align}

These results are supported by upper bounds on the matrix elements of the
dominant penguin operators from the large-$N_c$ dual-QCD approach
\cite{Buras:2015xba, Buras:2016fys}, which allows to derive an upper bound on
$\epe$ \cite{Buras:2015yba},
\begin{align}
  \label{eq:epe:BoundBGJJ}
  (\epe)_\text{SM} \le (8.6 \pm 3.2) \times 10^{-4} ,
\end{align}
still $2\,\sigma$ below the experimental data. In particular it has been
demonstrated in Ref.~\cite{Buras:2016fys} that final state interactions are much less
relevant for $\epe$ than previously claimed in Refs.~\cite{Antonelli:1995gw,
  Bertolini:1995tp, Frere:1991db, Pallante:1999qf, Pallante:2000hk,
  Buchler:2001np, Buchler:2001nm, Pallante:2001he}. These findings diminish
significantly hopes that improved lattice QCD calculations will be able to bring
the SM prediction for $\epe$ to agree with the experimental data in
\refeq{eq:epe:EXP}, motivating additionally to search for NP models capable of
alleviating this tension.

In fact it has been demonstrated that in general models with flavour-changing
$Z$ and $Z^\prime$ exchanges \cite{Buras:2015yca, Buras:2015jaq}, in the
Littlest Higgs model with $T$-parity \cite{Blanke:2015wba}, 331 models
\cite{Buras:2015kwd, Buras:2016dxz} and supersymmetric models
\cite{Tanimoto:2016yfy, Kitahara:2016otd, Endo:2016aws} agreement with the data
for $\epe$ can be obtained, with interesting implications for other flavour
observables.

We will see in \refsec{sec:numerics} that also in VLQ models large NP
contributions to $\epe$ are possible, such that agreement with the data in
\refeq{eq:epe:EXP} can be obtained with a significant impact not only on rare
$K$ decays but also $B$ decays.

%
%
%

\section{Patterns of flavour violation}
\label{sec:Comparison}

Our analysis involves three model variants $\GSM$, $\GSMUpr(S)$ and
$\GSMUpr(\Phi)$, with up to five VLQ representations.  In this section we
describe the patterns of flavour violation in $|\Delta F|=1,2$ FCNC processes in
the Kaon and $B_{d,s}$-meson sectors that can be expected in these models, based
on our results in Sections~\ref{sec:VLQ-decoupling}
and~\ref{sec:EFT-down-pheno}. The quantitative phenomenology depends in
addition to the NP parameters on the CKM and hadronic ones and will be discussed
in the next section.  However, on the basis of the information collected so far,
some general patterns of flavour violation emerge and it is possible to state
whether in a given model relevant NP contributions to a given observable can be
expected. We hope that the collection of observations below will be useful in
monitoring the numerical analysis of the next section.

%
%

\subsection[$|\Delta F|=2$]
{\boldmath $|\Delta F|=2$}

In all models local VLQ contributions to $|\Delta F|=2$ operators are generated
at the VLQ-scale $\mu_M$ via one-loop box diagrams. The contributions from
tree-level exchanges of $Z$ and $Z'$ at the scale $\mu_{\rm EW}$ are
power-suppressed due to the hierarchy~\refeq{eq:scale-hierarchy-1} and should be
therefore numerically subleading, at least for large VLQ masses.  This property
decouples $|\Delta F|=1$ and $|\Delta F|=2$ contributions to some extent,
rendering it easier to accommodate potential tensions \cite{Bazavov:2016nty,
Blanke:2016bhf} in $\Delta F=2$ processes.

In $\GSM$ models additional contributions from four-fermion operators are generated
through Yukawa RG evolution from  $\mu_M$ to $\mu_{\rm EW}$. In the case of models
$Q_V$ and $Q_d$ these contributions turn out to be dominant for $\mu_M\ge 1\tev$ in
the $K$ meson system and very important in the $B_{d,s}$ meson systems.
In the following we compare the various contributions one by one.

The $|\Delta F|=2$ box contributions given in Eq.~\refeq{eq:DF2-GSM-wilson} and
\refeq{eq:DF2-GSMprS-wilson} depend only on the VLQ mass(es)~$M$ and their
Yukawa couplings $\lambda_i^{\rm VLQ}$, but neither on the gauge couplings nor
on the scalar sector. Moreover for a given VLQ-representation, they are equal in
$\GSM$ and $\GSMUpr(\Phi)$ models owing to the equality of \refeq{eq:Yuk:H} and
\refeq{eq:Yuk:Phi} upon $H \leftrightarrow \Phi$.  Hence the measurements of
$|\Delta F|=2$ observables will result for a given $M$ in the very same
constraints on $\lambda_i^{\rm VLQ}$ in both $\GSM$ and $\GSMUpr(\Phi)$ models.

Using \refeq{eq:Zprime1}, the relative size of box-to-$Z$ exchange in $\GSM$ and
$\GSMUpr(\Phi)$ models is
\begin{align}
  \frac{(\Delta S)_\text{Box}}{(\Delta S)_{Z}} &
  = a\,\eta_{LL}\frac{g_Z^2}{8\pi^2} 
    \left[\frac{\eta_6^{2/7}}{r_Z}\right] \frac{M^2}{M^2_{Z}}
    \times \left\{ \begin{array}{lll}
       1 & \qquad & \GSM
       \\[0.2cm]
       c_\beta^{-4} & \qquad & \GSMUpr(\Phi)
      \end{array} \right. ,
\end{align}
with $\eta_{LL}$ collected in Table~\ref{tab:DF2-GSM-matching}, $r_Z\approx 1$,
and $a=4$ for $T_d$ and unity otherwise. While the $Z$ contribution is
comparable to the box contribution for $M\approx 1-2$~TeV, it amounts only to a
few percent for $M = 10$~TeV in $\GSM$ models, whereas in $\GSMUpr(\Phi)$
models the $Z$-contributions are suppressed by $c_\beta^4$. In $\GSMUpr(\Phi)$
models we have furthermore
\begin{align}
  \frac{(\Delta S)_{Z'}}{(\Delta S)_{Z}} &
  = (r')^2
    \left[\frac{r_{Z'}}{r_Z}\right] \frac{M_Z^2}{M^2_{Z'}} ,
    \qquad \GSMUpr(\Phi)
\end{align}
with $r_{Z'}\approx r_Z\approx 1$. Therefore $Z$ exchange might be more
important w.r.t. the $Z'$ contribution for $M_Z < M_{Z'}$, depending on $r'$,
see \refeq{eq:def-rprime} but both are suppressed w.r.t. the box 
contribution.

In the $\GSMUpr(S)$ models the same picture holds qualitatively, however a
$Z$-exchange is absent and the relative size of box-to-$Z'$ exchange is
different,
\begin{align}
  \frac{(\Delta S)_\text{Box}}{(\Delta S)_{Z'}} &
  = \frac{(X g')^2}{(4\pi)^2}
  \left[\frac{\eta_6^{2/7}}{r_{Z'}}\right] \frac{M^2}{M^2_{Z'}} ,
  \qquad \GSMUpr(S)
\end{align}
which for $X=1$ reduces to the result in Ref.~\cite{Altmannshofer:2014cfa}.  In
contrast to $\GSM$ and $\GSMUpr(\Phi)$ models, we note the particular structure
of $Z^\prime$ couplings, not being suppressed by $M^2_Z/M^2_{Z'}$.  A lower
bound on $|X|\, v_S = M_{Z'}/g' \gtrsim 750$~GeV exists in $\GSMUpr(S)$ models,
mainly from a combination of $Z\to 4 \mu$ and the neutrino trident production
\cite{Altmannshofer:2014cfa}. This implies that only for $M \gtrsim 9$~TeV the
ratio $(\Delta S)_\text{Box} /(\Delta S)_{Z'} \gtrsim 1$ and shows the numerical
importance of the $Z'$ contributions, unless one considers much larger VLQ
masses.

With only these contributions taken into account the $|\Delta F|=2$ observables 
are not sensitive to the chirality of the VLQ interactions as long as only one 
VLQ representation is present, because the contributions are additive as can be 
seen in \refeq{eq:DF2:S-sum}. However, the inclusion of RG Yukawa effects 
and NLO contributions discussed in \cite{Bobeth:2017xry} changes this picture 
drastically in the case of $\GSM$ models with flavour changing RH currents
($Q_d, Q_V$) and has also significant impact in the remaining three models with
LH currents.

In the case of $D$, $T_d$ and $T_u$ models we find
\begin{align}
  \label{eq:DF2-ratio-box-LL}
  \left[\frac{(\Delta S)_\text{RG}}{(\Delta S)_\text{Box}}\right]^{ij} &
  = \frac{\kappa_m}{\eta_{mm}} \frac{\lambda_{ij}^t}{\Lambda^m_{ij}}
  \frac{2 m_t^2}{v^2\eta_6^{2/7}}
  \left[ \ln\frac{\mu_M}{\mu_{\rm EW}} 
       + \frac{F_{\rm NLO}(x_t,\mu_{\rm EW})}{\kappa_m \Lambda^m_{ij}}\right]\,
\end{align}
with $\kappa_m$ given in \refeq{eq:kappam} and $\eta_{mm}$ in 
\reftab{tab:DF2-GSM-matching}. The NLO correction
\begin{equation}
\begin{aligned}
  F_{\rm NLO}(x_t,\mu_{\rm EW}) &
  = \wc[(1)]{Hq}{ij} H_1(x_t,\muEW) - \wc[(3)]{Hq}{ij} H_2(x_t,\muEW)
\\ & 
  + \frac{2 S_0(x_t)}{x_t} \sum_m \left( 
     \lambda_t^{im} \wc[(3)]{Hq}{mj} + \wc[(3)]{Hq}{im} \lambda_t^{mj} \right) 
\end{aligned}
\end{equation}
has been calculated in \cite{Bobeth:2017xry}, where also the $x_t$-dependent 
functions $H_{1,2}$ can be found. The result for $H_1(x_t, \muEW)$ in 
\cite{Bobeth:2017xry} has been confirmed in \cite{Endo:2016tnu}  where NLO 
corrections in the context of a general analysis of $Z$-mediated NP have been
calculated, however in contrast to \cite{Bobeth:2017xry} leaving out RG effects
above the electroweak scale represented by $\ln \mu_M/\muEW$ 
in~\refeq{eq:DF2-ratio-box-LL} and \refeq{eq:DF2-ratio-box-LR}.

In the case of $Q_d$ and $Q_V$ models the box and RG contributions yield
coefficients to different operators, hence a meaningful comparison of their
impact on observables has to include their QCD running between $\muEW$ and the
light flavour scales (we choose 3~GeV for Kaons and $M_B$ for $B_{d,s}$) as well
as the corresponding matrix elements. We find
\begin{align}
  \label{eq:DF2-ratio-box-LR}
  \left[\frac{(M_{12}^*)_\text{RG}}{(M_{12}^*)_\text{Box}}\right]^{ij} 
  = \left[\frac{(M_{12}^*)_\text{LR}}{(M_{12}^*)_\text{VRR}^\text{box}}\right]^{ij} 
  = \frac{\kappa_m}{\eta_{mm}} 
    \frac{\lambda_{ij}^t}{\Lambda^m_{ij}} \frac{2 m_t^2}{v^2\eta_6^{2/7}}
    \left[\ln\frac{\mu_M}{\mu_{\rm EW}} +H_1(x_t,\mu_{\rm EW})\right]\, R^{ij}\,,
\end{align}
with $R^{ij}$ including RG factors and the ratio of the hadronic matrix elements. 
From Eqs. (60) and (61) in \cite{Bobeth:2017xry} we obtain 
\begin{equation}
  R^{sd}\approx -80\qquad\qquad\mbox{and}\qquad\qquad R^{b(d,s)} \approx -3\,.
\end{equation}
This large chiral enhancement in the Kaon system renders the RG contribution
dominant, while in the $B_{d,s}$ systems the contribution remains comparable
with the box contribution.

%
%

\subsection[$|\Delta F|=1$]
{\boldmath $|\Delta F|=1$}

In semi-leptonic $|\Delta F|=1$ processes governed by
$d_j \to d_i + (\ell\bar\ell, \nu\bar\nu)$, the VLQ contributions arise from
tree-level $Z$ exchange in $\GSM$ models, $Z'$ exchange in $\GSMUpr(S)$ models
and both in $\GSMUpr(\Phi)$ models.

It is instructive to begin the discussion with $\GSMUpr(S)$ models considered
already in Ref.~\cite{Altmannshofer:2014cfa}, as they involve only $Z^\prime$
contributions to $\Delta F=1$ processes and the leptonic $Z^\prime$ couplings
have a special structure as given in Eq.~\refeq{eq:Zpr-lepton-coupl}. Moreover, as
pointed out in that paper, the $|\Delta F|=1$ contributions of VLQs in these
models are independent of the scalar- and gauge-sector parameters, in contrast
to $|\Delta F|=2$ contributions that depend on $v_S$. We find the following
pattern in NP contributions:
\begin{itemize}
\item Due to the equality of the LH and RH $Z^\prime$ couplings to leptons in
  \refeq{eq:Zpr-lepton-coupl}, $Z^\prime$ exchange does neither contribute to
  $B_{s,d}\to \mu\bar\mu$ nor to $K_L \to \mu\bar\mu$. If future improved data
  will show the need for NP contributions to $B_{s,d}\to \mu\bar\mu$, this will
  be a problem for this scenario.
\item The crucial virtue of $\GSMUpr(S)$ models, pointed out in
  \cite{Altmannshofer:2014cfa}, is the possibility of solving the LHCb
  anomalies; in particular, they can accommodate violation of lepton-flavour
  universality (LFU).
\item In $B\to K(K^*)\nu\bar\nu$ only small contributions are possible due to
  cancellations among muon and tau contributions when averaging over neutrino
  flavours as a consequence of the $\UonePr$ symmetry.
\item These cancellations are less efficient in $\kpn$ due to interference
  with the charm component, see \refapp{app:d->dvv}.
\end{itemize}

Considering next $\GSM$ and $\GSMUpr(\Phi)$ models in which tree-level $Z$
contributions to $\Delta F=1$ processes dominate, the most notable feature comes
from the tree-level decoupling of the VLQs depicted in
\reffig{fig:tree-decoupl-2}, which implies a relationship between the
flavour-changing $Z$ and $Z'$ couplings in these models, again owing to the
equality of \refeq{eq:Yuk:H} and \refeq{eq:Yuk:Phi} upon
$H \leftrightarrow \Phi$.  Below the scale $\mu_M$ in both models a
$\psi^2 \varphi^2 D$ operator is generated, with the same Wilson coefficient,
where $\varphi = H,\Phi$ in $\GSM$ and $\GSMUpr(\Phi)$ models, respectively.
The covariant derivative is the same in both models, up to the additional
$\UonePr$ part in $\GSMUpr(\Phi)$ models. Upon spontaneous symmetry breaking at
the scale $\mu_{\rm EW}$, this operator becomes $\propto v^2$ in $\GSM$ models
and $\propto v_1^2 = c^2_\beta\, v^2$ in $\GSMUpr(\Phi)$ models.  Consequently,
in $\GSMUpr(\Phi)$ models all $Z$ and $Z'$ couplings
$\propto c^2_\beta \Delta^{ij}$ are suppressed by
$c^2_\beta = (1 + \tan^2\beta)^{-1}$ w.r.t. $Z$ couplings $\propto \Delta^{ij}$
in $\GSM$ models, see \refeq{eq:def:Gij-Kij}, \refeq{eq:Dij} and
\reftab{tab:GSMP}.

Note that the additional modifications from $Z-Z'$ mixing in $\GSMUpr(\Phi)$
models do not affect the dependence on the $\lambda_i^{\rm VLQ}$.  The
suppression by $c^2_\beta$ can be only softened by going to very small
$\tan\beta$. In order to guarantee perturbativity of the top-quark Yukawa
coupling $0.3 \lesssim \tan \beta$ \cite{Branco:2011iw}. In
\refapp{app:scalar:S+H+Phi} we discuss further constraints on $\tan\beta$ in
$\GSMUpr(\Phi)$ models from the measured $Z$ mass and partial widths to leptons,
which for $M_Z < M_{Z'}$ allow at most $2 \lesssim \tan\beta$, \emph{i.e.}
$c^2_\beta \lesssim 0.2$. Depending on the choice of $g'$ and $v_S$, this bound
becomes even stronger.  Therefore, VLQ effects in $|\Delta F|=1$ FCNC processes
are generically suppressed in $\GSMUpr(\Phi)$ models w.r.t. $\GSM$ models. As an
example one might consider the Wilson coefficient $C_9^{ij}$ given in
\refeq{eq:DF1:bsll-C9}, governing $d_j \to d_i \ell\bar\ell$. The suppression
factor in $\GSMUpr(\Phi)$ versus $\GSM$ models is
\begin{align}
  \frac{(C_9^{ij})_{\GSMUpr(\Phi)}}{(C_9^{ij})_{\GSM}} &
  = c^2_\beta \left[ 
    1 - r' \xi_{ZZ'} - \frac{g'}{g_Z} \frac{4 Q'_\ell}{(1 - 4 s_W^2)} \xi_{ZZ'}
  -  \frac{g'}{g_Z} \frac{4 Q'_\ell}{(1 - 4 s_W^2)}\frac{M_Z^2}{M_{Z'}^2}\right] .
\end{align} 
The mixing angle $\xi_{ZZ'} \sim M_Z^2/M_{Z'}^2$ is small in most of the
parameter space, such that $(1 - 4 s_W^2)^{-1} \sim 10$ is overcompensated.  The
comparison of the first three terms with the last one in the brackets also shows
the relative size of the $Z'$ to $Z$ contribution in $\GSMUpr(\Phi)$ models,
which is also suppressed by $M_Z^2/M_{Z'}^2$. Consequently VLQ contributions to
semileptonic $|\Delta F|=1$ FCNC decays are in most cases suppressed in 
$\GSMUpr(\Phi)$ w.r.t. $\GSM$ models.

However, there are exceptions related to the fact that with the parametric 
suppression of the $Z$ and $Z^\prime$ couplings, the values of Yukawa couplings 
are weaker constrained by $\Delta F=1$ transitions than in $\GSM$ models and 
the constraints on Yukawas are governed this time by $\Delta F=2$ processes. 
A detailed numerical analysis in the next section then shows that the allowed 
NP effects in $\Delta M_K$ are in fact significantly larger than in $\GSM$
models.

For a given flavour-changing transition the correlations between different
$|\Delta F|=1$ observables depend on whether $Z^{(\prime)}$ have LH or RH
flavour-violating quark couplings and the size of the corresponding leptonic
$Z^{(\prime)}$ couplings.  A summary is given in \reftab{tab:DNA:bsll-WC}, where in
addition to $\GSM$ and $\GSMUpr(\Phi)$ models we include $\GSMUpr(S)$ models
discussed already above. The generically small NP contributions in
$C_9^{(\prime) ij,\ell}$ compared to $C_{10}^{(\prime)ij, \ell}$ and
$C_{L(R)}^{ij,\nu}$ in $\GSM$ models are due to the smallness of leptonic vector
$Z$ couplings relative to the axial-vector ones. The additional generic
suppression of NP effects in $\GSMUpr(\Phi)$ w.r.t. $\GSM$ is due to the
aforementioned suppression by $c^2_\beta$.

We observe that in $\GSM$ models significant NP effects in $\kpn$, $\klpn$,
$B_{s,d}\to \mu\bar\mu$, $B\to K^{(*)} \mu\bar\mu$ and $B\to K^{(*)}\nu\bar\nu$
are possible, but the LHCb anomalies in angular observables in
$B\to K^* \mu\bar\mu$ cannot be explained in these models because the vector
coupling of $Z$ to muons is suppressed by $(1 - 4 s_W^2) \sim 0.1$ w.r.t. the
axial-vector coupling of the $Z$.  LFU of $Z$ couplings precludes also the
explanation of the violation of this universality in $R_K$, hinted at by LHCb
data.

\begin{table}[t]
\addtolength{\arraycolsep}{4pt}
\renewcommand{\arraystretch}{1.5}
\centering
\begin{tabular}{|l|c|c|c|c|c||c|c||c|c|c|c|}
\hline
  & \multicolumn{5}{|c||}{$\GSM$}
  & \multicolumn{2}{c||}{$\GSMUpr(S)$}
  & \multicolumn{4}{c|}{$\GSMUpr(\Phi)$}
\\
& $D$    & $Q_V$  & $Q_d$  & $T_d$  & $T_u$  & $D$    & $Q_V$  & $D$     & $Q_d$   & $T_d$   & $T_u$
\\
\hline\hline
  $C_9^{ij,\ell}$
& \red   &  ---   &  ---   & \red   & \red   &  ---   & \green & \sred    &  ---    & \sred   & \sred
\\
\hline
  $C_9^{\prime\, ij,\ell}$ 
&  ---   & \red   & \red   &  ---   &  ---   & \green &  ---   &  ---    & \sred   &  ---    &  --- 
\\  
\hline
  $C_{10}^{ij,\ell}$ 
& \green &  ---   &  ---   & \green & \green &  ---   &  ---   & \sgreen &  ---    & \sgreen & \sgreen 
\\
\hline
  $C_{10}^{\prime\, ij,\ell}$ 
&  ---   & \green & \green &  ---   &  ---   &  ---   &  ---   &  ---    & \sgreen &  ---    &  ---  
\\
\hline\hline
  $C_L^{ij,\nu}$
& \green &  ---   &  ---   & \green & \green &  ---   & \green & \sgreen &  ---    & \sgreen & \sgreen
\\
\hline
  $C_R^{ij,\nu}$ 
&  ---   & \green & \green &  ---   &  ---   & \green &  ---   &  ---    & \sgreen  &  ---   &  --- 
\\  
\hline
\end{tabular}
\renewcommand{\arraystretch}{1.0}
\setlength{\tabcolsep}{2pt}
\caption{\small
  ``DNA'' table for NP contributions to the $b \to s \mu^+ \mu^-$ Wilson coefficients
  $C_{9,10}^{(\prime)}$ and to the $d_j \to d_i \nu\bar\nu$ ones $C_{L,R}^{\nu}$.
  \green~means that the NP contribution is potentially large, while \red~stands for a
  generically small contribution, due to the suppressed vector couplings of the $Z$
  to leptons compared to its axial-vector couplings. Smaller symbols in the $\GSMUpr(\Phi)$
  models indicate the general suppression by $c^2_\beta$ w.r.t. $\GSM$ models.
}
\label{tab:DNA:bsll-WC}
\end{table}

Due to the particular structure of $Z^\prime$ couplings, the general pattern of
NP contributions to $\kpn$, $\klpn$, $B_{s,d}\to \mu\bar\mu$,
$B\to K^{(*)} \mu\bar\mu$ and $B\to K^{(*)}\nu\bar\nu$ in $\GSMUpr(\Phi)$ models
is dominated by tree-level $Z$ contributions as in $\GSM$ models, but because of
the aforementioned suppression by $c^2_\beta$ these contributions are 
smaller, with few exceptions mentioned above, than in the latter models.
On the other hand, the presence of $Z^\prime$
with only vector lepton couplings allows in principle to address the LHCb
anomalies more easily; however, given the generic suppression of the $Z^\prime$
couplings, this is harder than in $\GSMUpr(S)$ models. 

Hadronic $|\Delta F|=1$ processes governed by $d_j \to d_i q\bar{q}$ receive VLQ
contributions only from tree-level $Z$ exchange in $\GSM$ and $\GSMUpr(\Phi)$
models. The suppression of VLQ effects by $c^2_\beta$ in $\GSMUpr(\Phi)$ models
w.r.t. $\GSM$ models is the same as discussed previously for semileptonic
$|\Delta F|=1$ processes. Such contributions are entirely absent in $\GSMUpr(S)$
models and $\epe$ is generated for example in the case of $d_j \to d_i d\bar{d}$
either by $Z'$ double insertions or via box diagrams, which are both
additionally suppressed by $|\lambda_d|^2$ compared to $\GSM$ models.

%
%

\subsection[Determination of $M$]
{\boldmath Determination of $M$ \label{sec:mass}}

There is a common claim that from flavour-violating processes it is only
possible to measure the ratio $g_\text{NP}/M_\text{NP}$, where $g_\text{NP}$ is
the coupling present in a given theory, while $M_\text{NP}$ is the NP scale.
The scale tested by a given observable is typically quoted at the value of
$M_{\text{NP}}$ when setting $g_{\text{NP}}=1$, and correspondingly changes when
the latter is suppressed by some mechanism, as in the case of MFV.

Here we would like to point out that in concrete models with correlations between
$|\Delta F|=2$ and $|\Delta F|=1$ processes, it is in general possible to determine
$M_\text{NP}$ without making any assumptions on the couplings involved. This is
in particular important if $M_\text{NP}$ should turn out to be beyond the reach
of direct searches at the LHC.

In the context of 331 models the relevant correlations that allow the
determination of $M_{Z^\prime}$ can be found in Section~7.2 of
\cite{Buras:2012dp}, although this point has not been made there. In order to
illustrate this in the case of VLQ models we consider the $\GSM$-models.
Let us consider the example of $\Delta M_s$ and first take into account
for the shift $\Delta S$ only box contributions with VLQ exchanges.
On the other hand, $\Delta Y$
entering the branching ratio for $B_s\to\mu\bar\mu$ is governed by tree-level
$Z$ exchange. Then we find independently of Yukawa couplings and CKM parameters
a useful formula:
\begin{align}
  \label{eq:DETM}
  \frac{\sqrt{(\Delta S)^*}}{\Delta Y} & 
  = 1.90\,b\,  \sqrt{\eta_{mm}}\left[\frac{M}{10\,\tev}\right], 
    \qquad {(\rm Boxes)},
\end{align}
where $\eta_{mm}$ are given in \reftab{tab:DF2-GSM-matching} and $b=1$ for $D$
and $Q_V$, $b=-1$ for $T_u$ and $Q_d$ and $b=1/2$ for $T_d$.  Note that
$\Delta S$ and $\Delta Y$ are generally complex but their phases are related so
that r.h.s of this equation is real valued.  Extracting $\Delta S$ and
$\Delta Y$ from experiment, a range for $M$ can be determined. 

This formula is modified in the presence of Yukawa RG effects and when the simple
tree-level $Z$ contributions cannot be neglected:
\begin{itemize}
\item
For sufficiently large $M$ the Yukawa RG effects become important. As these
contributions have the same dependence on the couplings as $\Delta F=1$ amplitudes 
and the dependence on the VLQ mass differs only by a logarithm, the determination
of $M$ will not be possible if the RG contribution dominates. However, we expect
this situation only for RH $\GSM$ models in the Kaon sector, as explained above.
If RG and box contributions are comparable, the determination of $M$ will be possible, 
although the relevant expressions will be more involved than \refeq{eq:DETM}.
\item
For sufficiently low $M$ the tree-level $Z$ contributions to $|\Delta F|=2$ 
could become important and again dilute the sensitivity to $M$. However,
if VLQs are not found at the LHC, the value of $M$ is sufficiently large so that
these contributions are numerically irrelevant. 
On the other hand, if VLQs are discovered at the LHC, we will know their
masses and this determination will not be necessary --- instead, the determination of the
couplings would improve. 
\end{itemize}

In summary the determination of $M$ outside the reach of the LHC will depend
on the relevance of box contributions relative to the RG Yukawa effects. Unless
RG contributions are clearly dominant, which is only the case in the Kaon sector
for RH scenarios, this determination should be possible by means of a formula
like \refeq{eq:DETM}. The determination is expected to work best for LH scenarios,
but also for RH scenarios it should remain possible for $b\to d,s$ transitions, as
discussed in the following section.

%
%

\subsection[Kaon and $B$-meson systems]
 {\boldmath Kaon and $B$-meson systems}

\begin{table}[t]
\addtolength{\arraycolsep}{4pt}
\renewcommand{\arraystretch}{1.5}
\centering
\begin{tabular}{|l|c|c|c|c|c||c|c||c|c|c|c|}
\hline
  & \multicolumn{5}{|c||}{$\GSM$}
  & \multicolumn{2}{c||}{$\GSMUpr(S)$}
  & \multicolumn{4}{c|}{$\GSMUpr(\Phi)$}
\\
  & $D$ & $Q_V$ & $Q_d$ & $T_d$ & $T_u$& $D$ & $Q_V$ & $D$ & $Q_d$ & $T_d$ &$T_u$
\\
\hline\hline
$|\Delta F|=2$& \rSt & \bSt & \bSt & \rSt & \rSt & \bSt & \rSt & \rSt & \bSt & \rSt & \rSt
\\
\hline
$B_{s,d}\to\mu\bar\mu$ & \rSt &\bSt  & \bSt & \rSt & \rSt &  &  & \rst & \bst & \rst & \rst
\\
\hline
$B\to K\mu\bar\mu$ & \rSt &\bSt  & \bSt & \rSt & \rSt & \bSt & \rSt & \rst & \bst & \rst & \rst
\\
\hline
$B\to K^*\mu\bar\mu$ &  &  &  &  &  & \bSt & \rSt & \rst & \bst & \rst & \rst
\\
\hline
$B\to K(K^*)\nu\bar\nu$ & \rSt &\bSt  & \bSt & \rSt & \rSt &  &  & \rst & \bst & \rst & \rst
\\
\hline
$\kpn$ & \rSt &\bSt  & \bSt & \rSt & \rSt &  &  & \rst & \bst & \rst & \rst
\\
\hline
$\klpn$ & \rSt & \bSt & \bSt & \rSt & \rSt &  &  &  &  &  & 
\\
\hline
$\epe$ & \rSt & \bSt & \bSt & \rSt & \rSt &  &  & \rst & \bst & \rst & \rst
\\
\hline
\end{tabular}
\renewcommand{\arraystretch}{1.0}
\caption{\small
  ``DNA'' of flavour effects in VLQ models. A star indicates that significant
  effects in a given model and given process are in principle possible, but 
  could be reduced (see \refsec{sec:numerics}) through correlation among several
  observables. Empty space means
  that the given model does not predict sizeable effects in that observable.
  The star \rSt{} indicates left-handed currents and the star \bSt{}
  right-handed ones, smaller stars indicate the suppression of $|\Delta F|=1$
  decays in $\GSMUpr(\Phi)$ models. 
}
\label{tab:DNA}
\end{table}

The correlations between flavour observables in different meson systems are
governed by the Yukawa structure of the model in question, as will be elaborated
quantitatively in \refsec{sec:numerics}. The important property of VLQ models is
that the products defined in Eq.~\refeq{eq:Lambda},
\begin{align}
  \label{eq:LambdaAbsPhi}
  \Lambda_{ij}^m & = |\Lambda_{ij}^m| e^{i\varphi_{ij}^m} ,
\end{align} 
together with the VLQ mass $M$ determine at the same time the flavour-violating
$j\to i$ couplings of $Z$ \emph{and} $Z'$, as well as the flavour-diagonal $Z'$
couplings to quarks.  The relevant flavour-changing parameters are hence
$\Lambda_{ds}^m$ in Kaon decays, and $\Lambda_{db}^m,\,\Lambda_{sb}^m$ in
$b\to d,s$ transitions of $B$ mesons, respectively. Since only the
\emph{relative} phases of the $\lambda_i^{\rm VLQ}$ enter the $\Lambda_{ij}^m$,
the phases $\varphi_{ij}^m$ fulfill the relation
\begin{align}
  \label{eq:phiij}
  \varphi_{bs} & = \varphi_{bd}-\varphi_{sd} ,
\end{align}
dropping the index $m$ of the VLQ representation for convenience. This leaves us
with five parameters for the three complex quantities $\Lambda_{ij}$.  The
phases $\varphi_{ij}$ can vary in the full range $[-\pi,\pi]$, implying the
occurrence of discrete ambiguities when determining them from experiment, as
explicitly seen in the plots in Ref.~\cite{Buras:2012jb} and in the plots in the next
section. They can be resolved using observables where interference with the SM
occurs. The absolute values $\lambda_i^{\rm VLQ}$ can be determined via
\begin{align}
  |\lambda_d| & = \sqrt{\frac{\Lambda_{bd}\Lambda_{sd}^*}{\Lambda_{bs}}} , &
  |\lambda_s| & = \sqrt{\frac{\Lambda_{bs}\Lambda_{sd}}{\Lambda_{bd}}} , &
  |\lambda_b| & = \sqrt{\frac{\Lambda_{bd}\Lambda_{bs}^*}{\Lambda_{sd}}} .
\end{align}
One might expect the strongest constraints numerically to stem from $s\to d$
processes, because of the strong suppression of the SM contribution by
$V_{td}^{} V_{ts}^*$.

In a sense, as more explicitly seen in the next section, the flavour structure
of VLQ models has some parallels to the one in 331 models \cite{Buras:2012dp,
  Buras:2013dea, Buras:2015kwd, Buras:2016dxz}. However, in 331 models the NP
contributions are dominated by $Z'$ tree-level exchanges and once the
constraints from $B_{s,d}$ observables are taken into account, NP effects in the
$K$ system are found to be small, with the exception of $\epe$. In the present
analysis important $Z$ boson contributions are present and this allows for more
interesting NP effects than in 331 models in $\kpn$ and $\klpn$.
Furthermore, the partial decoupling of $|\Delta F|=1$ and $|\Delta F|=2$
processes due to the presence of important box diagram contributions to 
$|\Delta F|=2$ processes
in VLQ models discussed above modifies the corresponding correlations
derived in Ref.~\cite{Buras:2012jb}, increasing the impact of $|\Delta F|=2$
constraints on $|\Delta F|=1$ processes relative to the one found in
\cite{Buras:2012jb}. The latter is also true for RG effects in $\GSM$
models, specifically for RH scenarios, where the importance of $\Delta F=2$ can
be drastically enhanced.
In \reftab{tab:DNA} we summarize the patterns discussed above.

%
%
%

\section{Numerics}
\label{sec:numerics}

In this section we perform the numerical analysis of the VLQ models presented
above.  For this purpose we start by constraining the VLQ couplings by the
available flavour data and if applicable also by data from other sectors.  We
proceed by presenting the predictions for a number of key observables given
these constraints, including their correlations where they are sizeable. These
fits are performed for different VLQ masses, in order to illustrate the explicit
mass dependence of flavour observables discussed in Section~\ref{sec:mass}.

Model-independent constraints on $\psi^2\varphi^2D$ operators have been derived
from $Z$- and $W$-boson observables \cite{Efrati:2015eaa}, which are applicable
to $\GSM$ models. Although these constraints are not entirely independent from
other operators, in VLQ-models the latter are loop-suppressed and can be
neglected.  The constraints on the modulus of the couplings are weak and of the
order $|\lambda_i|\lesssim M/(1\, \mbox{TeV})$.\footnote{There is one tension
  from $[\hat{c}_{Hd}]_{33} = (-4.6 \pm 1.6) \times 10^{-2}$
  \cite{Efrati:2015eaa} (A.9) for the VLQ representation $Q_V$.  }

More stringent constraints derive from $|\Delta F| = 2, 1$ flavour observables
\cite{Ishiwata:2015cga}.  We constrain the five parameters $|\Lambda_{ij}|$ and
$\varphi_{ij}$ \refeq{eq:LambdaAbsPhi} with the $|\Delta F| = 2, 1$ processes
listed in \reftab{tab:exp-flavor-constraints}.  Master formulae used in these
constraints are collected in \refapp{app:master}. The SM predictions in
\reftab{tab:exp-flavor-constraints} are based on the determination of CKM
parameters from a tree-level fit given in \reftab{tab:input}. Some comments
regarding the included observables are in order:
\begin{itemize}
\item The observable $\Delta M_K$ does not provide constraints in $\GSM$ models
  and is omitted due to too large
  uncertainties from long-distance contributions in $\GSMUpr(\Phi)$ models. The
  prospects for controlling this long-distance part by lattice calculations are
  good \cite{Bai:2014cva} and in the future this constraint could play an
  important role.
\item We find that huge NP effects in $\epe$ are not excluded by the constraints
  listed in \reftab{tab:exp-flavor-constraints} in $\GSM$- and 
  $\GSMUpr(\Phi)$-models, such that we impose bounds on the NP contribution 
  $(\epe)_{\rm NP}$ itself
  \begin{align}
    \label{eq:epe-constraint}
    (\epe)_{\rm NP} & \in [0,\, 20] \times 10^{-4},
  \end{align}
  in order to avoid showing predictions for other observables that are easily
  excluded by $\epe$, and to analyse its influence on the correlations of
  observables. This range roughly corresponds to NP required assuming present
  predictions from lattice QCD. We have checked that decreasing this range to
  $[5,\, 10] \times 10^{-4}$ as expected from the dual approach to QCD
  \cite{Buras:2015xba, Buras:2016fys} would have only minor impact on the global
  fit as what matters is the unique selection of the sign of the relevant phase
  required for the enhancement of $\epe$.
\item Due to the sizeable experimental uncertainties, $Br(B_d\to \mu\bar\mu)$
  does not constrain the VLQ parameters further.  It is thus omitted from the
  fit and we compare its prediction in our models to the present measurement.
\item A full analysis of $B\to K^*\ell\bar\ell$ is beyond the scope of this
  work. We do therefore not include the LHCb anomalies \cite{Altmannshofer:2014rta, 
  Altmannshofer:2015sma, Descotes-Genon:2015uva, Hurth:2016fbr} in our fits.
  The analysis of $b\to s\ell\bar\ell$ in $\GSMUpr(S)$ models has been already 
  presented in \cite{Altmannshofer:2014cfa, Altmannshofer:2015mqa} and
  we have nothing to add here. In $\GSM$
  models the shift in $C_9$ is too small to be relevant, while in $\GSMUpr(\Phi)$
  models the effects are only moderately interesting and we will not address
  them here.
\end{itemize}

The three sectors $s\to d$, $b\to d$ and $b\to s$ are not independent, due to
relation~\refeq{eq:phiij}. In our analysis we show first the results separately
for the three quark transitions and demonstrate in a global fit that
$K$-physics constraints have an impact on $B$ physics but not vice versa.

\begin{table}[t]
\addtolength{\arraycolsep}{4pt}
\renewcommand{\arraystretch}{1.5}
\centering
\begin{tabular}{|cc|cc|c|}
\hline
 \boldmath $i\to j$ & observable & measurement & ref. & SM (c.v. [$95\%$~CL]) 
\\
\hline\hline
 \multirow{4}{*}{\boldmath $s\to d$}
 & $\varepsilon_K$ 
 & $2.228(11) \times 10^{-3}$
 & \cite{Agashe:2014kda} 
 & $(2.21\,[1.57,2.98] ) \times 10^{-3}$
\\ 
 & $Br(K^+\to\pi^+\nu\bar\nu)$ 
 & $(17.3^{+11.5}_{-10.5}) \times 10^{-11}$ 
 & \cite{Artamonov:2008qb}
 & $(8.5\,[7.3,9.5]) \times 10^{-11}$
\\
 & $Br(K_L\to\mu\bar\mu)_{\rm SD}$ 
 & $<2.5 \times 10^{-9}$ 
 & \cite{Isidori:2003ts}
 & $\chi_{\rm SD}: 1.81\,[1.65,1.94]$
\\
 \cline{2-5}
 & $(\epe)_{\rm NP}$
 & $[0,\, 20] \times 10^{-4}$
 & ${}^\dagger$ 
 & $0 \times 10^{-4}$ 
\\
\hline \hline
 \multirow{4}{*}{\boldmath $b\to d$}
 & $\Delta M_d$  [ps$^{-1}$]
 & $0.5055(20)$
 & \cite{Amhis:2014hma}
 & $0.62\,[0.45,0.78]$
\\
 & $\sin(2 \beta_d)$ 
 & $0.691(17)^{*}$  
 & \cite{Amhis:2014hma}
 & $0.734\,[0.686,0.796]$
\\ 
 & $Br(B^+\to\pi^+\mu\bar\mu)_{[15,22]}$ & $3.29(84)\times 10^{-9}$ 
 & \cite{Aaij:2015nea}
 & $(5.0\,[3.8,7.2])\times 10^{-9}$
\\
\hline \hline
 \multirow{4}{*}{\boldmath $b\to s$}
 & $\Delta M_s$ [ps$^{-1}$]
 & $17.757(21)$ 
 & \cite{Amhis:2014hma}
 & $19.0\,[16.2,21.9]$
\\
 & $\sin(2 \beta_s)$
 & $-0.034(33)^{*}$  
 & \cite{Amhis:2014hma}
 & $-0.040\,[-0.044,-0.036]$
\\ 
 & $Br(B_s\to \mu\bar\mu)$ & $(2.8^{+0.7}_{-0.6})\times 10^{-9}$ 
 & \cite{CMS:2014xfa}
 & $(3.41\,[3.01,3.81])\times 10^{-9}$
\\
 & $Br(B^+\to K^+\mu\bar\mu)_{[15,22]}$ & $8.47(50)\times 10^{-8}$ 
 & \cite{Aaij:2014pli}
 & $(11.0\,[6.4,15.6])\times 10^{-8}$
\\
\hline
\end{tabular}
\renewcommand{\arraystretch}{1.0}
\caption{\small The list of $|\Delta F|=2,1$ flavour observables in $i\to j$
  down-type transitions that are used to constrain the VLQ couplings. SM predictions
  are obtained with CKM parameters determined from the tree-fit. ${}^\dagger$We impose
  this conservative range on the NP contribution of $\epe$ to avoid
  values excluded by this observable in the predictions for other observables, 
  see text for more details. ${}^{*}$ Note that we neglect potential 
  ``penguin pollution'' in $b\to c\bar cs$ transitions, which
  have been shown in recent analyses to be at most of the size of the present experimental
  uncertainties~\cite{Jung:2012mp, DeBruyn:2014oga, Frings:2015eva}.
}
\label{tab:exp-flavor-constraints}
\end{table}

%
%

\subsection[$\GSM$ models]
 {\boldmath $\GSM$ models}

In $\GSM$ models the absence of additional scalars allows to vary the mass of
the VLQ's down to about $1$~TeV without violating the hierarchy
\refeq{eq:scale-hierarchy-1}.  The fits of the $\Lambda_{ij}$ for the three
types of transitions $j\to i = \{s\to d,\, b\to d,\, b\to s\}$ in $\GSM$ models
are shown in \reffig{fig:GSM-A_ij-fits-10TeV} for $M_{\rm VLQ} = 10$~TeV and in
\reffig{fig:GSM-A_ij-fits-1TeV} for $M_{\rm VLQ} = 1$~TeV for the single-VLQ
scenarios $D$ and $Q_V$ with LH and RH couplings, respectively.  The plots for
LH scenarios $T_{u,d}$ are qualitatively similar to $D$ whereas the RH scenario
$Q_d$ is similar to $Q_V$. Quantitative differences arise due to changes of the
sign in couplings and a factor $1/2$ for $T_d$ w.r.t. $D$ and $T_u$, which are shown
in \reftab{tab:GSM}. The statistical approach for these fits is detailed in 
\refapp{app:num-input}. We make the following observations:

\begin{figure}
  \centering
     \includegraphics[width=0.4\textwidth]{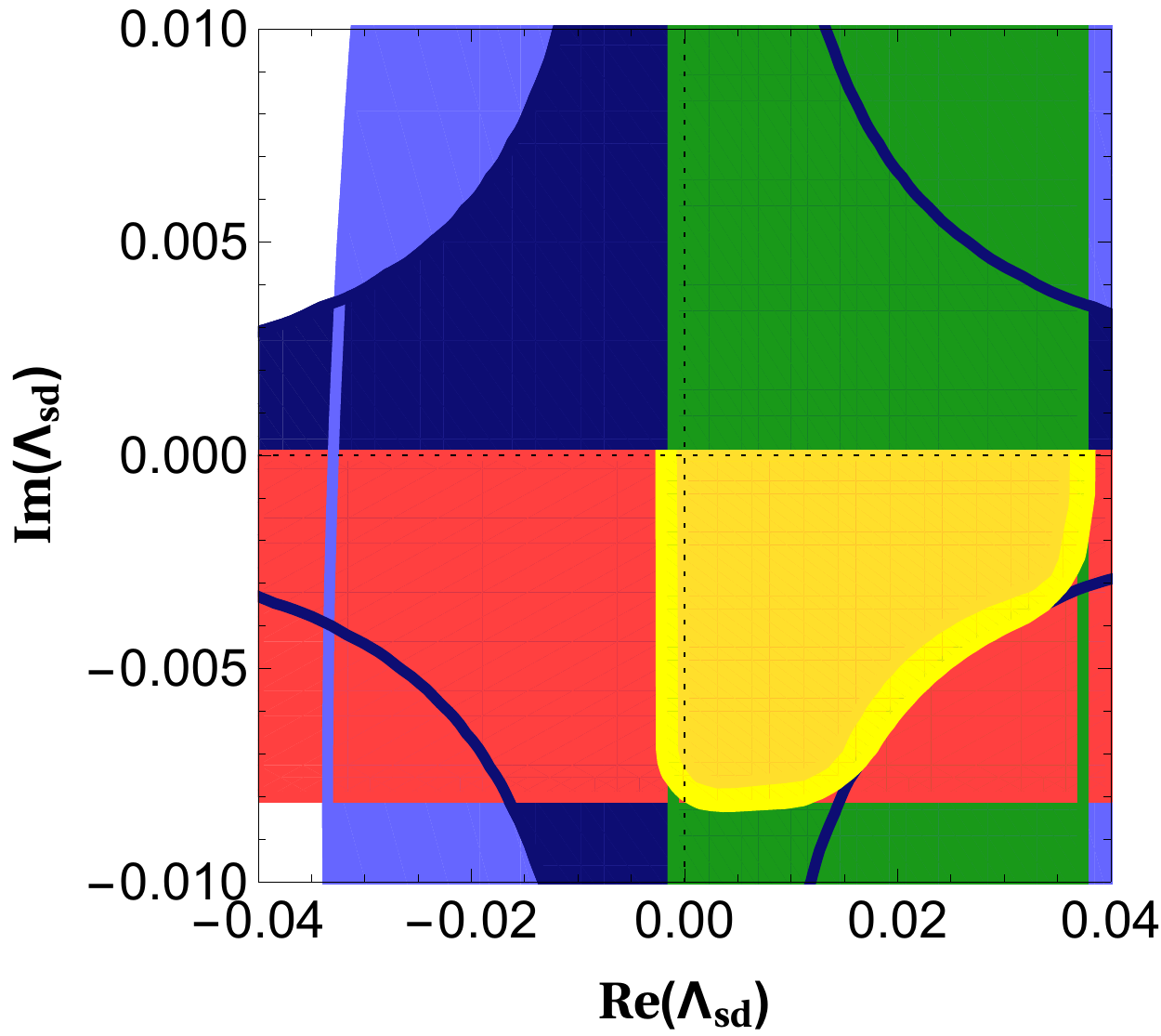} 
     \hskip 0.05\textwidth
     \includegraphics[width=0.4\textwidth]{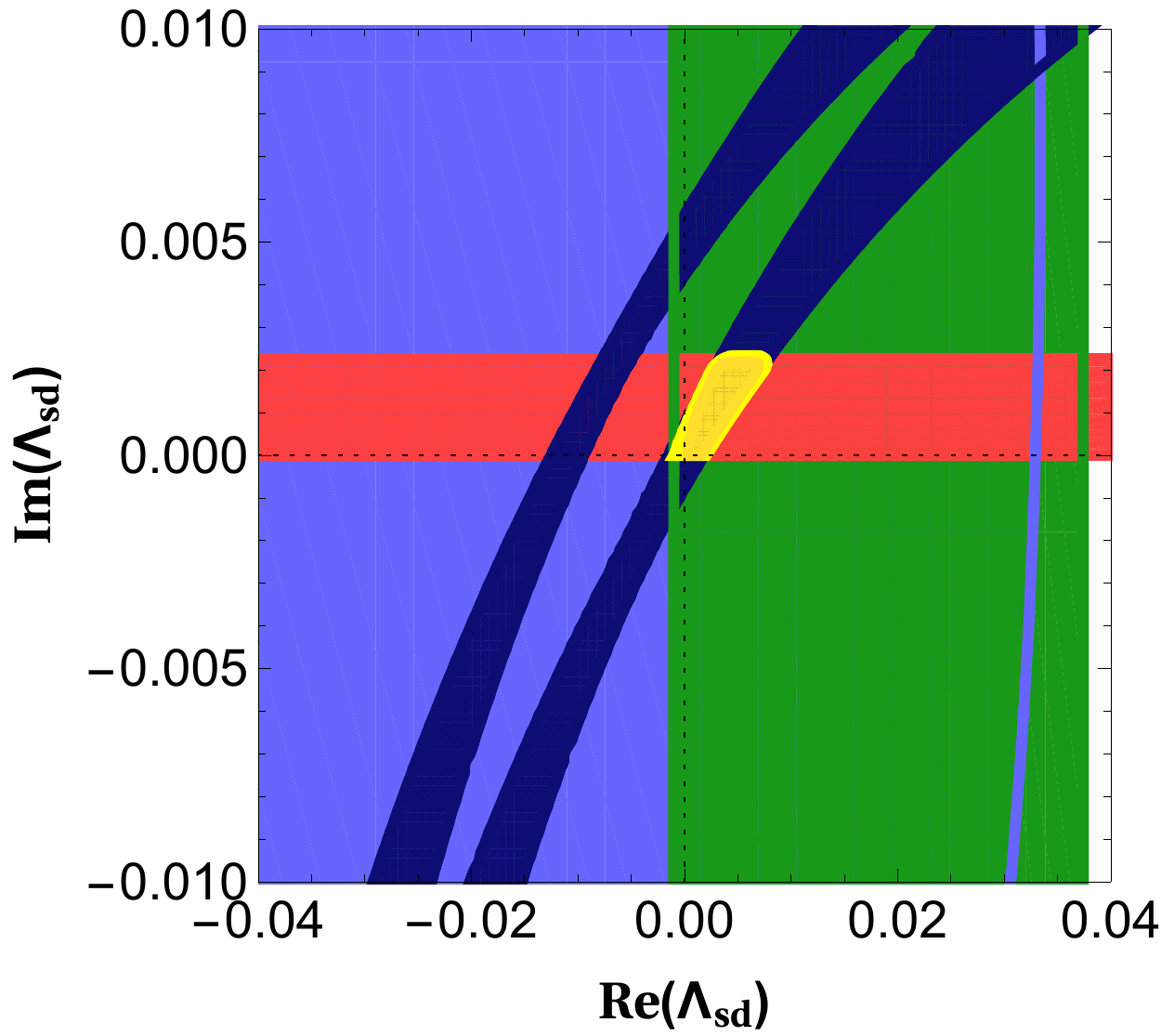}    
   \vskip 0.1cm
     \includegraphics[width=0.39\textwidth]{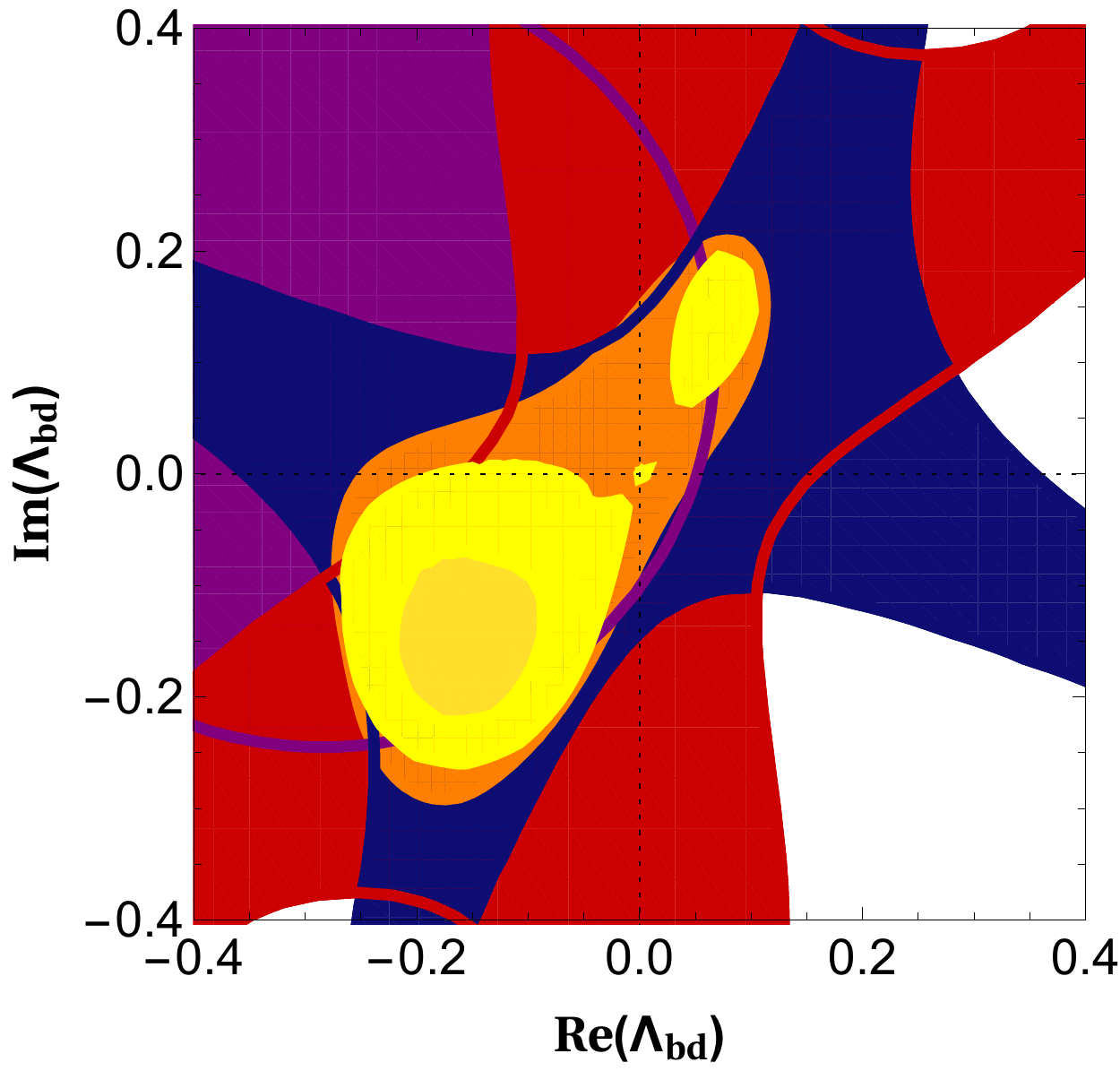} 
     \hskip 0.05\textwidth
     \includegraphics[width=0.39\textwidth]{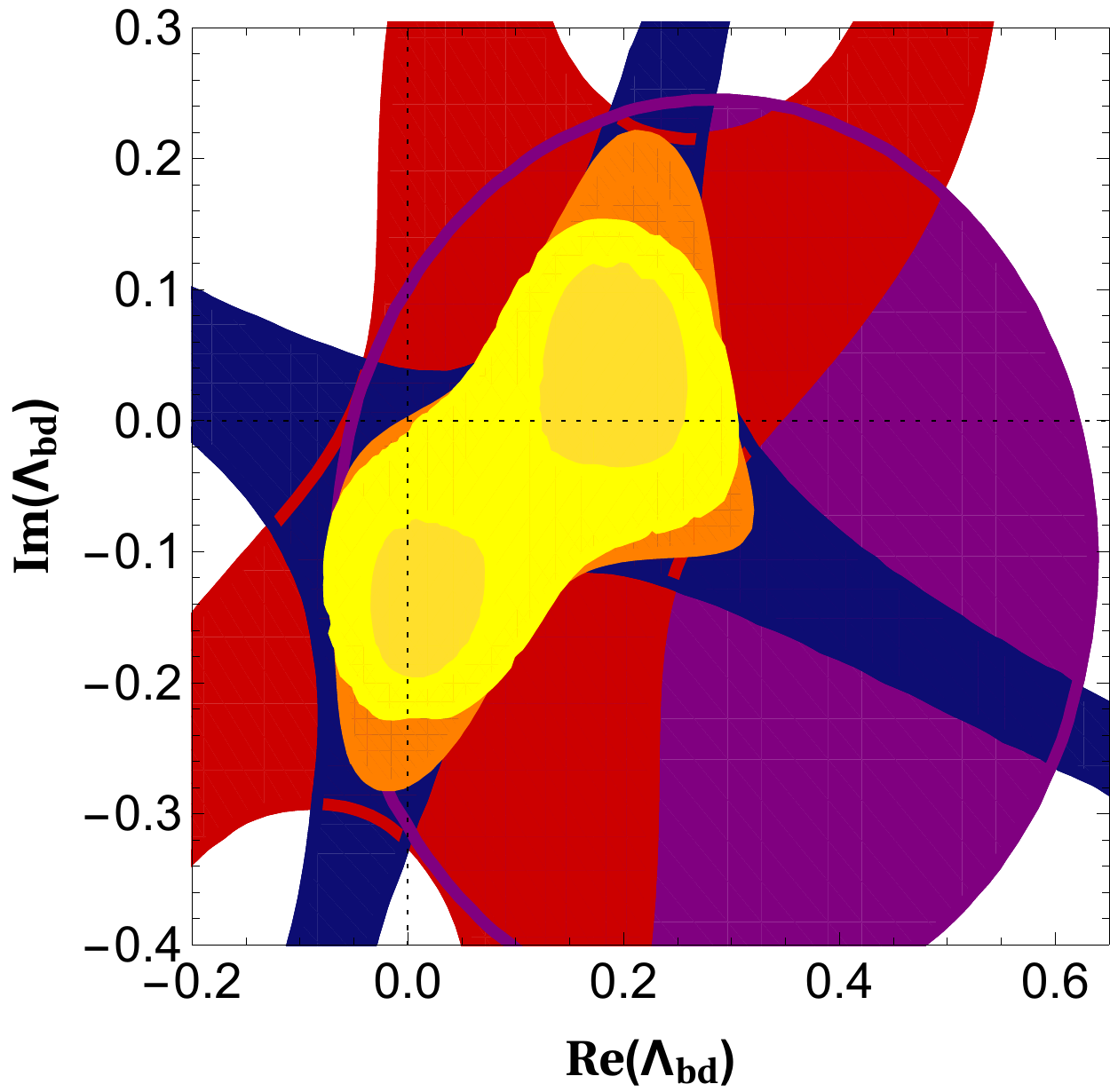}
   \vskip 0.1cm    
     \includegraphics[width=0.38\textwidth]{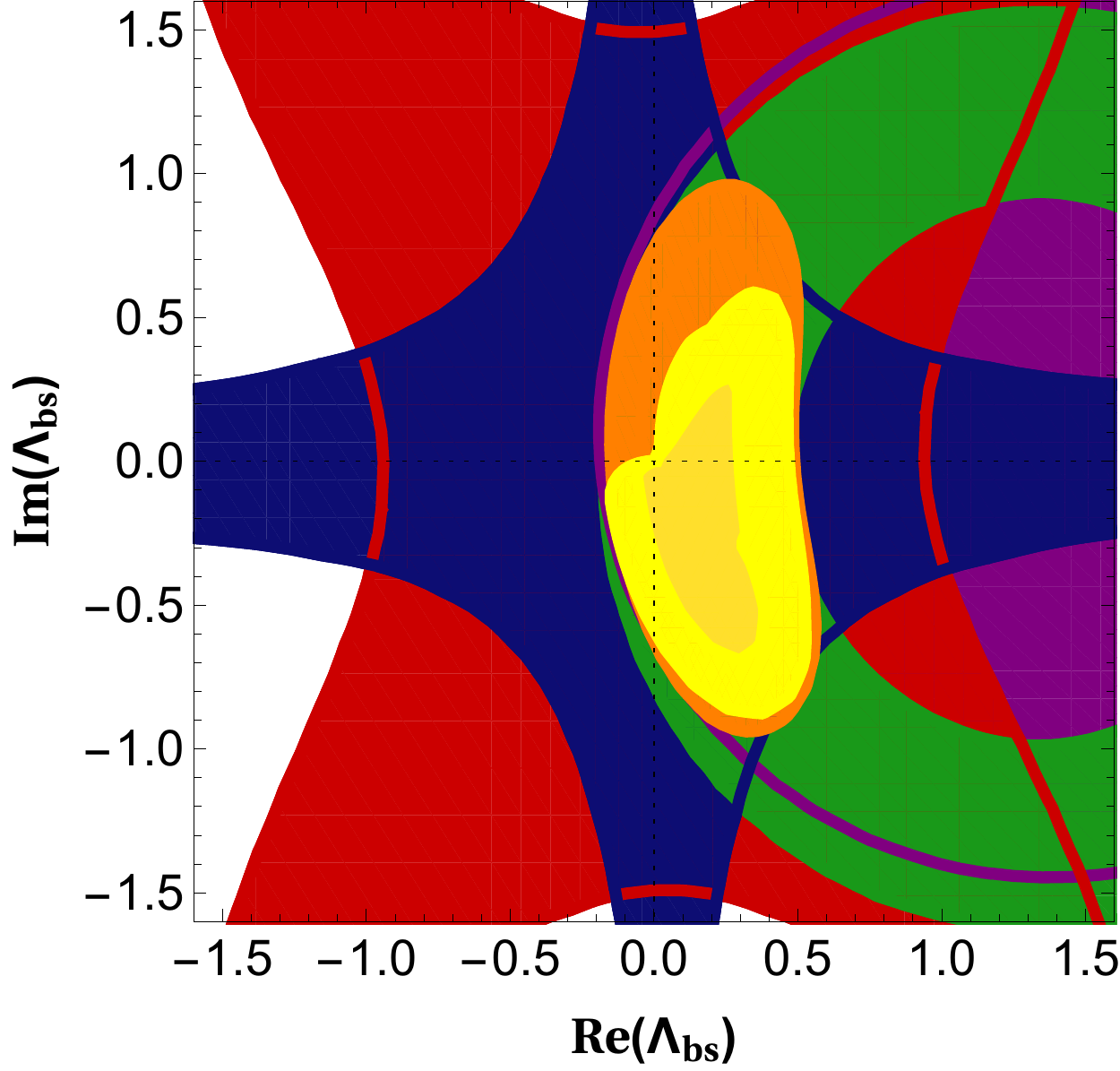} 
     \hskip 0.05\textwidth
     \includegraphics[width=0.38\textwidth]{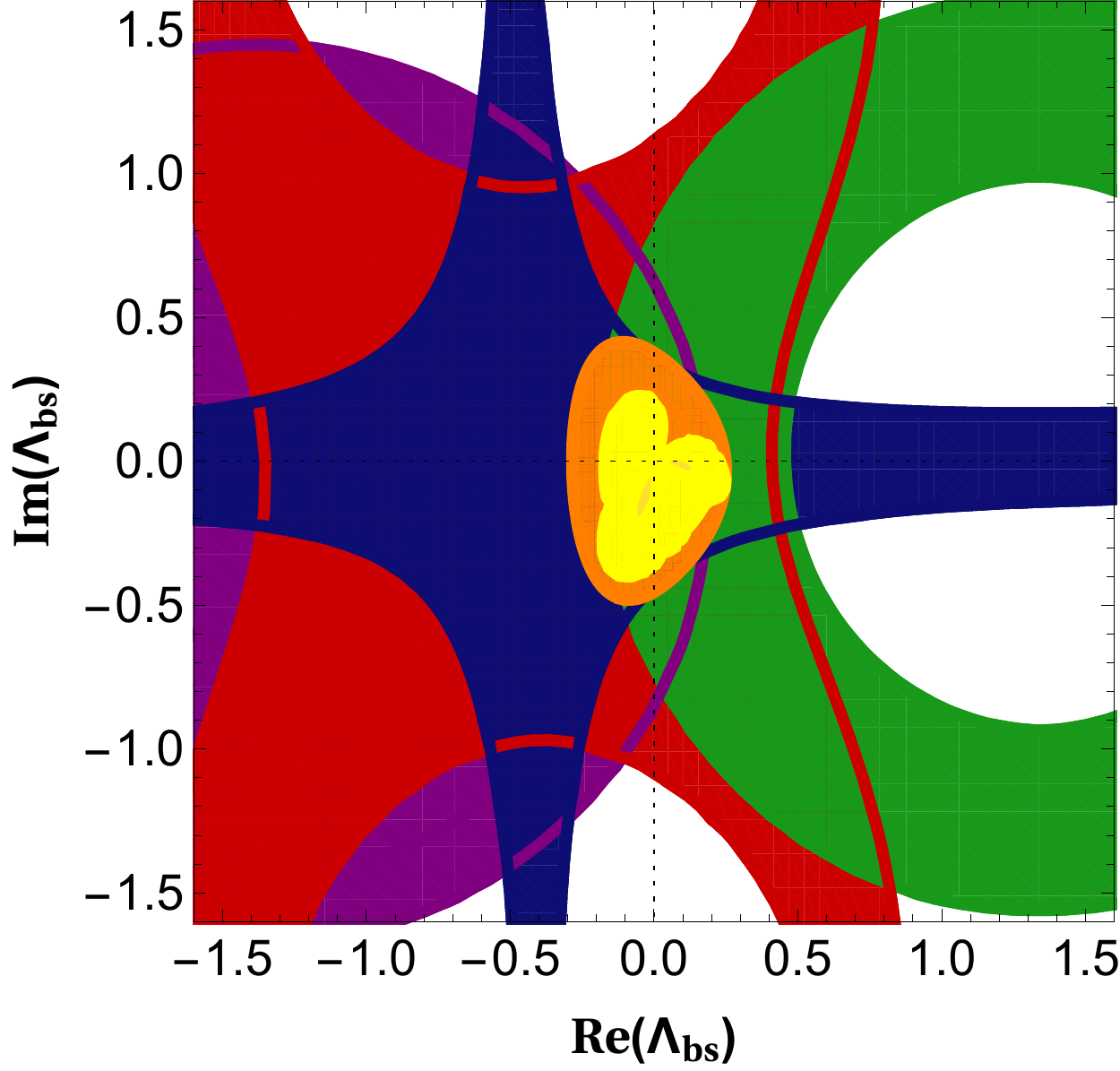}        
\caption{
  \small 
  \label{fig:GSM-A_ij-fits-10TeV}
  Fits of $\mbox{Im}(\Lambda_{ij})$ vs. $\mbox{Re}(\Lambda_{ij})$ for
  $ij=sd,\, bd,\, bs$ [upper, middle, lower] in $\GSM$-scenarios $D$~[left] and
  $Q_V$[right] for $M_{\rm VLQ} = 10$~TeV. Constraints from single observables
  and the combined fit for each separate sector [orange] are shown at $95\%$~CL,
  the global fit [yellow] at $68\%$ and $95\%$.  For $ij=sd$: $\varepsilon_K$
  [dark blue], $Br(K^+\to \pi^+\nu\bar\nu)$ [blue],
  $Br(K_L\to \mu\bar\mu)_{\rm SD}$ [green], and $(\epe)_{\rm NP}$~[red].  For $ij=bd$:
  $\Delta M_d$ [dark red], $\sin(2\beta_d)$ [dark blue] and
  $Br(B^+\to\pi^+\mu\bar\mu)_{[15,\,22]}$ [purple].  For $ij=bs$: $\Delta M_s$
  [dark red], $\sin(2\beta_s)$ [dark blue], $Br(B_s\to \mu\bar\mu)$ [green] and
  $Br(B^+\to K^+\mu\bar\mu)_{[15,\,22]}$ [purple]. 
}
\end{figure}

\begin{figure}
  \centering
    \includegraphics[width=0.4\textwidth]{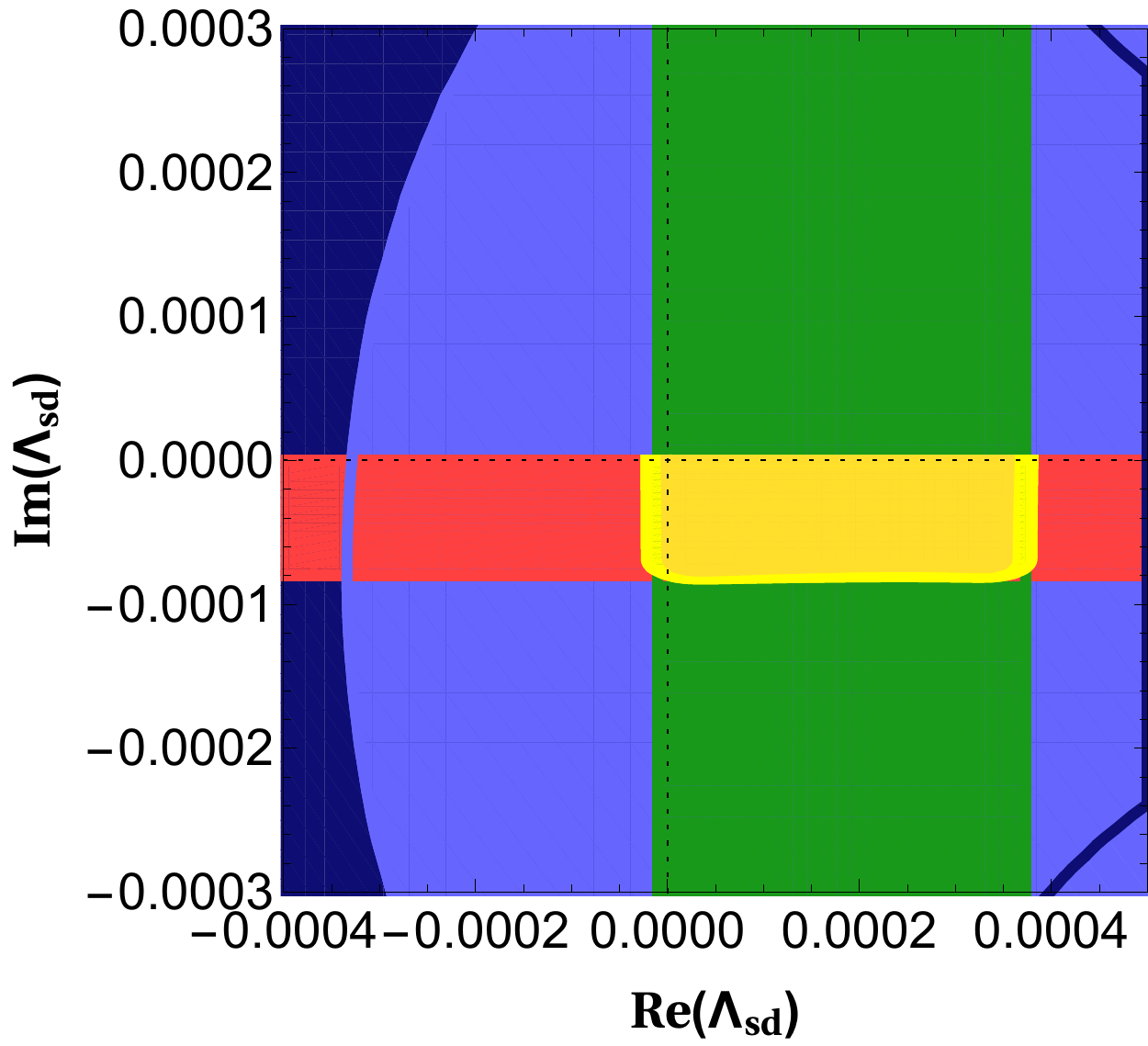} 
    \hskip 0.05\textwidth
    \includegraphics[width=0.4\textwidth]{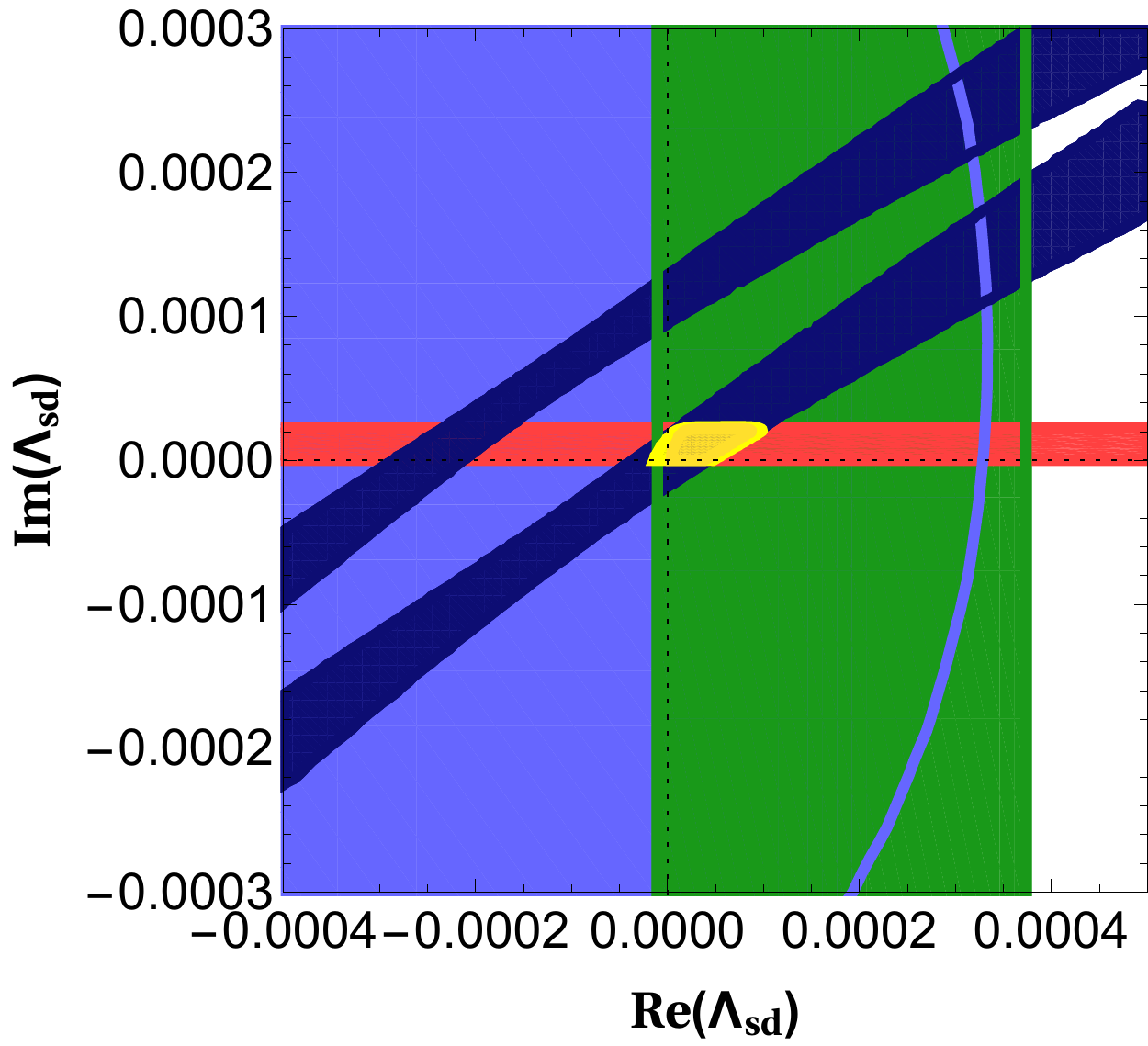}    
  \vskip 0.1cm    
    \includegraphics[width=0.4\textwidth]{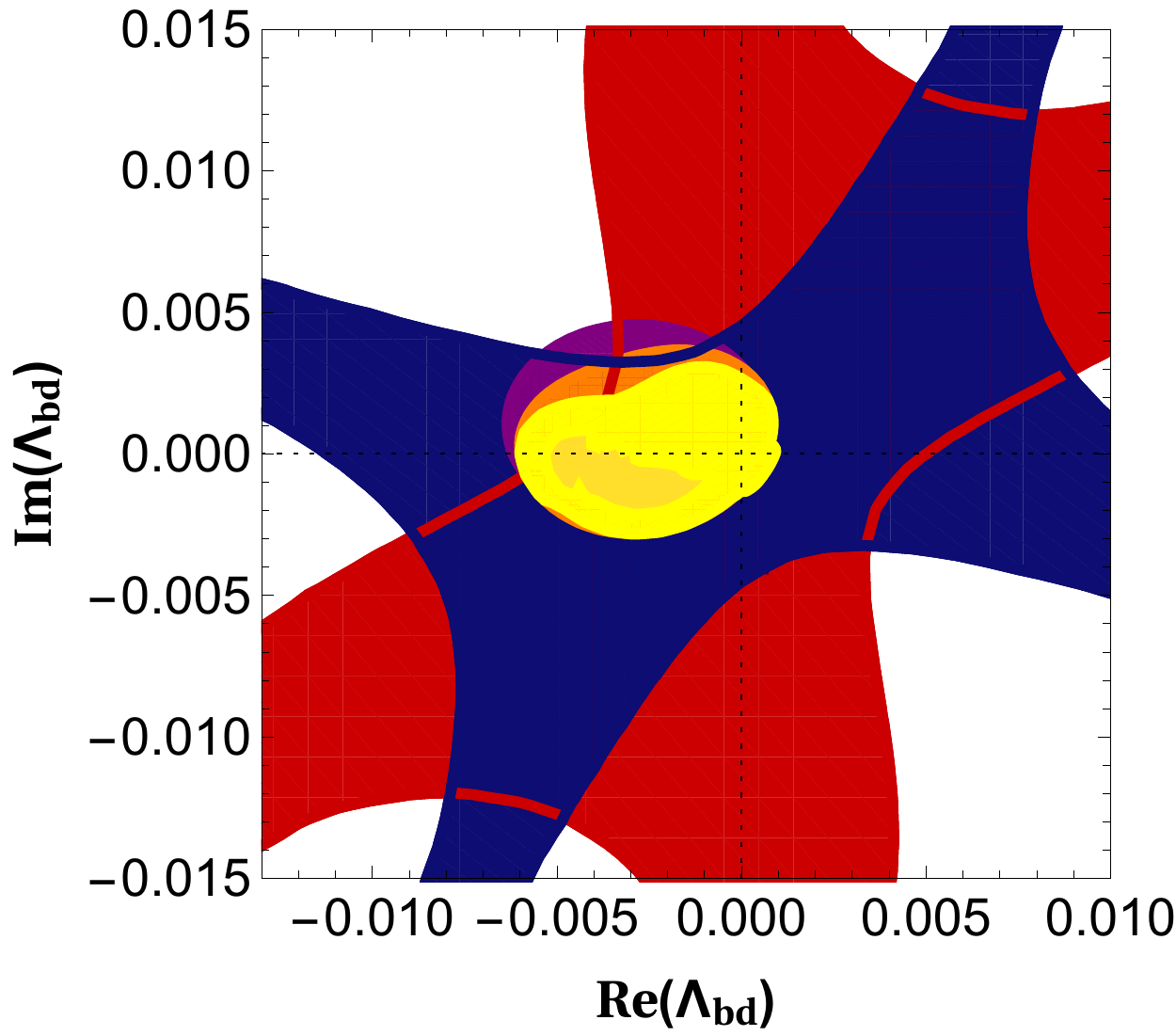} 
    \hskip 0.05\textwidth
    \includegraphics[width=0.4\textwidth]{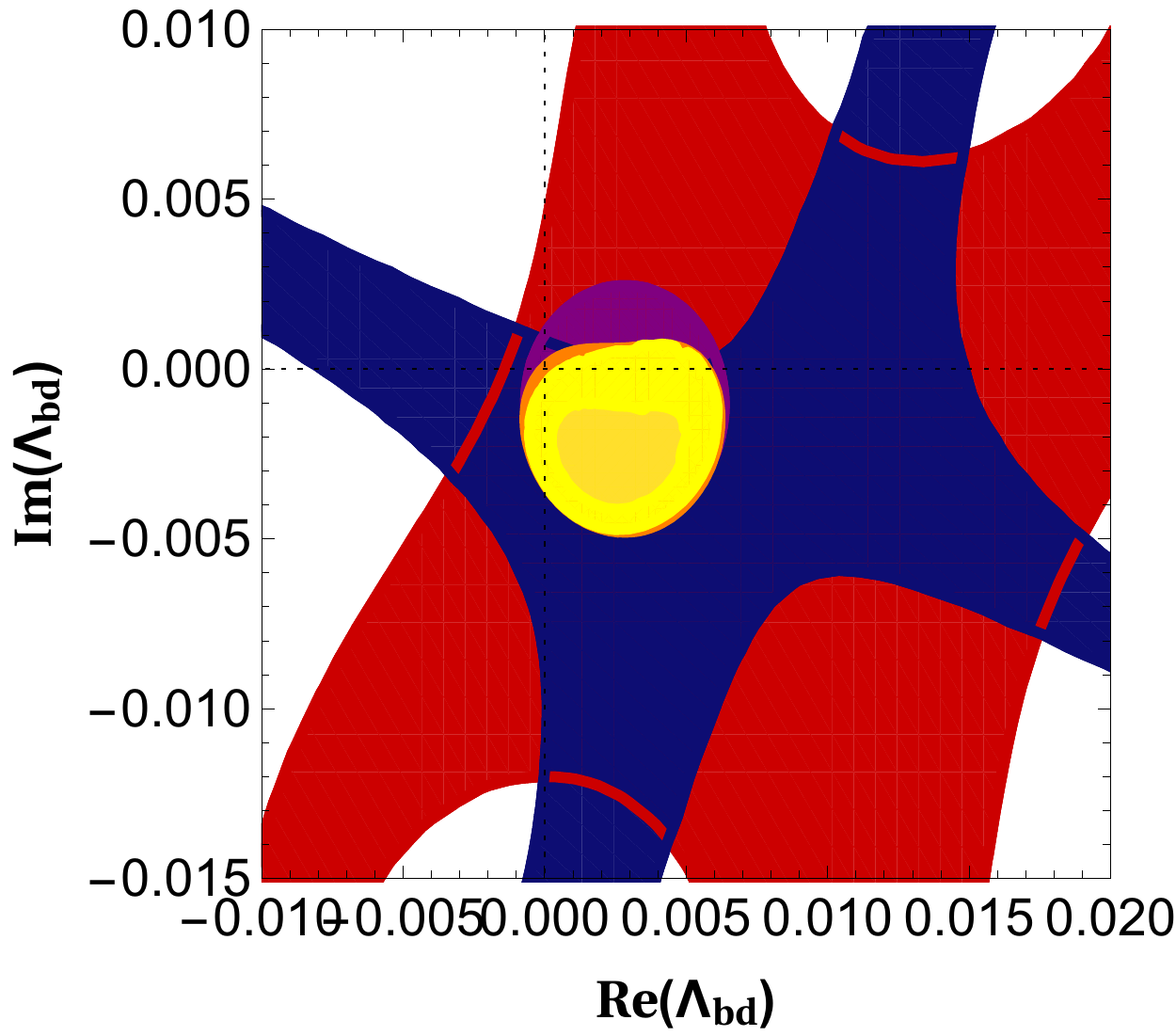}
  \vskip 0.1cm    
    \includegraphics[width=0.4\textwidth]{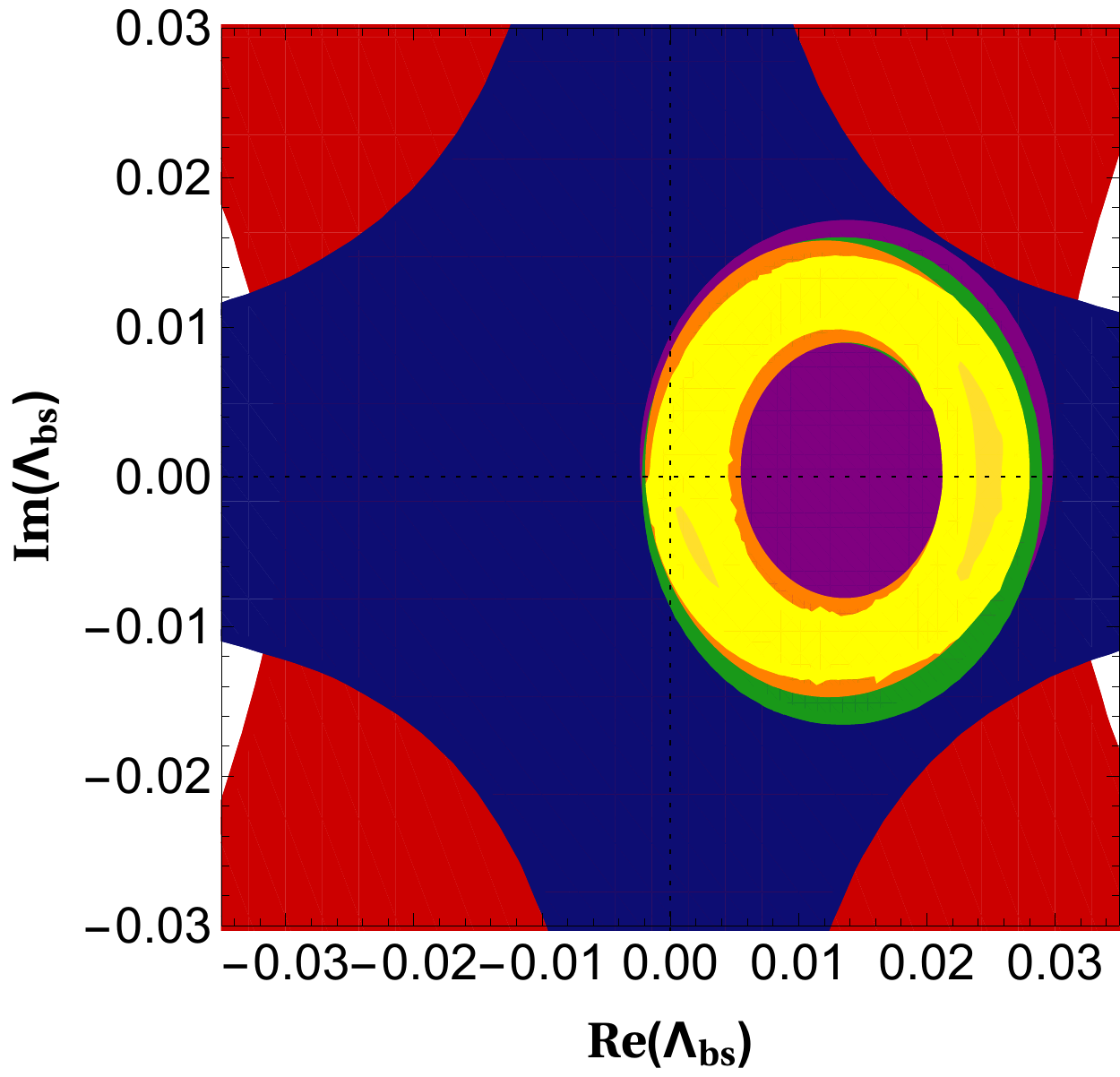} 
    \hskip 0.05\textwidth
    \includegraphics[width=0.4\textwidth]{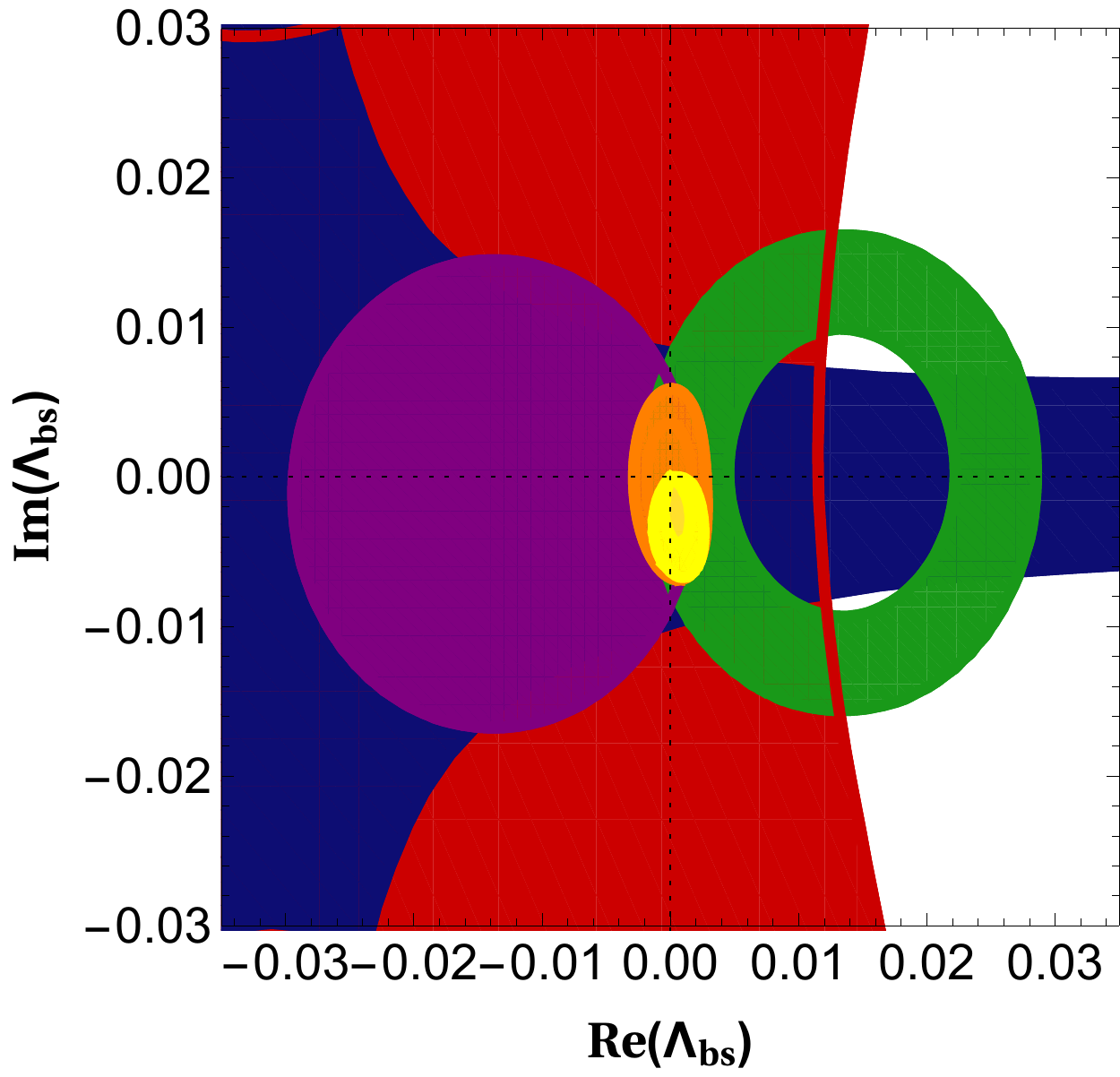}    
\caption{
  \small 
  \label{fig:GSM-A_ij-fits-1TeV}
  Fits of $\mbox{Im}(\Lambda_{ij})$ vs. $\mbox{Re}(\Lambda_{ij})$ for
  $ij=sd,\, bd,\, sd$ [upper, middle, lower] in $\GSM$-scenarios $D$~[left] and
  $Q_V$[right] for $M_{\rm VLQ} = 1$~TeV. The colour scheme is as in
  \reffig{fig:GSM-A_ij-fits-10TeV}.
}
\end{figure}

\begin{figure}
  \includegraphics[width=0.46 \textwidth]{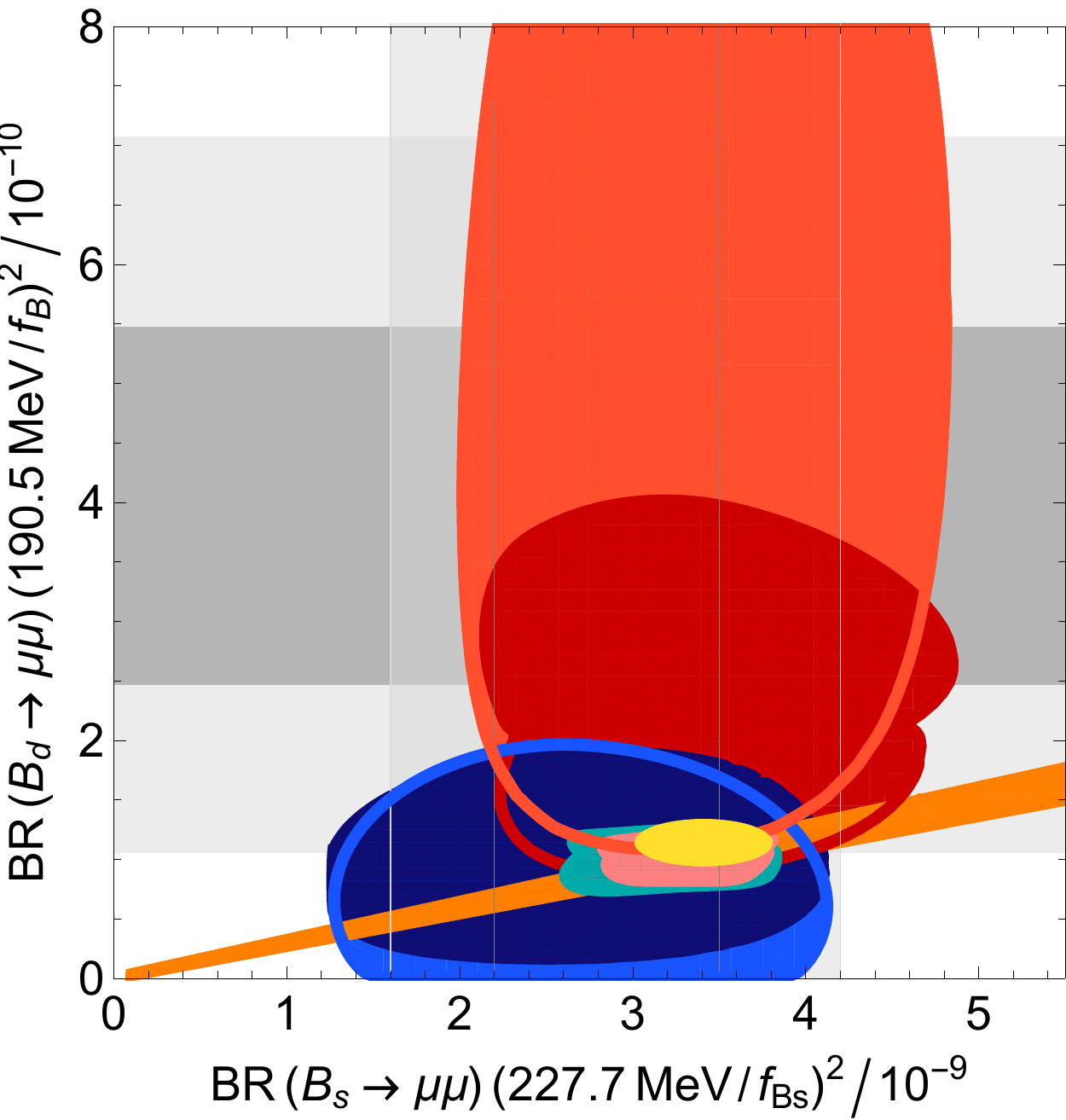} 
  \hfill\parbox[b]{0.42\textwidth}{
\caption{\small \label{fig:B-mumu-crr}
  $Br(B_d\to\mu\bar\mu)$ vs. $Br(B_s\to\mu\bar\mu)$ in $\GSM$ at 10~TeV [darker 
  colours] and 1~TeV (lighter colours), as well as $\GSMUpr(\Phi)$ models at 10~TeV. 
  The colours correspond to the representations $Q_V=Q_d$ [red and pink for $\GSM$ 
  and $\GSMUpr(\Phi)$] and $D$ [similar to $T_{u,d}$, blue and cyan]. The orange 
  band corresponds to a scenario of constrained minimal flavour violation (CMFV)
  \cite{Buras:2014fpa}.}
}
\end{figure}

\begin{itemize}
\item All included observables are compatible with the SM prediction at
  $95\%$~CL.  Correspondingly also the global fit allows for the SM solution at
  $95\%$~CL in all planes in both scenarios, except for $\Lambda_{bd}^{Q_V}$
  with $M^{Q_V}_{\rm VLQ} = 1, \;10$~TeV, where the SM is slightly outside that
  region. This is due to the slight tensions of $\Delta M_d$ and
  $Br(B^+\to \pi^+\ell\bar\ell)$ with their SM predictions, which fortify each
  other in this case.
\item For $M_{\rm{VLQ}}=10$~TeV, $|\Delta F|=2$ constraints are competitive to the
    $|\Delta F|=1$ ones, and their interplay determines the global fit regions. 
    For $M_{\rm{VLQ}}=1$~TeV the global fit is almost
    completely determined by $|\Delta F|=1$ processes in LH scenarios, but also in 
    RH scenarios for $b\to d,s$. On the other hand, $\epsilon_K$ is a very powerful
    constraint in RH scenarios also for 1~TeV, due to the RG effects discussed above. 
    This is in accordance with our previous discussion of the mass-dependence of these
    transitions. Specifically for $K^+\to\pi^+\nu\bar\nu$, 
    large effects are excluded by $\epsilon_K$ in combination with $\epe$ and 
    $K_L\to\mu\bar\mu$. Without the RG contributions, enhancements up to the present
    experimental limit would have been possible.
\item In $b\to s$, the $|\Delta F|=1$ observables distinguish between scenarios
  with LH and RH currents due to their different dependences on the
  corresponding Wilson coefficients, most importantly $C_{10}$ and $C'_{10}$,
  \begin{align}
    Br(B_s\to \mu\bar\mu) & \propto |C_{10}^{} - C'_{10}|^2, &
    Br(B^+\to K^+ \mu\bar\mu) & \propto |C_{10}^{} + C'_{10}|^2.  
  \end{align} 
  The consequence is shown in \reffig{fig:GSM-A_ij-fits-10TeV} and
  \reffig{fig:GSM-A_ij-fits-1TeV} where allowed regions almost overlap for LH
  scenarios, but intersect only around the SM for RH scenarios, thereby
  diminishing the size of potential VLQ effects in other $b\to s$ observables.
  The same observation holds for $b\to d$ transitions, which will help once
  $B_d\to \mu\bar\mu$ is measured more precisely.
  In \reffig{fig:B-mumu-crr} we illustrate how $Br(B_{d,s}\to \mu\bar\mu)$
  can be used in a large region of parameter space to discriminate between 
  LH and RH models. 
\item In $s\to d$ transitions, the constraints from $\epsilon_K$, 
  $(\epe)_{\rm NP}$ and $Br(K_L\to \mu\bar\mu)_{\rm SD}$ constrain the allowed 
  values for $\varphi_{sd}$. This in combination with the slight tensions especially 
  in $b\to d$ leads to stronger constraints in the global fit compared to the fits
  for the individual transitions in $b\to d,s$. As a consequence correlations
  between different transitions arise, but at the moment they are not very
  strong yet. This would change with significant measurements away from the SM
  for at least two of the transitions.
\item The $|\Delta F|=2$ CP-asymmetric observables $\epsilon_K$ and
  $\sin(2\beta_{d,s})$ impose constraints in the complex
  $\Lambda_{ij}$-planes, which are not limited along the direction corresponding
  to the SM phase. Such a limit is provided by $\Delta M_{d,s}$, whereas in the
  case of $s\to d$ the one from $\Delta M_K$ is very weak and outside of the
  ranges shown.
\item There is a complementarity in the constraints from $Br(K^+\to \pi^+\nu\bar\nu)$
  and $Br(K_L\to \mu\bar\mu)_{\rm SD}$ for every VLQ representation. Thus an improved
  measurement of $Br(K^+\to \pi^+\nu\bar\nu)$ by NA62, which will operate until the
  LHC shut down in 2018 and aims at a 10\% uncertainty \cite{Rinella:2014wfa,
  Ceccucci:2016xpd}, will provide stronger cuts into the allowed parameter space. 
  On the other hand, while the constraints from $(\epe)_{\rm NP}$ and
  $Br(K_L\to \mu\bar\mu)_{\rm SD}$ are theoretically limited at present, they 
  could become very powerful in the future if theory improves.
\end{itemize}

\begin{table}[htbp]
\addtolength{\arraycolsep}{4pt}
\renewcommand{\arraystretch}{1.5}
\centering
\resizebox{\textwidth}{!}{
\begin{tabular}{|c|r|cccc|}
\hline
 SM & measurement & $D$  & $Q_d,\,Q_V$ & $T_u$ & $T_d$
\\
\hline\hline
\multicolumn{6}{|c|}{$10^{11} \times Br(K_L\to\pi^0\nu\bar\nu)$, $\quad 10^{11} \times Br(K^+\to\pi^+\nu\bar\nu)$}
\\
\hline
  3.2\,[2.5,\,4.3]
& $\leq 2600$ \hfill \cite{Ahn:2009gb}
& $^{[0,\, 4.3]}_{[0,\, 4.3]}$ 
& $^{[1.3,\, 3.3]}_{[1.2,\, 3.5]}$
& $^{[0,\, 4.3]}_{[0,\, 4.3]}$
& $^{[0,\, 4.2]}_{[0,\, 4.2]}$
\\
  8.5\,[7.3,\,9.5]
& $17.3^{+11.5}_{-10.5}$ \hfill \cite{Ahn:2009gb}
& $^{[0.8,\, 9.2]}_{[0.8,\, 9.2]}$
& $^{[7.7,\, 13.6]}_{[7.8,\, 13.0]}$
& $^{[0.7,\,8.9]}_{[0.8,\, 9.2]}$ 
& $^{[0.7,\,8.9]}_{[1.2,\,9.2]}$
\\
\hline\hline
\multicolumn{6}{|c|}{$10^{10} \times Br(B_d\to\mu\bar\mu)$}
\\
\hline
  $1.14\,[0.94, 1.32]$
& $\leq 6.3$ \hfill \cite{Aaij:2013aka}
& $^{[0.0,\, 1.7]}_{[0.1,\, 1.8]}$ 
& $^{[1.2,\, 10.4]}_{[1.0,\, \phantom{1}4.0]}$
& $^{[0.0,\, 1.7]}_{[0.1,\, 1.9]}$
& $^{[0.0,\, 1.7]}_{[0.5,\, 1.7]}$
\\
\hline\hline
\multicolumn{6}{|c|}{$A_{\Delta\Gamma}(B_s\to \mu\bar\mu)$, $\quad S(B_s\to \mu\bar\mu)$}
\\\hline
  $1$
& ---
& $^{[-1.00,\, 1.00]}_{[\phantom{-}0.12,\, 0.99]}$
& $^{[ 0.67,\, 1.00]}_{[0.87,\, 1.00]}$
& $^{[-1.00,\, 1.00]}_{[\phantom{-}0.46,\, 1.00]}$ 
& $^{[-0.28,\, 1.00]}_{[\phantom{-}0.86,\, 1.00]}$
\\
  $0$
& ---
& $^{[-1.00,\, 1.00]}_{[-0.99,\, 0.99]}$ 
& $^{[-0.63,\, 0.74]}_{[-0.41,\, 0.48]}$
& $^{[-1.00,\, 1.00]}_{[-0.87,\, 0.89]}$
& $^{[-1.00,\, 1.00]}_{[-0.49,\, 0.51]}$
\\
\hline
\multicolumn{6}{|c|}{$10^2 \times A_{7,\,8,\,9}(B\to K^*\mu\bar\mu)_{[1,6]}$}
\\\hline
  $<0.1$
& $\phantom{-}4.5 \pm 5.0$ \hfill \cite{Aaij:2015oid}
& $^{[-23.4,\, 23.3]}_{[-14.5,\, 14.1]}$
& $^{[-8.9,\,  7.4] }_{[-5.9, \, 5.0]}$
& $^{[-23.7,\, 23.7]}_{[-12.0,\, 11.9]}$
& $^{[-18.3,\, 17.3]}_{[-6.1, \, 5.8]}$
\\
  $<0.1$
& $-4.7 \pm 5.8$ \hfill \cite{Aaij:2015oid}
& $^{[-0.9,\, 0.9]}_{[-0.5,\, 0.5]}$
& $^{[-6.9,\, 5.8]}_{[-4.6,\, 3.9]}$
& $^{[-0.9,\, 0.8]}_{[-0.4,\, 0.4]}$
& $^{[-0.6,\, 0.6]}_{[-0.2,\, 0.2]}$
\\
  $<0.1$
& $\phantom{-}3.3 \pm 4.2$ \hfill \cite{Aaij:2015oid}
& SM 
& $^{[-3.5,\, 4.2]}_{[-2.4,\, 2.8]}$
& SM & SM
\\
\hline
\multicolumn{6}{|c|}{$10^2 \times A_{8,\,9}(B\to K^*\mu\bar\mu)_{[15,19]}$}
\\\hline
  $<0.1$
& $\phantom{-}2.5 \pm 4.8$ \hfill \cite{Aaij:2015oid}
& SM
& $^{[-5.2,\, 4.3]}_{[-3.5,\, 2.9]}$
& SM & SM
\\
  $<0.1$
& $-6.1 \pm 4.3$ \hfill \cite{Aaij:2015oid}
& SM 
& $^{[-7.8,\, 9.4]}_{[-5.2,\, 6.3]}$ 
& SM & SM
\\
\hline
\multicolumn{6}{|c|}{
  ${\cal R}_{B\to K \nu\bar\nu}$, $\quad{\cal R}_{B\to K^* \nu\bar\nu}$, $\quad{\cal R}_{F_L}$}
\\\hline
  $1$
& $\leq 4.3$ \hfill \cite{Lees:2013kla}
& $^{[0.02,\, 1.10]}_{[0.63,\, 1.10]}$
& $^{[0.79,\, 1.24]}_{[0.78,\, 1.20]}$
& $^{[0.01,\, 1.10]}_{[0.65,\, 1.11]}$
& $^{[0.62,\, 1.11]}_{[0.71,\, 1.12]}$
\\
  $1$
& $\leq 4.4$ \hfill \cite{Lutz:2013ftz}
& $^{[0.02,\, 1.10]}_{[0.63,\, 1.10]}$
& $^{[0.87,\, 1.17]}_{[0.88,\, 1.17]}$
& $^{[0.01,\, 1.10]}_{[0.65,\, 1.11]}$
& $^{[0.62,\, 1.11]}_{[0.71,\, 1.12]}$
\\
  $1$
& ---
& SM 
& $^{[0.92,\, 1.07]}_{[0.93,\, 1.07]}$ & 
SM & SM
\\
\hline
\end{tabular}
}
\renewcommand{\arraystretch}{1.0}
\caption{\small 
  Ranges still allowed for observables when taking the constraints from
  \reftab{tab:exp-flavor-constraints} for the individual $s\to d$, $b\to d$
  and $b\to s$ sectors into account, fitting at the time same CKM and hadronic
  parameters. Upper and lower intervals are for $M_{\rm VLQ} = 1$~TeV and $10$~TeV, 
  respectively. Entries denoted as ``SM'' have tiny or no deviations from the SM.
  Experimental upper bounds are given at 90\% CL.
}
\label{tab:allowed-rngs-GSM}
\end{table}

Using the above constraints, we obtain allowed ranges for observables that are
yet to be measured (precisely), listed in \reftab{tab:allowed-rngs-GSM}. 
We furthermore analyze patterns for each transition, that will help to distinguish
VLQ models from other NP scenarios, and different VLQs from each other.
In this respect we point out that models $Q_d$ and $Q_V$ have the same experimental 
signatures in down-type quark FCNC transitions and are hence indistinguishable.
Such a distinction might be possible after invoking additional constraints from
up-type quark FCNC transitions, where both models differ from each other as indicated
in Eq.~\refeq{eq:Yuk:H}. Still, in $Q_V$ models strong correlations between the
up- and down-type sectors are not expected due to the in principle independent
up- and down-type Yukawa couplings. 

\begin{figure}
\centering
  \includegraphics[width=0.325\textwidth]{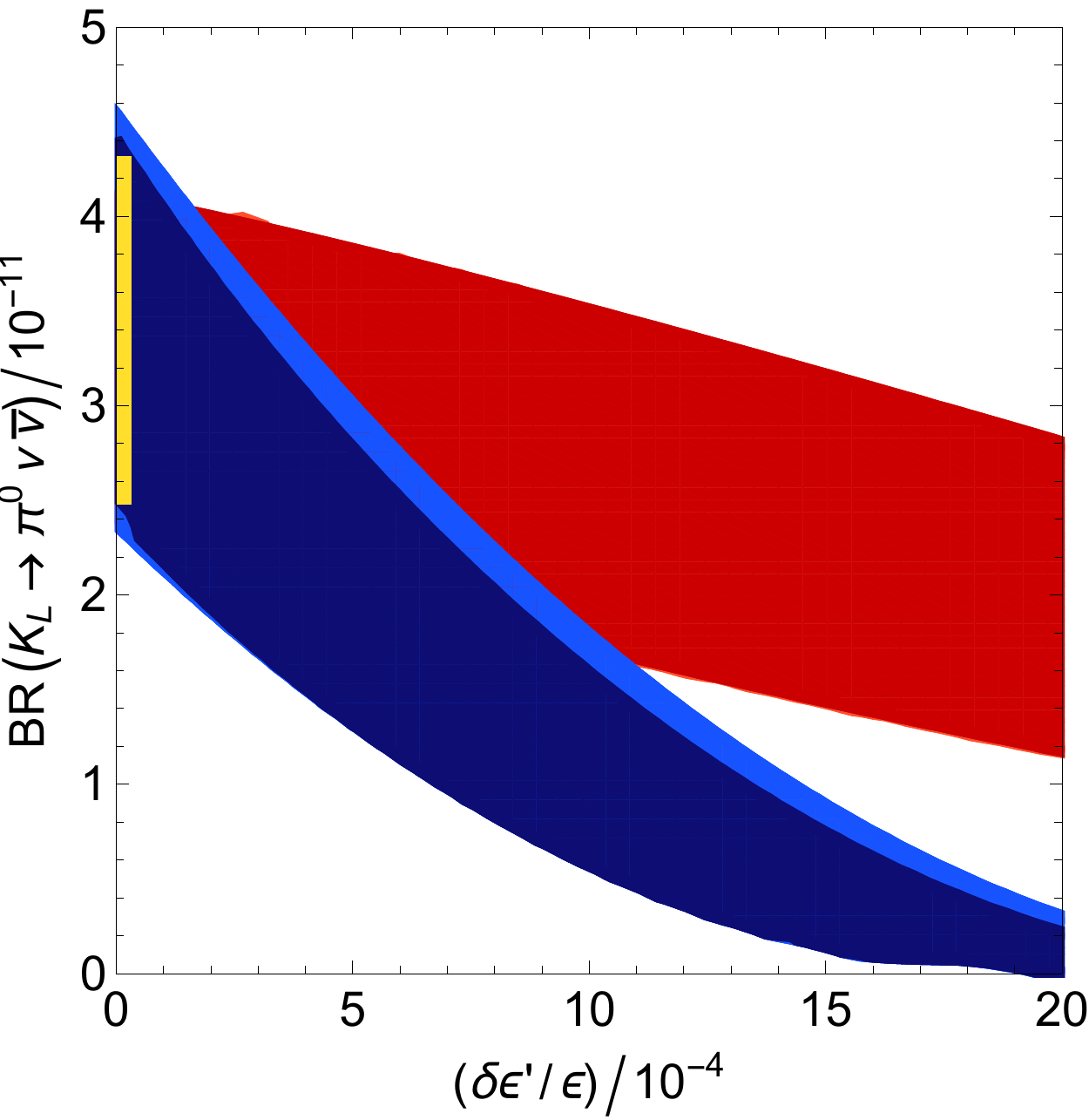} 
  \includegraphics[width=0.31\textwidth]{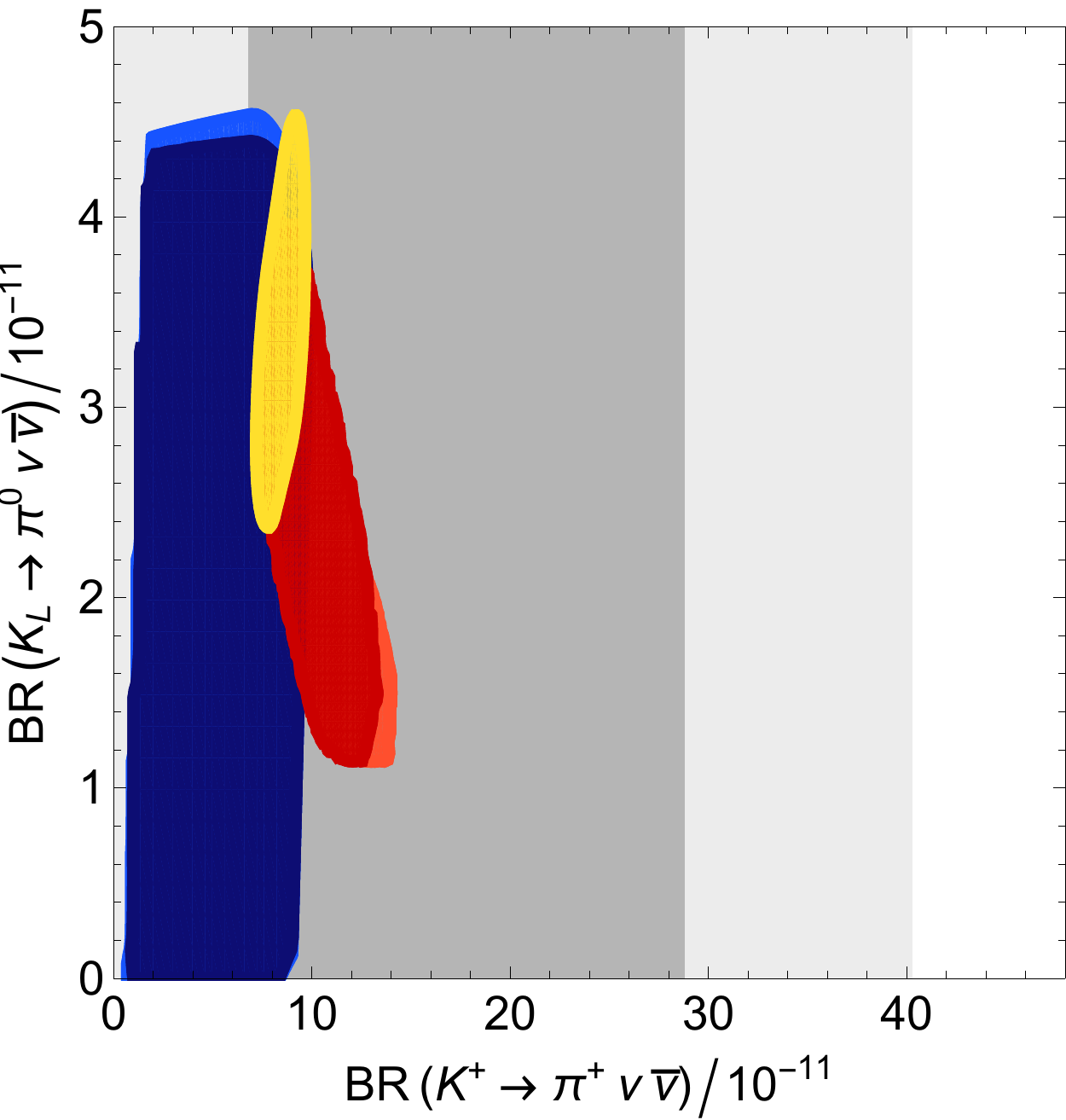}
  \includegraphics[width=0.315\textwidth]{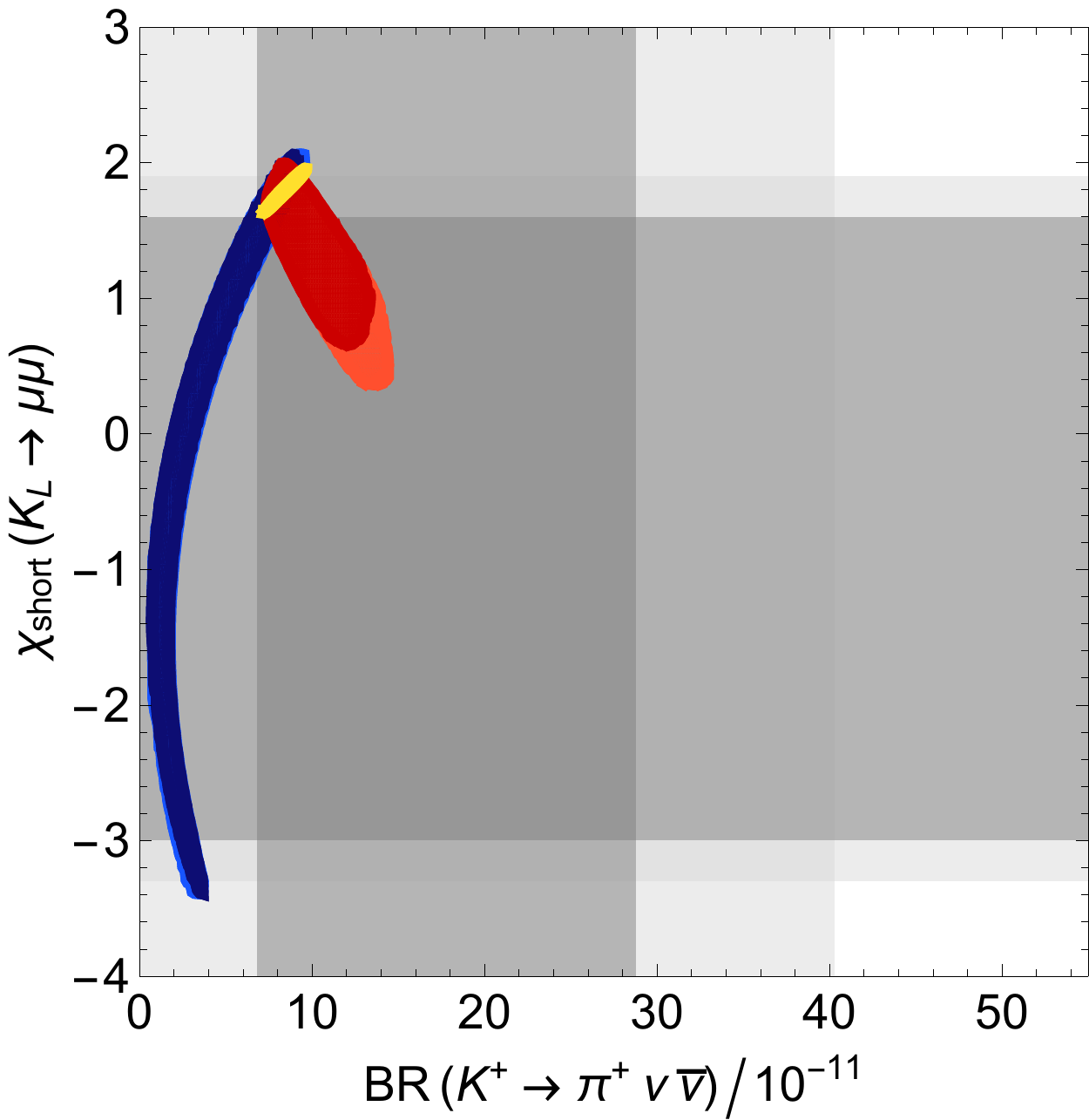}
\caption{\small
  The correlations of observables in the Kaon sector in $\GSM$ scenarios
  at $95\%$~CL for the SM, as well as for $M_{\rm VLQ} = 10$~TeV 
  [darker colours] and $M_{\rm VLQ} = 1$~TeV [lighter colours].
  The colours correspond to the SM prediction [Yellow] and the VLQ-representations
  $D$ [Blue] and $Q_V$ [Red]; the results for $T_{u,d}$ and $Q_d$ are very
  similar to the former and the latter, respectively. 
  Dark and light grey bands show experimental measurements at 1- and 
  2$\sigma$. 
}
\label{fig:GSM-crr-kaon}
\end{figure}

In the Kaon sector, we make the following observations, see also
\reffig{fig:GSM-crr-kaon}:
\begin{itemize}
\item The VLQ models allow to enhance $\epe$ significantly, thereby addressing
  the apparent gap between the SM prediction and data, at the expense of
  suppressing $Br(\klpn)$. This suppression is
  significantly weaker for $Q_V$ and $Q_d$ models (RH currents) than for $D$,
  $T_d$ and $T_u$ (LH currents), in accordance with the general study in
  \cite{Buras:2015jaq}. Simultaneous agreement with the data for
  $\varepsilon_K$ and $\epe$ can be obtained without fine-tuning of parameters.
\item While the impact of $\epe$ on $\klpn$ is large as stated above, $\kpn$ and
  $\epe$ are only weakly correlated.
  However, in RH models $\epsilon_K$ prevents large enhancements of 
  $Br(\kpn)$, the maximal enhancement is about $50\%$ of its SM value. 
  In models with LH currents, a strong suppression is possible, and
  the SM value corresponds to an upper bound in this case when a stricter 
  bound from $K_L\to\mu\bar\mu$ is used. This implies that a measurement
  of a significantly enhanced $Br(\kpn)$, as presently still allowed by data, could exclude
  all $\GSM$ models with a single VLQ representation, although in models 
  with LH currents a more conservative bound from  $K_L\to\mu\bar\mu$ would 
  presently still allow the enhancement of  $Br(\kpn)$ up to a factor of two.
\item In this context it should be again emphasized that the modes 
  $\kpn$ and $K_L\to\mu\bar\mu$ are strongly correlated in VLQ
  models, however, again differently so for LH and RH currents. While for RH
  currents one can easily infer the allowed range in one mode from a
  determination of the other, within the limited range allowed by $\epe$ 
  and $\epsilon_K$, LH-current models are more strongly
  constrained from $K_L\to\mu\bar\mu$. Progress for the latter mode depends solely on the
  capability to separate the long-distance contributions to this mode from the
  short-distance ones, since the relevant data are already very precise, see
  \refapp{app:master}. Note that there is basically no correlation between
  $\epe$ and $K_L\to\mu\bar\mu$, as they are governed by imaginary and real
  parts of the corresponding couplings, respectively.
\item The VLQ mass does not have a large impact on all these correlations, as
  can be seen by comparing the lighter and darker areas in \reffig{fig:GSM-crr-kaon}. 
  The reason is in LH models that $|\Delta F|=1$ transitions are the dominant
  constraints at both masses, rendering the allowed ranges for other $|\Delta F|=1$ 
  processes mass-independent. For RH models, the same conclusion is reached
  by considering additionally the fact that $\epsilon_K$ is dominated by RG-induced
  contributions which scale similarly to $|\Delta F|=1$ ones.
\end{itemize}

\begin{figure}
  \centering
  \includegraphics[width=0.305\textwidth]{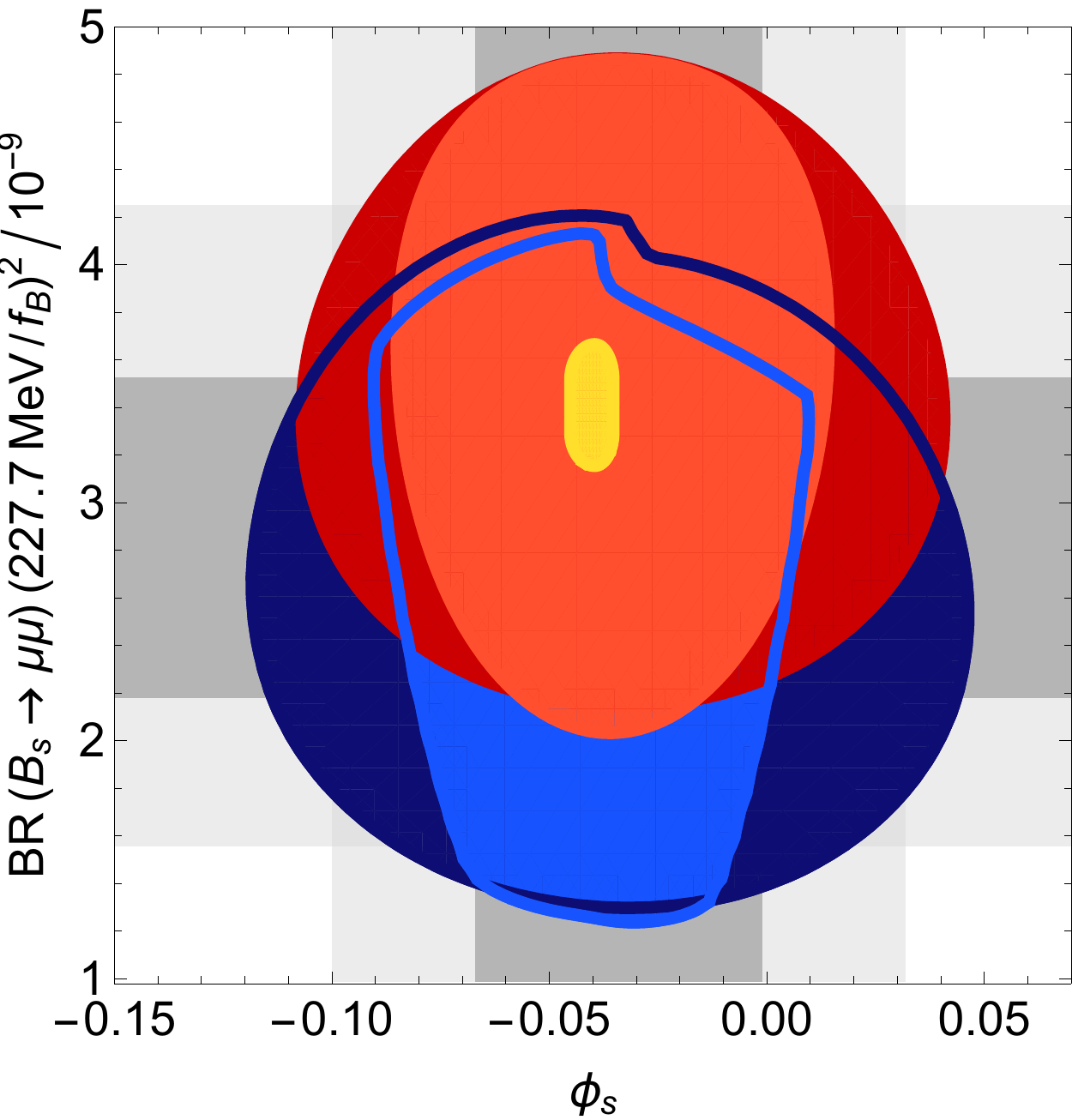} 
  \includegraphics[width=0.33\textwidth]{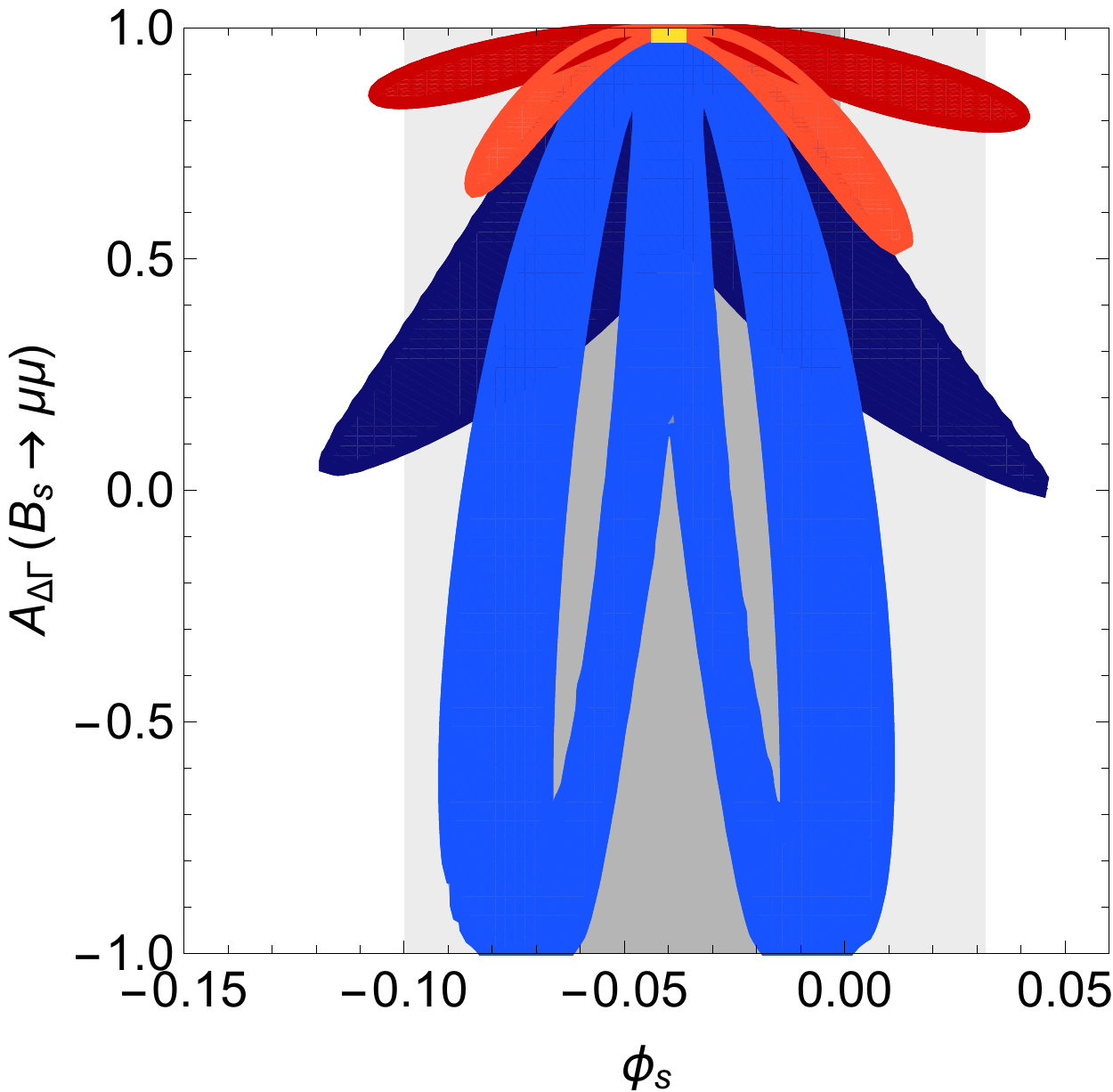}
  \includegraphics[width=0.33\textwidth]{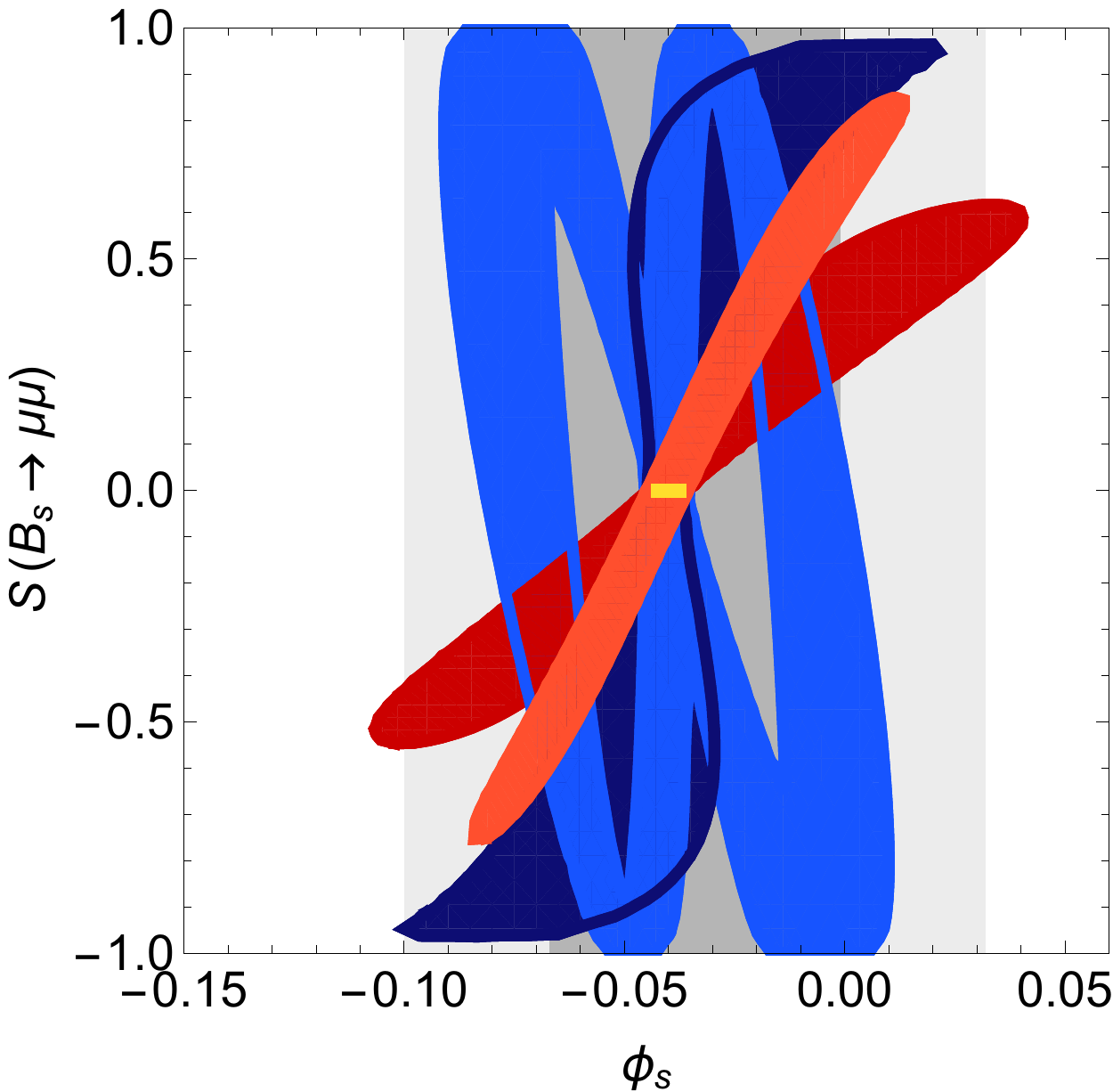}
  \vskip 1.0cm 
    \includegraphics[width=0.328\textwidth]{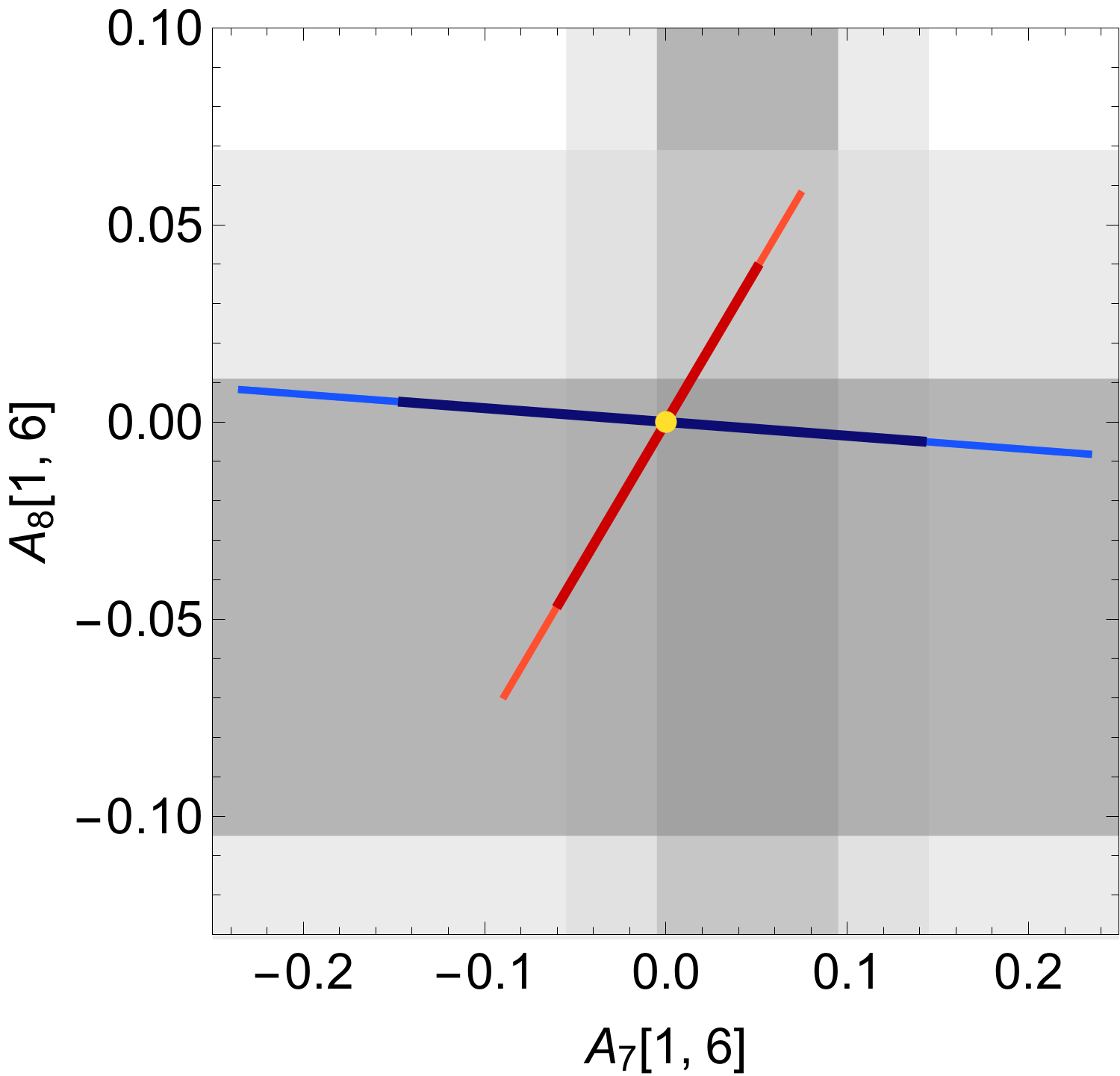} 
    \includegraphics[width=0.32\textwidth]{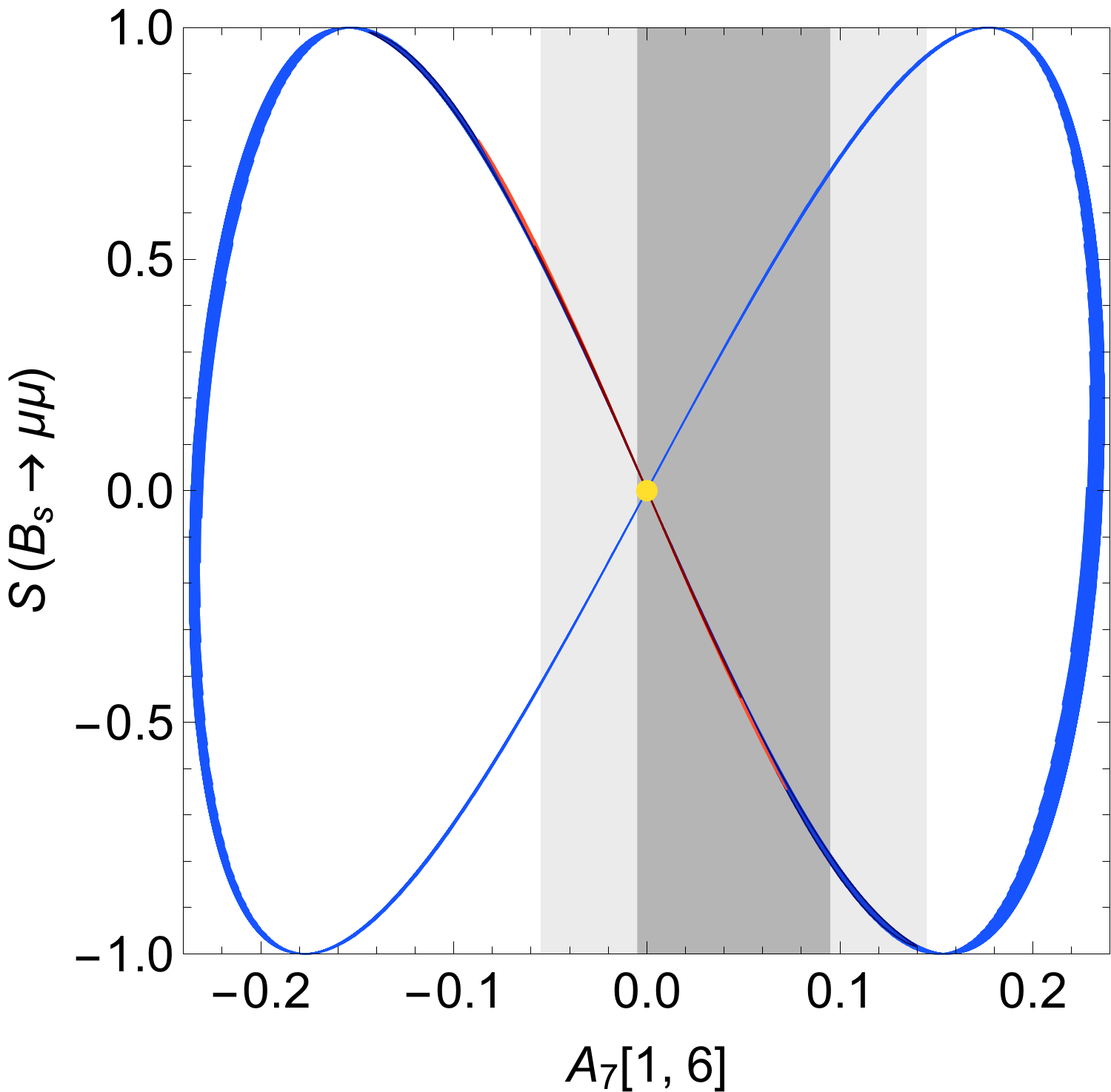} 
    \includegraphics[width=0.335\textwidth]{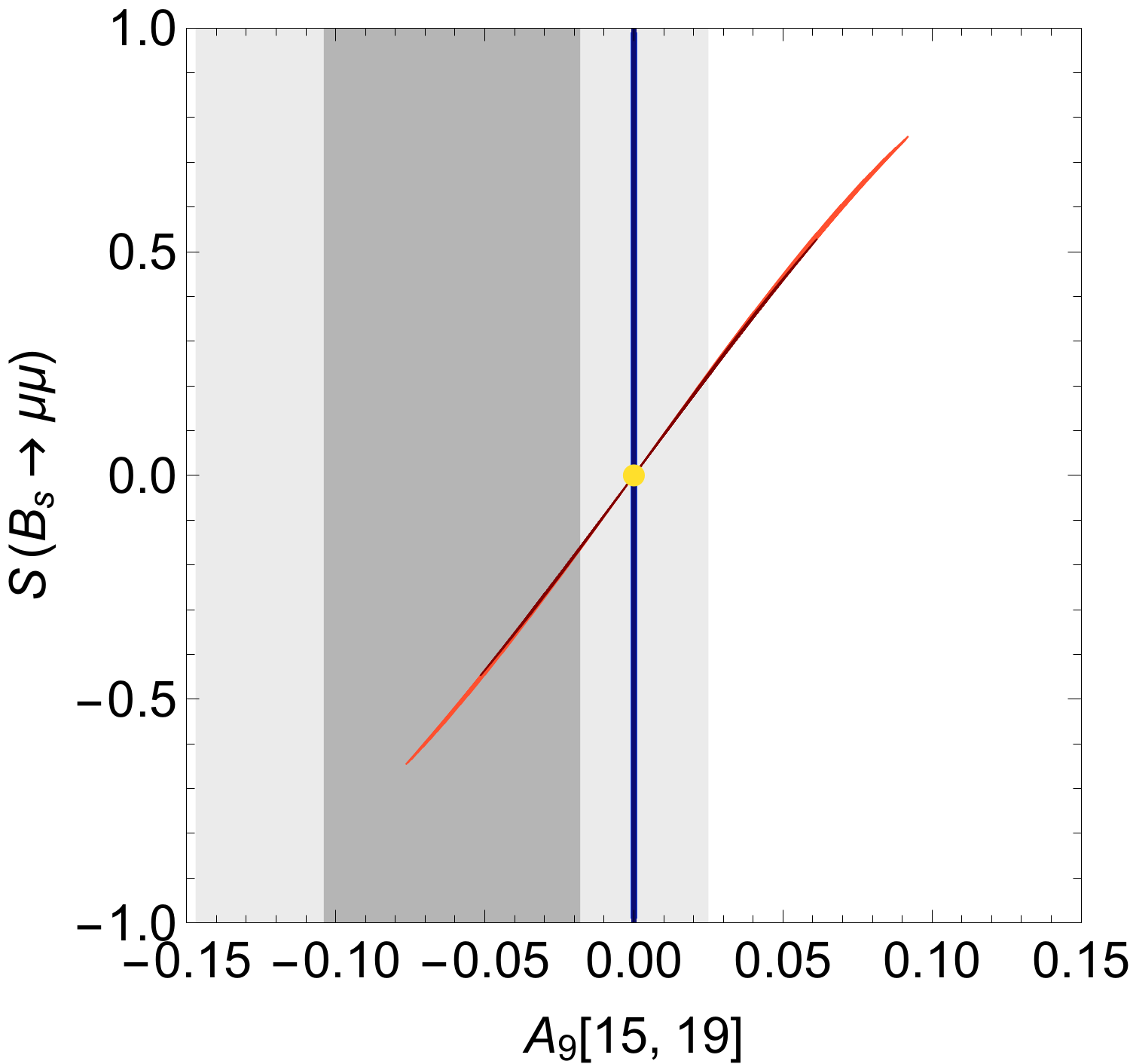} 
\caption{ \small
  The correlations of observables in the $b\to s$ sector in $\GSM$-scenarios for
  $M_{\rm VLQ} = 10$~TeV [darker colours] and $1$~TeV [lighter colours]
  within the 95\% CL regions. The colours are for VLQ-representations $D$ [Blue] (similar to $T_{u,d}$), 
  $Q_V=Q_d$ [Red]. Grey bands show experimental measurements at 1- and $2\sigma$ and the
  yellow dots are the SM predictions.
}
\label{fig:GSM-crr-bs}
\end{figure}

Correlation plots for observables in $b\to s$ processes are shown in
\reffig{fig:GSM-crr-bs}. We observe the following patterns:
\begin{itemize}
\item Since NP effects in all three quark transitions are governed by different
  parameters, the slight tensions in $|\Delta F|=2$ observables hinted at by new
  lattice data~\cite{Bazavov:2016nty} can easily be removed in VLQ models. This is
  in contrast to constrained-MFV models, where $\varepsilon_K$ prohibits
  large effects in $\Delta M_{d,s}$ \cite{Blanke:2016bhf}.
\item $Br(B_s\to\mu\bar\mu)$ can be strongly suppressed below its SM value, as 
  slightly favoured by experiment, while still allowing for sizeable NP effects
  in $\sin(2\beta_s)$, in particular in the case of models with LH currents. For
  $M_{\rm VLQ} = 1$~TeV $|\Delta F|=1$ observables constrain the NP effects
  in $\phi_s$ to be smaller than for larger VLQ masses.
\item Sizeable deviations from the SM prediction are still possible for the
  mass-eigenstate rate asymmetry $A_{\Delta\Gamma}(B_s\to \mu\bar\mu)$ and the
  mixing-induced CP-asymmetry $S(B_s\to \mu\bar\mu)$. 
  Indeed, both can essentially vary in the full range $[-1,\, 1]$ for
  LH models for $M_{\rm VLQ} = 1$~TeV. For RH models, $A_{\Delta\Gamma}(B_s\to \mu\bar\mu)
  \geq 50\%$ for $M_{\rm VLQ} = 1$~TeV, but still $|S(B_s\to \mu\bar\mu)|$ can
  reach up to $80\%$. For $M_{\rm VLQ} = 10$~TeV, the former is restricted to
  positive values in both LH and RH models, the latter slightly stronger constrained
  in RH models, but not in LH ones. Of course, the experimental measurements are
  very challenging for $S(B_s\to \mu\bar\mu)$. We note that to very good accuracy 
  $A_{\Delta\Gamma}^2 + S^2 = 1$, since the direct CP-asymmetry $C(B_s\to \mu\bar\mu)$
  is negligible.
\item CP-violating quantities are almost $100\%$ correlated in $b\to s$ transitions
  as long as only one representation is considered. The reason is that the SM
  predictions are tiny and all NP contributions therefore directly proportional to
  the imaginary part of $\Lambda_{bs}$, which hence cancels in the ratio of two
  CP-violating quantities. For small NP contributions, the asymmetries are simply
  proportional to each other, for larger effects the relation depends on the normalisation
  of the asymmetry. These statements hold not only in VLQ models, but in all models
  that provide only a single new phase in $b\to s$ transitions, only the proportionality
  constant changes in other models.
\item The imaginary parts of $b\to s\mu\bar\mu$ Wilson coefficients
  $C_{9,9',10,10'}$ can give rise to naive T-odd CP-asymmetries $A_{7,8,9}$ in
  $B\to K^* \mu\bar\mu$ that are tiny in the SM.\footnote{Note that we use
    different convention of angles w.r.t. LHCb:
   $A_{7,9}^{} = -A_{7,9}^{\rm LHCb}$ and $A_{8}^{} = A_{8}^{\rm LHCb}$.}  The
  rough dependences on the Wilson coefficients are \cite{Bobeth:2008ij}
  \begin{align}
    A_7 & 
    \propto \mbox{Im}\big[ (C_{10}^{}-C_{10}') C_7^* \big], &
    A_{8,9} & 
    \propto \mbox{Im}\big[ C_9^{} C_9^{\prime\ast} + C_{10}^{} C_{10}^{\prime\ast} 
    + \ldots  \big], 
  \end{align}
  where the dots indicate other numerically suppressed interference terms of
  $C_{9,9'}$ with $C_7$ that are included in the numerical evaluation.  The
  $A_7$ remains tiny at high dilepton invariant mass $q^2$ \cite{Bobeth:2012vn}.
  These CP-asymmetries have been measured in various $q^2$-bins by LHCb
  \cite{Aaij:2015oid} and we choose $q^2 \in [1,\,6]$ and $[15,\,19]$ GeV$^2$,
  which have smallest experimental and theoretical uncertainties. As can be seen in
  \reftab{tab:allowed-rngs-GSM}, the largest VLQ-effects in $A_{8,9}$ arise in
  RH $\GSM$-scenarios $Q_d$ and $Q_V$, almost independent from the VLQ mass and
  with a strong anti-correlation shown in \reffig{fig:GSM-crr-bs}. The potential 
  size of VLQ effects exceeds slightly the current experimental uncertainties,
  specifically for the CP asymmetry $A_7$ in LH scenarios, such that
  improved measurements will provide additional bounds on VLQ couplings in the
  future, especially on their imaginary parts. $A_7$ is correlated with $A_8$ and
  anti-correlated with $A_9$ in RH scenarios, whereas in LH scenarios $A_{8,9}$ remain
  SM-like.
\end{itemize}

\begin{figure}
  \includegraphics[width=0.46\textwidth]{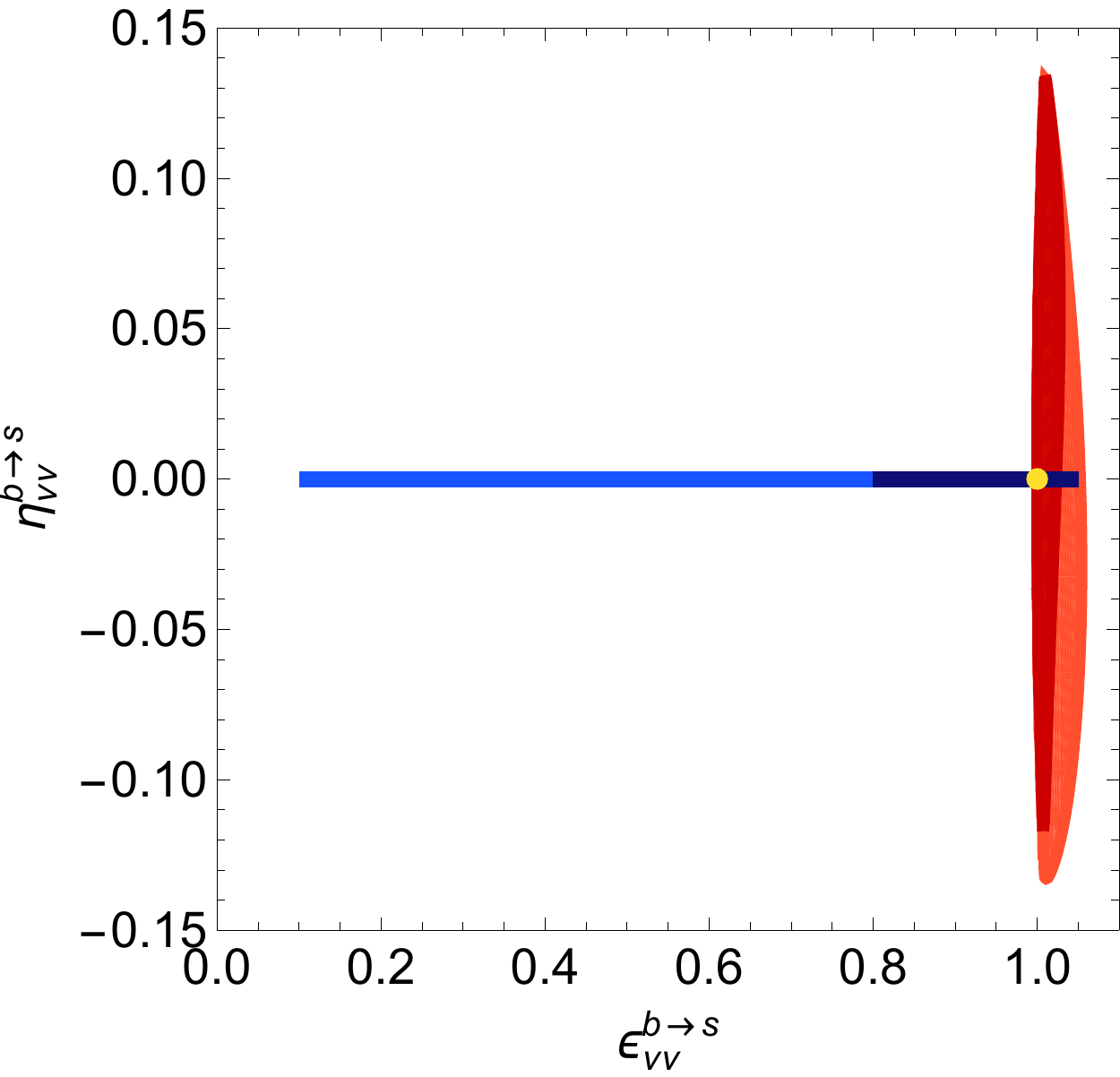}
  \qquad\parbox[b]{0.42\textwidth}{
\caption{ \small 
  \label{fig:GSM-crr-bs-2}
  The correlations of the quantities $\eta$ and $\epsilon$, which
  determine the observables in the $b\to s \nu\bar\nu$ sector in
  $\GSM$-scenarios for $M_{\rm VLQ} = 10$~TeV [darker] and $1$~TeV [lighter]
  within the 95\% CL regions. The colours are for VLQ-representations
  $D$ [Blue] (similar to $T_{u,d}$), $Q_V=Q_d$ [Red].}
}
\end{figure}

\begin{itemize}  
\item The decays $B\to K^{(*)}\nu\bar\nu$ are also sensitive probes of LH and RH
  NP effects due to $Z$-exchange and in order to exhibit these effects
  we consider the ratios \cite{Altmannshofer:2009ma}
  \begin{align}
    \epsilon & = \frac{\sqrt{|C_L|^2+|C_R|^2}}{|C_L^{\rm SM}|} 
    & & \mbox{and} & 
    \eta & = \frac{-{\rm Re}(C_LC_R^*)}{|C_L|^2+|C_R|^2} ,
  \end{align}
  which are unity and zero in the SM, respectively, and which determine the observables
  \begin{align}
    \label{eq:b->svv:obs}
    {\cal R}_{B\to K^{(*)} \nu\bar\nu} & 
      = \frac{Br(B\to K^{(*)} \nu\bar\nu)}{Br(B\to K^{(*)} \nu\bar\nu)_{\rm SM}}, &
    {\cal R}_{F_L} & 
      = \frac{F_L(B\to K^* \nu\bar\nu)}{F_L(B\to K^* \nu\bar\nu)_{\rm SM}}
  \end{align}
  via\cite{Buras:2014fpa}
  \begin{align}
    {\cal R}_{B\to K \nu\bar\nu} & = (1-2\eta)\epsilon^2, & 
    {\cal R}_{B\to K^* \nu\bar\nu} & = (1+\kappa_\eta\eta)\epsilon^2, & 
    {\cal R}_{F_L} & =  \frac{1+2\eta}{1+\kappa_\eta \eta} ,
  \end{align}
  where $\kappa_\eta$ is form-factor dependent and given in Ref.~\cite{Buras:2014fpa}.
  The Belle~II experiment is expected to measure these branching ratios with 30\%
  uncertainty \cite{Aushev:2010bq} if they are of the size as predicted in the SM. 
  In RH scenarios large VLQ effects are excluded due to the strong complementarity of the
  $|\Delta F|=1$ constraints from $Br(B_s\to \mu\bar\mu)$ and $Br(B^+\to K^+ \mu\bar\mu)$
  as mentioned above. $\epsilon$ has to be larger than one in these cases.
  The VLQ effects for $M_{\rm VLQ} = 1$~TeV can lead to a rather large suppression
  in LH scenarios for $\epsilon$ while $\eta = 0$, leading to maximally correlated
  ${\cal R}_{B\to K^{(*)} \nu\bar\nu}$. The suppression is smaller for $M_{\rm VLQ}
  = 10$~TeV, whereas ${\cal R}_{F_L} = 1$. The correlation plot is shown in 
  \reffig{fig:GSM-crr-bs-2}. It will be challenging to distinguish the small
  deviations from SM predictions in RH scenarios; however, large (suppression) effects
  are possible and LH and RH scenarios are well distinguishable. A measurement 
  of $\epsilon$ significantly larger than one would challenge all $\GSM$ scenarios with
  a single VLQ representation.
\end{itemize}

\begin{figure}
  \includegraphics[width=0.46 \textwidth]{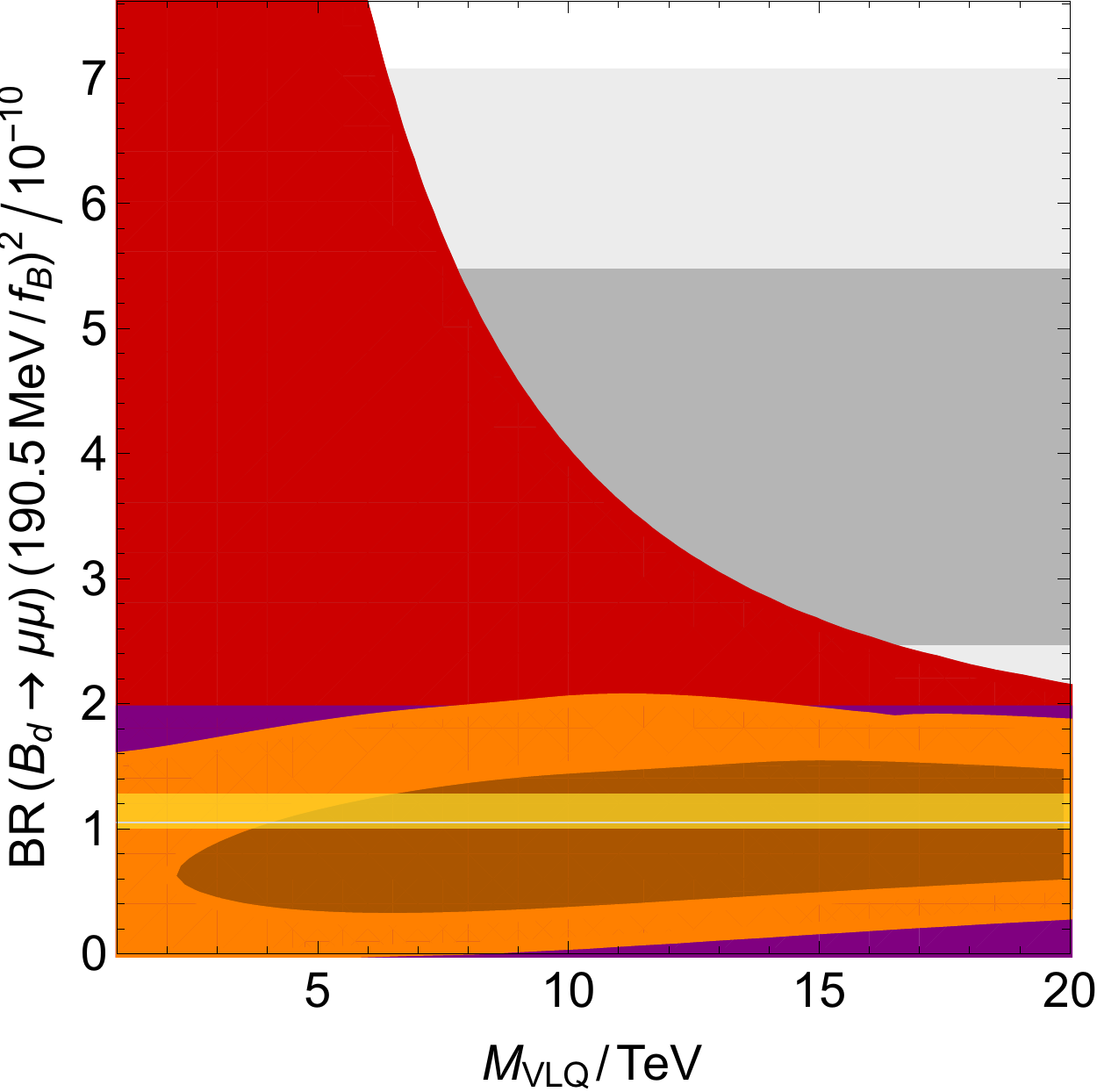} 
  \hfill
  \includegraphics[width=0.46\textwidth]{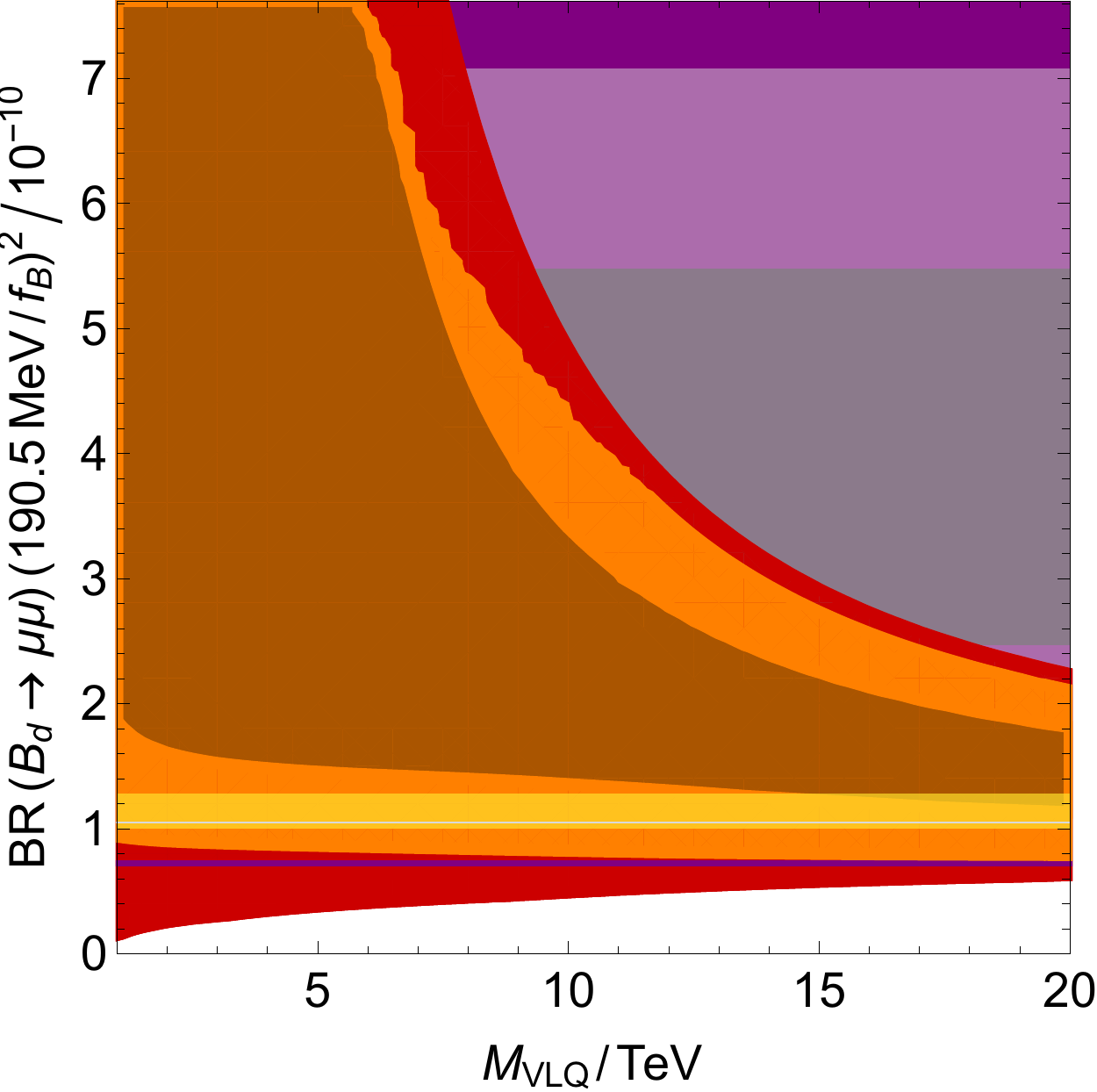}
\caption{\small
   Predictions for $Br(B_d\to \mu\bar\mu)$ for the LH $\GSM$ scenario $D$ [left]
   and  RH $\GSM$ scenario  $Q_V$ [right], in dependence on the VLQ mass. In dark
   red the constraint from $|\Delta F|=2$ processes is shown, \emph{i.e.} 
   $\Delta M_d$ and $\sin 2\beta$,
   in purple the constraint from $B^+\to \pi^+\mu\bar\mu$, and in orange their
   combination. The yellow band corresponds to the SM prediction, the grey one
   to the measurement by the CMS and LHCb collaborations~\cite{CMS:2014xfa}.
   All constraints correspond to $95\%$~CL, only inner darker bands to $68\%$~CL.
}
\label{fig:bdmumu}
\end{figure}

Similar correlation plots exist for $b\to d$ processes; however, given the CKM
suppression of these modes compared to $b\to s$, precision measurements in 
$b\to d\ell\bar\ell$ and significant measurements of $b\to d \nu\bar\nu$ processes
are not expected in the next couple of years. Nevertheless, we illustrate in
\reffig{fig:bdmumu} the impact of more precise measurements in this sector
exemplarily for $Br(B_d\to\mu\bar\mu)$. All $|\Delta F|=1$ processes depend only
on the combination $\Delta_{ij}$, see \refeq{eq:Dij}, of NP parameters; the allowed 
range predicted from one $|\Delta F|=1$ process for another is therefore mass-independent,
in contrast to the prediction from $|\Delta F|=2$ processes. The present measurement
from the CMS and LHCb collaborations is about $2\sigma$ larger than the SM prediction.
As seen in \reffig{fig:bdmumu} a confirmation of the present central value with
higher precision would exclude LH $\GSM$ scenarios and yield at least an upper
limit on $M_{\rm VLQ}$ for the RH ones, in accordance with the discussion in
\refsec{sec:mass}.

%
%

\subsection[$\GSMUpr(\Phi)$ model]
{\boldmath $\GSMUpr(\Phi)$ model}

\begin{figure}
  \includegraphics[width=0.55\textwidth]{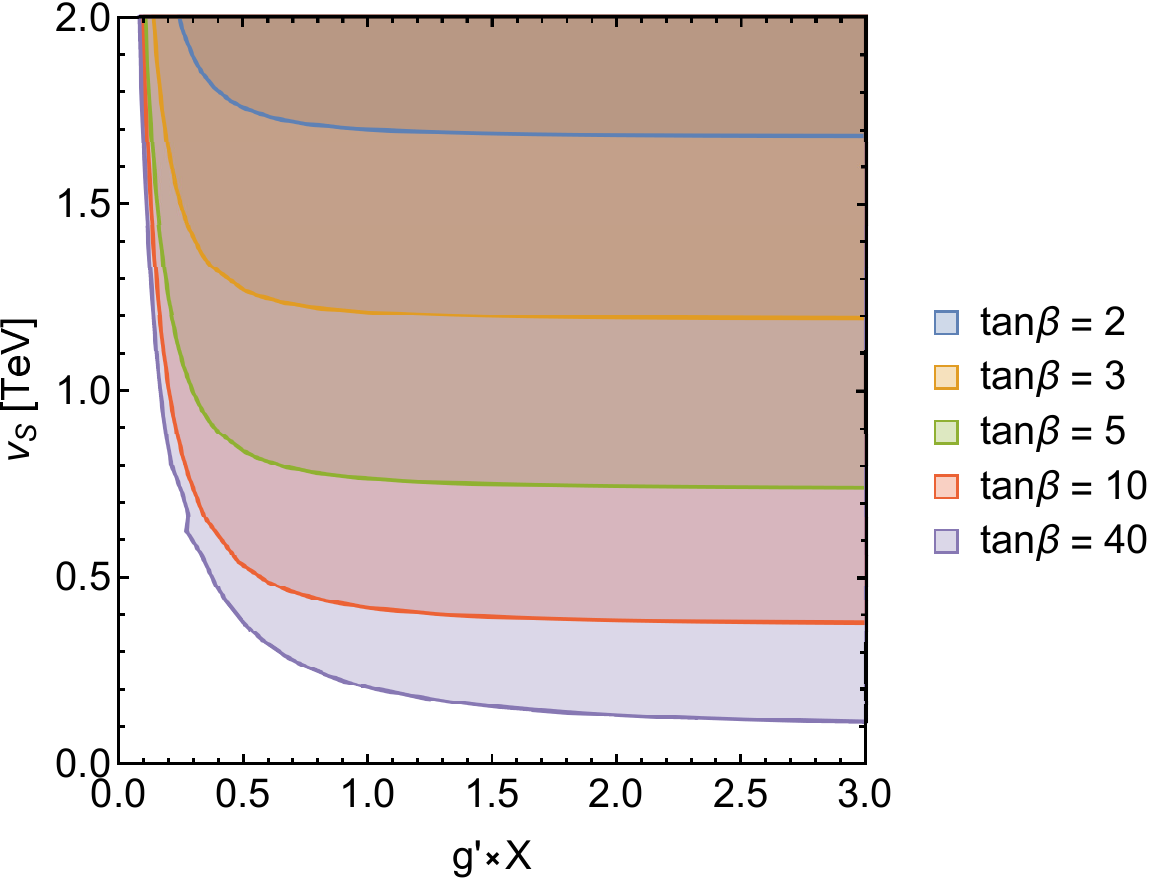} 
  \hfill\parbox[b]{0.42\textwidth}{
\caption{ \small 
  \label{fig:GSMprPhi-constr}
  The allowed 95\% CL regions in the $v_S$ versus $g' X$ plane for fixed 
  $\tan\beta = 2,\,3,\,5,\,10,\,40$ [Blue, Yellow, Green, Red, Purple] from
  the constraints $M_Z$ and partial widths $\Gamma[Z\to \ell\bar\ell]$ 
  ($\ell = e,\, \mu,\,\tau$), imposing $M_Z < M_{Z'}$.}
}
\end{figure}

In $\GSMUpr(\Phi)$ models $|\Delta F|=1$ transitions are suppressed by $\tan\beta$
compared to $\GSM$ models, such that $|\Delta F|=2$ transitions dominate via the
box contributions the constraints on VLQ couplings. In our numerical analysis of
$\GSMUpr(\Phi)$ models we fix the parameters
\begin{align}
  g' & = 1.5 , & X & = 1 , & M_{\rm VLQ} & = 10~{\rm TeV},
\end{align}
and choose two benchmark points BP1 and BP2:
\begin{align}
  \mbox{BP1:} && \tan\beta & = 2 , &  v_S & = 1.8\,\mbox{TeV} ,
\\
  \mbox{BP2:} && \tan\beta & = 3 , &  v_S & = 1.3\,\mbox{TeV} ,
\end{align}
in the lower range of possible values of $\tan\beta$ --- see also 
\reffig{fig:GSMprPhi-constr} --- from constraints described in 
\refapp{app:scalar:S+H+Phi} to maximally enhance VLQ contributions
in $|\Delta F|=1$ transitions. The corresponding $Z$ and $Z'$ masses
and mixing angles are
\begin{align}
  \mbox{BP1:} && 
  M_Z & = 91.51 \,\mbox{GeV} , &  
  M_{Z'} & = 1.36 \,\mbox{TeV} , &  
  \xi_{ZZ'} & = 0.0037 ;
\\ 
  \mbox{BP2:} &&
  M_Z & = 91.58 \,\mbox{GeV} , &  
  M_{Z'} & = 0.98 \,\mbox{TeV} , & 
  \xi_{ZZ'} & = 0.0035 .
\end{align}

The allowed regions of $\Lambda_{ij}$ in $\GSMUpr(\Phi)$ models correspond to the
regions allowed  by $|\Delta F|=2$ constraints in $\GSM$ models given in
\reffig{fig:GSM-A_ij-fits-10TeV}. We find
that $|\Delta F|=1$ processes in \reftab{tab:exp-flavor-constraints} provide 
only tiny additional constraints in $b\to d,s$ and small ones in 
$s\to d$, allowing thus in $\GSMUpr(\Phi)$ models much 
larger values for $\Lambda_{ij}$ compared to $\GSM$ models.

The ranges still allowed for different observables with $|\Delta F|=1,2$ transitions 
are listed in \reftab{tab:allowed-rngs-GSMprPhi}, obtained by varying $\Lambda_{ij}$
within the 95\%~CL regions, neglecting theory uncertainties. For this purpose $(\epe)_{\rm NP}$
has been restricted as given in Eq.~\refeq{eq:epe-constraint} and we used
here $Br(K_L \to \mu\bar\mu)_{\rm SD} < 2.5 \times 10^{-9}$. Notable features 
for the benchmark points are:
\begin{itemize}
\item $\epe$ can also be enhanced in $\GSMUpr(\Phi)$ models and thereby decrease 
  the tension with the measurement. Especially in RH scenarios the constraint
  \refeq{eq:epe-constraint} is saturated, such that even larger effects are
  possible. The enhancement of $\epe$ falls off fast for larger
  values of $\tan\beta$ and $v_S$ than in the benchmark points.
\item Whereas VLQ effects in $Br(\klpn)$ are small, $Br(\kpn)$ can still be 
  enhanced over the SM prediction by a factor of two for LH and five for RH 
  scenarios, while even larger effects are excluded by the upper bound on 
  $Br(K_L \to \mu\bar\mu)_{\rm SD}$.
  Most notably, $(\Delta M_K)_{\rm SD}$ can also be enhanced by a factor of more
  than two, in contradistinction to $\GSM$ models, where VLQ effects are tiny.
  The reason for this enhancement is the absence of strong constraints from 
  $|\Delta F|=1$ on the real part of $\Lambda_{sd}$. Thus large $(\Delta M_K)_{\rm SD}$
  is independent of $\epe$, since the latter is sensitive to the imaginary part
  of $\Lambda_{sd}$. This effect is enhanced with decreasing VLQ effects in
  $|\Delta S|=1$ transitions as can be seen by comparing the results for BP1 and BP2.
  For these benchmark points large effects in $Br(K^+\to \pi^+ \nu\bar\nu)$ 
  in RH models remain independently of whether a conservative or stricter bound on
  $Br(K_L \to \mu\bar\mu)_{\rm SD}$ is used, although the stricter bound would force
  $Br(K^+\to \pi^+ \nu\bar\nu)$ to be above the SM prediction. But the large effects
  in $Br(K^+\to \pi^+ \nu\bar\nu)$ in RH models would be constrained by an improved
  theoretical prediction of long-distance effects in $\Delta M_K$ from lattice. 
  In LH models the upper and especially the lower bounds on $Br(K^+\to \pi^+ 
  \nu\bar\nu)$ would also be constrained by the latter improvements and also
  improved bounds on $Br(K_L \to \mu\bar\mu)_{\rm SD}$.
\item The VLQ effects are small for $Br(B_{s,d}\to \mu\bar\mu)$ and 
  $A_{\Delta\Gamma}(B_s\to \mu\bar\mu)$, as can be seen from \reffig{fig:B-mumu-crr} and
  \reftab{tab:allowed-rngs-GSMprPhi}, respectively, but can be still sizeable for 
  $S(B_s\to \mu\bar\mu)$.
  The CP asymmetries $A_{7,8,9}(B\to K^*\mu\bar\mu)$ can
  still be significantly enhanced over the SM to the percent level, but are
  a factor 2-3 smaller for BP1 than in $\GSM$ models, see 
  \reftab{tab:allowed-rngs-GSM}.
\item VLQ effects in $B\to K^{(*)}\nu\bar\nu$ in $\GSMUpr(\Phi)$ models are
  smaller than in $\GSM$ models, at the level of only $(10-20)$\%
  deviation from the SM predictions.
\end{itemize}
We provide a summary of enhancements and/or suppressions w.r.t. the SM predictions
of the observables discussed above due to VLQ effects in \reftab{tab:DNA-after-fit}. 

\begin{table}[htbp]
\addtolength{\arraycolsep}{4pt}
\renewcommand{\arraystretch}{1.5}
\centering
\resizebox{\textwidth}{!}{
\begin{tabular}{|c|r|cccc|}
\hline
 SM & measurement & $D$ & $Q_d$ & $T_u$ & $T_d$
\\
\hline\hline
\multicolumn{6}{|c|}{$10^{4} \times (\epe)_{\rm NP}$}
\\
  (\ref{eq:epe:LATTICE}, \ref{eq:epe:LBGJJ}, \ref{KNT})
& \refeq{eq:epe:EXP}
& $^{[0.0,\, 20.0]}_{[0.0,\, 20.0]}$
& $^{[0.0,\, 20.0]}_{[0.0,\, 19.9]}$ 
& $^{[0.0,\, 20.0]}_{[0.0,\, 20.0]}$
& $^{[0.0,\, 19.9]}_{[0.0,\, 20.0]}$
\\
\hline
\multicolumn{6}{|c|}{$10^{11} \times Br(K_L\to\pi^0\nu\bar\nu)$, 
  $\quad 10^{11} \times Br(K^+\to\pi^+\nu\bar\nu)$}
\\
3.2\,[2.5,\,4.3] 
& $\leq 2600$ \hfill \cite{Ahn:2009gb}
& $^{[0.1,\, 3.2]}_{[0.1,\, 3.2]}$
& $^{[1.8,\, 3.2]}_{[1.8,\, 3.2]}$
& $^{[0.1,\, 3.2]}_{[0.1,\, 3.2]}$
& $^{[0.1,\, 3.2]}_{[0.9,\, 3.2]}$
\\
8.5\,[7.3,\,9.5] 
& $17.3^{+11.5}_{-10.5}$ \hfill \cite{Artamonov:2008qb}
& $^{[1.5,\, 15.4]}_{[1.6,\, 15.4]}$ 
& $^{[3.9,\, 45.4]}_{[3.9,\, 45.4]}$ 
& $^{[1.5,\, 15.4]}_{[1.6,\, 15.4]}$ 
& $^{[1.5,\, 15.4]}_{[1.6,\, 15.4]}$
\\
\hline
\multicolumn{6}{|c|}{$\quad 10^9 \times Br(K_L \to \mu\bar\mu)_{\rm SD}$}
\\
& $\leq 2.5$ \hfill \cite{Isidori:2003ts}
& $^{[0.0,\, 2.5]}_{[0.0,\, 2.5]}$
& $^{[0.0,\, 2.5]}_{[0.0,\, 2.5]}$
& $^{[0.0,\, 2.5]}_{[0.0,\, 2.5]}$
& $^{[0.0,\, 2.5]}_{[0.0,\, 2.5]}$ 
\\
\hline
\multicolumn{6}{|c|}{$\quad 10^4 \times (\Delta M_K)_{\rm SD}$ [ps$^{-1}$]}
\\
& $52.93 \pm 0.09$ \hfill \cite{Agashe:2014kda}
& $^{[46.0,\, 58.8]}_{[44.7,\, 95.7]}$ 
& $^{[46.4,\, 71.1]}_{[46.1,\, 144.9]}$
& $^{[45.9,\, 61.9]}_{[44.2,\, 105.9]}$
& $^{[44.2,\, 105.9]}_{[44.5,\, 161.0]}$
\\
\hline\hline
\multicolumn{6}{|c|}{$10^{10} \times Br(B_d\to\mu\bar\mu)$}
\\
  $1.14\,[0.94, 1.32]$
& $\leq 6.3$ \hfill \cite{Aaij:2013aka}
& $^{[0.74,\, 1.32]}_{[0.90,\, 1.21]}$
& $^{[0.90,\, 1.34]}_{[0.97,\, 1.19]}$
& $^{[0.77,\, 1.30]}_{[0.92,\, 1.19]}$
& $^{[0.92,\, 1.20]}_{[1.00,\, 1.13]}$
\\
\hline\hline
\multicolumn{6}{|c|}{$A_{\Delta\Gamma}(B_s\to \mu\bar\mu)$, $\quad S(B_s\to \mu\bar\mu)$}
\\
  $1$
& ---
& $^{[0.88,\, 1.00]}_{[0.97,\, 1.00]}$
& $^{[0.95,\, 1.00]}_{[0.98,\, 1.00]}$ 
& $^{[0.91,\, 1.00]}_{[0.98,\, 1.00]}$ 
& $^{[0.97,\, 1.00]}_{[0.99,\, 1.00]}$
\\
  $0$
& ---
& $^{[-0.47,\, 0.46]}_{[-0.25,\, 0.25]}$ 
& $^{[-0.34,\, 0.34]}_{[-0.18,\, 0.18]}$
& $^{[-0.43,\, 0.42]}_{[-0.22,\, 0.22]}$
& $^{[-0.22,\, 0.22]}_{[-0.11,\, 0.11]}$
\\
\hline
\multicolumn{6}{|c|}{$10^2 \times A_{7,\,8,\,9}(B\to K^*\mu\bar\mu)_{[1,\,6]}$}
\\
  $<0.1$
& $\phantom{-}4.5 \pm 5.0$ \hfill \cite{Aaij:2015oid}
& $^{[-5.0,\, 5.1]}_{[-2.6,\, 2.7]}$
& $^{[-3.7,\, 3.7]}_{[-1.9,\, 1.9]}$
& $^{[-4.5,\, 4.6]}_{[-2.4,\, 2.4]}$ & $^{[-2.4,\, 2.4]}_{[-1.2,\, 1.2]}$
\\
  $<0.1$
& $-4.7 \pm 5.8$ \hfill \cite{Aaij:2015oid}
& $^{[-0.6,\, 0.5]}_{[-0.5,\, 0.4]}$ & $^{[-3.0,\, 2.9]}_{[-1.6,\, 1.6]}$ & $^{[-0.5,\, 0.5]}_{[-0.4,\, 0.4]}$ & $^{[-0.3,\, 0.2]}_{[-0.2,\, 0.2]}$
\\
  $<0.1$
& $\phantom{-}3.3 \pm 4.2$ \hfill \cite{Aaij:2015oid}
& SM & $^{[-1.7,\, 1.7]}_{[-0.9,\, 0.9]}$ & SM & SM
\\
\hline
\multicolumn{6}{|c|}{$10^2 \times A_{8,\,9}(B\to K^*\mu\bar\mu)_{[15,\,19]}$}
\\
  $<0.1$
& $\phantom{-}2.5 \pm 4.8$ \hfill \cite{Aaij:2015oid}
& SM & $^{[-2.4,\, 2.4]}_{[-1.4,\, 1.4]}$ & SM & SM 
\\
  $<0.1$
& $-6.1 \pm 4.3$ \hfill \cite{Aaij:2015oid}
& SM & $^{[-4.4,\, 4.4]}_{[-2.4,\, 2.5]}$ & SM & SM 
\\
\hline
\multicolumn{6}{|c|}{
  ${\cal R}_{B\to K \nu\bar\nu}$, $\quad{\cal R}_{B\to K^* \nu\bar\nu}$, $\quad{\cal R}_{F_L}$}
\\
  $1$
& $\leq 4.3$ \hfill \cite{Lees:2013kla}
& $^{[0.78,\, 1.13]}_{[0.88,\, 1.09]}$ & $^{[0.87,\, 1.15]}_{[0.93,\, 1.08]}$ & $^{[0.80,\, 1.13]}_{[0.89,\, 1.08]}$ & $^{[0.90,\, 1.08]}_{[0.95,\, 1.05]}$ 
\\
  $1$
& $\leq 4.4$ \hfill \cite{Lutz:2013ftz}
& $^{[0.78,\, 1.13]}_{[0.88,\, 1.09]}$ & $^{[0.91,\, 1.10]}_{[0.95,\, 1.05]}$ & $^{[0.80,\, 1.13]}_{[0.90,\, 1.08]}$ & $^{[0.90,\, 1.08]}_{[0.95,\, 1.05]}$
\\
  $1$
& ---
& SM & $^{[0.95,\, 1.04]}_{[0.97,\, 1.02]}$ & SM & SM 
\\
\hline
\end{tabular}
}
\renewcommand{\arraystretch}{1.0}
\caption{\small Ranges still allowed for observables when varying $\Lambda_{ij}$ of
  $\GSMUpr(\Phi)$ models in the 95\% CL ranges for individual $s\to d$, $b\to d$ and 
  $b\to s$ sectors for benchmark points BP1/BP2 [upper/lower]. Moreover $(\epe)_{\rm NP}$
  is restricted as given in Eq.~\refeq{eq:epe-constraint}. Entries
  denoted as ``SM'' have tiny or no deviations from the SM.
  Experimental upper bounds are given at 90\% CL. 
}
\label{tab:allowed-rngs-GSMprPhi}
\end{table}

\newcommand{\uA}{{$\Uparrow$}}
\newcommand{\dA}{{$\Downarrow$}}
\newcommand{\udA}{{$\Updownarrow$}}

\begin{table}[htbp]
\addtolength{\arraycolsep}{4pt}
\renewcommand{\arraystretch}{1.5}
\centering
\begin{tabular}{|l|cccc|cccc|}
\hline
& \multicolumn{4}{|c|}{$\GSM$}
& \multicolumn{4}{c|}{$\GSMUpr(\Phi)$}
\\
& $D$ & $Q_{V,d}$ & $T_d$ & $T_u$ & $D$ & $Q_d$ & $T_d$ & $T_u$
\\
\hline\hline
$\Delta M_K$ 
&            &      &      &      & \uA  & \uA  & \uA  & \uA
\\
\hline
$\epe$ 
& \uA  & \uA  & \uA  & \uA  & \uA  & \uA  & \uA  & \uA
\\
\hline
$\kpn$ 
& \dA  & \uA  & \dA  & \dA  & \udA & \udA & \udA & \udA
\\
$\klpn$ 
& \dA  & \dA  & \dA  & \dA  & \dA  & \dA  & \dA  & \dA
\\
\hline\hline
$Br(B_d \to\mu\bar\mu)$ 
& \udA & \udA & \udA & \udA &      &      &      &
\\
\hline\hline
$Br(B_s \to\mu\bar\mu)$
& \udA & \udA & \udA & \udA &      &      &      &
\\
$A_{\Delta\Gamma}(B_s \to\mu\bar\mu)$
& \dA  & \dA  & \dA  & \dA  &      &      &      &
\\
$S(B_s \to\mu\bar\mu)$
& \udA & \udA & \udA & \udA & \udA & \udA & \udA & \udA
\\
\hline
$A_{7}(B\to K^*\mu\bar\mu)_{[1,6]}$
& \udA & \udA & \udA & \udA & \udA & \udA & \udA & \udA
\\
$A_{8,\,9}(B\to K^*\mu\bar\mu)_{[1,6]}$
&      & \udA &      &      &      & \udA &      &
\\
$A_{8,\,9}(B\to K^*\mu\bar\mu)_{[15,19]}$
&      & \udA &      &      &      & \udA &      &
\\
\hline
${\cal R}_{B\to K \nu\bar\nu}$
& \dA  & \udA & \dA  & \dA  &      &      &      & 
\\
${\cal R}_{B\to K^* \nu\bar\nu}$
& \dA  & \udA & \dA  & \dA  &      &      &      &
\\
${\cal R}_{F_L}$ 
&      & \dA  &      &      &      &      &      &
\\
\hline
\end{tabular}
\renewcommand{\arraystretch}{1.0}
\caption{\small
  Summary of allowed VLQ effects in $\GSM$- and $\GSMUpr(\Phi)$-models in
  flavour observables after the fit using experimental measurements of 
  \reftab{tab:exp-flavor-constraints}. Possible enhancement, suppression or
  both w.r.t. SM predictions are indicated by according \uA, {\dA} or \udA. 
  Empty space means that the given model does not predict sizeable
  effects in that observable. Note that $(\epe)_{\rm NP}$ has been restricted
  \refeq{eq:epe-constraint}, affecting other $s\to d$ observables.
}
\label{tab:DNA-after-fit}
\end{table}

%
%
%

\section{Summary and Conclusions}
\label{sec:summary}

In this paper we have analysed flavour-violation patterns in the $K$ and
$B_{s,d}$ sectors in eleven models with vector-like quarks (VLQs). Five of them,
called $\GSM$-models, contain only VLQs as new particles. Two of them, called 
$\GSMUpr(S)$-models, have in addition a heavy $Z^\prime$ and a scalar $S$. The final four 
of them, called $\GSMUpr(\Phi)$-models, contain a heavy $Z^\prime$, a scalar $S$ and
a scalar doublet $\Phi$. Our summary of patterns of flavour violation in these 
models in Section~\ref{sec:Comparison}, accompanied by two DNA tables \ref{tab:DNA:bsll-WC}
and \ref{tab:DNA} and in particular our extensive numerical analysis in
Section~\ref{sec:numerics}, see specifically tables~\ref{tab:allowed-rngs-GSM}
and \ref{tab:allowed-rngs-GSMprPhi}, has shown that NP effects in several of
these models can be still very large and that simultaneous consideration of
several flavour observables should allow to distinguish between these models. 
This is also seen in \reftab{tab:DNA-after-fit}, which shows that models 
with LH currents can be distinguished from models with RH currents through
 several observables.

On the theoretical side our paper presents the first analysis of VLQ models
in the context of SMEFT, which allowed to include RG effects from the NP scale 
$M_{\rm VLQ}$ down to the electroweak scale, thereby identifying very important 
Yukawa enhancement of NP contributions to $|\Delta F|=2$ observables in the 
Kaon sector through the generation of left-right operators with smaller, but
significant effects in $B_{s,d}$ observables. These RG effects, relevant only 
in  $\GSM$-models, have been already identified in general $Z$ models in 
\cite{Bobeth:2017xry}, but in the present paper they could be studied explicitly
in concrete models. The relevant technology is described in detail in 
\cite{Bobeth:2017xry} and in \refsec{sec:VLQ-decoupling}, 
\refsec{sec:EFT-down-pheno} and \refapp{app:VLQ-decoupl} of the present paper.

As our results have been systematically summarized in the previous section, we
list here only the main highlights. Most interesting
NP effects are found in $\GSM$-models, even if they do not provide the
explanation of the present LHCb anomalies. In particular
\begin{itemize}
\item Tree-level $Z$ contributions to $\epe$ can be large, so that the apparent
  upward shift in $\epe$ can easily be obtained, bringing the theory to agree
  with data.
\item Simultaneously the branching ratio for $\kpn$ can be enhanced over 
  its SM prediction, but the size of the enhancement depends on whether RH 
  currents or LH currents are considered. In models with flavour-violating RH
  currents, the maximal enhancement is limited to $\sim 50\%$ of its SM value 
  because of the strong constraint from $\epsilon_K$, caused by RG-enhanced 
  contributions. In the LH current case an enhancement of  $\kpn$ is only possible if 
  the present conservative bound on $K_L\to\mu\bar\mu$ is used. With the stricter
  bound only suppression of  $\kpn$ is possible. On the other hand the positive
  shift in $\epe$ implies uniquely the suppression of the $\klpn$ branching ratio.
\item Potential tensions between $\Delta M_{s,d}$ and $\varepsilon_K$ can be easily
  removed in these models, since no MFV relation is imposed on the couplings.
\item Significant suppressions of the $Br(B_s\to\mu\bar\mu)$ and
  of $A_{\Delta\Gamma}(B_s\to\mu\bar\mu)$, in particular in models with LH currents,
  are possible. As far as  $Br(B_d\to\mu\bar\mu)$ is concerned, significant
  enhancements, in particular in the RH current scenarios, are still possible, as
  seen in \reffig{fig:bdmumu} and \reffig{fig:B-mumu-crr}.
  While such effects are also possible in 331 models, they cannot be as large as
  in VLQ models.
\item CP-violating effects for a given quark transition are strongly correlated
  in all of theses models, as long as only one representation is present, specifically
  for $b\to s$, where CP violation in the SM is tiny.
\end{itemize}

Having the LHCb anomalies in mind we have considered also VLQ models with a
heavy $Z^\prime$ related to $\UonePr$ symmetry. Our finding are as follows:
\begin{itemize}
\item The $\GSMUpr(S)$-models, considered already in
  Ref.~\cite{Altmannshofer:2014cfa}, can explain the LHCb anomalies by providing
  sufficient suppression of the coefficient $C_9$, but NP effects in
  $B_{s,d}\to \mu\bar\mu$ and $K_L\to \mu\bar\mu$ are absent, those in
  $b\to s\nu\bar\nu$ transitions small and the ones in $\kpn$ and $\klpn$ much
  smaller than in $\GSM$-models. Most importantly these models fail badly in
  explaining the $\epe$ anomaly.
\item In the $\GSMUpr(\Phi)$-models, the explanation of LHCb anomalies is more
  difficult than in $\GSMUpr(S)$-models, but this time, due to the presence of $Z$
  contributions, interesting effects in other observables can be found. 
\item In particular, in contrast to $\GSM$-models, the parametric suppression
  of $Z$ couplings by $\tan\beta$ allows for increased values of Yukawa
  couplings that are this time mainly bounded by $|\Delta F|=2$ transitions. 
\item
  We find that NP effects in $\epe$ \emph{and} $\kpn$ can be large, the latter in
  contrast to $\GSM$-models, and also the corresponding effects in $\Delta M_K$
  can be significantly larger than in $\GSM$-models. This could appear in 
  contradiction with the pattern in \reftab{tab:DNA} and is the result of 
  weaker constraints in these models. In particular if in the future the $\Delta M_K$
  constraint will be improved, such large enhancements of $Br(\kpn)$ are likely 
  to be excluded. On the other hand NP effects in $\klpn$, $K_L\to\mu\bar\mu$, 
  $B\to K(K^*)\nu\bar\nu$ and $B_{d,s} \to \mu\bar\mu$ are very  small and
  beyond the reach of even presently planned future facilities. While effects
  in the CP asymmetries $A_{7,8,9}(B\to K^*\mu\bar\mu)$ are smaller
  than in $\GSM$ models, they might be still within reach of LHCb.
\end{itemize}

Thus if NP will be found in $B_{s,d}\to\mu\bar\mu$ and the 
$\epe$-anomaly will be confirmed by future lattice data, $\GSM$-models would offer
the best explanation among VLQ models. If, on the other hand, the LHCb anomalies
will be confirmed in the future and no visible NP will be found in rare $K$ decays,
$\GSMUpr(S)$-models and $\GSMUpr(\Phi)$-models would be favoured over $\GSM$-models. 
A large enhancement of $Br(\kpn)$ would uniquely select RH $\GSMUpr(\Phi)$
models subject to the future status of $\Delta M_K$, although LH $\GSM$ and 
$\GSMUpr(\Phi)$ models could provide a moderate enhancement, in case of the latter
depending on the theoretical treatment of $K_L\to\mu\bar\mu$. On the other hand, 
a large enhancement of $Br(B\to K^{(*)}\nu\bar\nu)$ would disfavour all considered
models, at least with only one VLQ representation. Also the confirmation of all
anomalies in combination with sizeable effects in \emph{e.g.} $Br(B_{d,s}\to\mu
\bar\mu)$ would force us to extend the models analyzed by us by considering
several VLQ representations simultaneously. We have also pointed out that in
$\GSMUpr(\Phi)$-models significant NP effects in $\Delta M_K$ can be found, larger
than in $\GSM$ and $\GSMUpr(S)$-models.

While the discovery of VLQs at the LHC would give a strong impetus to the models
considered by us, non-observation of them at the LHC would not preclude their
importance for flavour physics. In fact, as we have shown, large NP effects in
flavour observables can be present for $M_{\rm VLQ} = 10$~TeV and in the flavour-precision
era one is sensitive to even  higher scales. In this context we have pointed out
that the combination of $|\Delta F|=2$ and $|\Delta F|=1$ observables in a given
meson system generally allows to determine the masses of VLQs in a given
representation independently of the size of Yukawa couplings.

%
%
%

\section*{Acknowledgements}
C.B. thanks Martin Gorbahn for numerical checks on $\epe$. We thank Sebastien
Descotes-Genon for providing us an update of a tree-level CKM fit from CKMfitter
\cite{Charles:2004jd}.  This research was done and financed in the context of the
ERC Advanced Grant project ``FLAVOUR''(267104) and was partially supported by the
DFG cluster of excellence ``Origin and Structure of the Universe''. The work of
A.C. is supported by the Alexander von Humboldt Foundation. This work is supported 
in part by the DFG SFB/TR~110 ``Symmetries and the Emergence of Structure in QCD". 

%
%
%

\appendix

\section{\boldmath Scalar sectors of $\GSMUpr$-models}
\label{app:models}

%
%

\subsection[$\GSMUpr(S)$ models]
 {\boldmath $\GSMUpr(S)$ models}
\label{app:scalar:S+H}

The scalar sector in $\GSMUpr(S)$-models with one complex scalar $S(1,0,X)$ and
the SM doublet $H(2,+1/2,0)$ is given by
\begin{align}
  \mathcal{L} & 
  = |{\cal D}_{\mu} H|^2 + |{\cal D}_{\mu} S |^2 - V
\end{align}
with the potential
\begin{align}
  V & 
  = m^2 H^{\dag} H + \frac{\lambda}{2} \left( H^{\dag} H \right)^2     
  + \frac{b_2}{2} |S|^2 + \frac{d_2}{4} |S|^4 
  + \frac{\delta }{2}\, H^{\dag} H |S|^2 .
\end{align}
We parametrise the SM Higgs doublet and the complex scalar as
\begin{align}
  H & 
  = \begin{pmatrix} H^+ \\ H^0 \end{pmatrix}
  = \begin{pmatrix} G^+ \\ \left(v + h^0 + i G^0\right)/\sqrt{2} \end{pmatrix} , &
  S & 
  =  \frac{(v_S + R_0 + i I_0)}{\sqrt{2}} . 
\end{align}
The neutral mass-eigenstates are given by $(h, H)^T \simeq (h^0, R_0)^T$ with
approximate masses
\begin{align}
  m_h^2 & 
  \approx v^2 \left(\lambda - \frac{\delta^2}{2 d_2} \right) , &
  m_H^2 &
  \approx v_S^2 \frac{d_2}{2} ,
\end{align}
up to terms ${\cal O}(v^2/v_S^2)$. The general expressions can be found in
\cite{Basso:2010jm}.

Kinetic mixing of $Z$ and $Z'$ is caused by VLQ-exchange and depends on the VLQ
masses $M$ and the $\UonePr$-gauge coupling. It will be neglected in the following,
see Ref.~\cite{Altmannshofer:2014cfa}. Mass mixing does not occur in $\GSMUpr(S)$
models.

%
%

\subsection[$\GSMUpr(\Phi)$ models]
 {\boldmath $\GSMUpr(\Phi)$ models}
\label{app:scalar:S+H+Phi}

The scalar sector in $\GSMUpr(\Phi)$-models with one complex scalar $S(1,0,X/2)$ and 
the two doublets $\Phi_1 \equiv \Phi(2,+1/2,X)$ and $\Phi_2 \equiv H(2,+1/2,0)$
is given by 
\begin{align}
  \mathcal{L} & 
  = | {\cal D}_{\mu} \Phi_1 |^2 + | {\cal D}_{\mu} \Phi_2 |^2 
  + | {\cal D}_{\mu} S |^2 - V ,
\end{align}
with the potential
\begin{equation}
\begin{aligned}
  V & = 
  m_a^2 \Phi_a^{\dag} \Phi_a 
  + \frac{\lambda_a}{2}  \left( \Phi_a^{\dag} \Phi_a \right)^2
  + \lambda_3 \left( \Phi_1^{\dag} \Phi_1 \right) \left( \Phi_2^{\dag} \Phi_2 \right)
  + \lambda_4 \left( \Phi_1^{\dag} \Phi_2 \right) \left( \Phi_2^{\dag} \Phi_1 \right)
\\
  & + \frac{b_2}{2} |S|^2  +  \frac{d_2}{4} |S|^4
  + \frac{\delta_{a}}{2}\, \Phi_{a}^{\dag} \Phi_a |S|^2
  - \frac{\delta_3}{4} \left[ \Phi_1^{\dag} \Phi_{2} S^2 
                        + \Phi_2^{\dag} \Phi_{1} (S^*)^2 \right] .
\end{aligned}
\end{equation}
We neglect kinetic mixing and parametrise the mass mixing via
\begin{align}  
  \label{eq:Zmixing}
  \begin{pmatrix}
  \hat Z_{\mu}      \\[0.1cm]
  \hat Z_{\mu}^{\prime} 
  \end{pmatrix} & = \begin{pmatrix}
   \cos\xi_{ZZ'} & - \sin\xi_{ZZ'}  \\[0.1cm]
   \sin\xi_{ZZ'} &   \cos\xi_{ZZ'}
   \end{pmatrix}
   \, \begin{pmatrix}
   Z_{\mu}      \\[0.1cm]
   Z_{\mu}^{\prime} 
  \end{pmatrix} .
\end{align}
After partial diagonalization of the neutral gauge boson system, the $Z$ and $Z'$ 
masses and their mass mixing are given by \cite{Babu:1997st}
\begin{align}
  \label{eq:ZZp-masses:GSMPhi:1}
  \hat M_Z^2 & 
  = g_Z^2 \frac{v^2}{4} , &
  \hat M_{Z^{\prime}}^2 & 
  =  (g^\prime X)^2 \frac{v_S^2}{4} \left(1 + 4 c_\beta^2 \frac{v^2}{v_S^2} \right) , & 
  \Delta^2 & 
  = - g_Z\, g^{\prime} X c_\beta^2 \frac{v^2}{2} ,
\end{align}
with $e = \sqrt{4 \pi \alpha} = g_2 \hat s_W = g_1 \hat c_W = g_Z \hat s_W \hat c_W$.
The $Z-Z'$ mixing angle
\begin{align}
  \label{eq:ZZp-mixing-angle}
  \tan 2\, \xi_{ZZ'} & 
  = \frac{2 \Delta^2}{\hat M_Z^2 - \hat M_{Z^{\prime}}^2} 
  = c_{\beta}^2 \frac{4 X g^{\prime}}{g_Z} \frac{\hat M_Z^2}{(\hat M_{Z^{\prime}}^2 - \hat M_Z^2)}
\end{align}
is small unless $X$  becomes large. The diagonalisation of the 
neutral gauge boson mass matrix gives mass eigenvalues
\begin{align}
  M_{Z, Z^{\prime}}^2 &
  = \frac{1}{2} \left[\hat M_{Z^{\prime}}^2 + \hat M_{Z}^2  
    \mp \sqrt{(\hat M_{Z^{\prime}}^2 - \hat M_{Z}^2)^2 + 4 \Delta^4} \right] ,
\end{align}
which differ from the ones in Eq.~\refeq{eq:ZZp-masses:GSMPhi:1} by terms
${\cal O}(v^2/v_S^2)$. Note that we present only the solution for which
$M_{Z} < M_{Z'}$, i.e. throughout we will implicitly impose that the lighter
mass eigenstate couples predominantly SM-like to quarks and leptons.  As a
consequence a lower bound on $g'$ will be obtained. On the other hand, the
decoupling limit $g'\to 0$ is not excluded, but it will lead to $M_{Z'} < M_Z$,
i.e. that the heavier mass-eigenstate couples predominantly to SM-like fermions.
The $\tan\beta$ dependence of $M_{Z'}$ becomes irrelevant once
$v_S \gtrsim 0.5$~TeV. The mixing angle $\xi_{ZZ'}$ can be suppressed with large
$\tan\beta$ and $M_{Z'}$, since we work in the part of the parameter space,
where the other possibility of $g'\to 0$ is not an option.

In $\GSMUpr(\Phi)$-models we make use of the fact that photon- and
$W^\pm$-interactions to leptons are SM-like in order to determine the values of
the fundamental gauge couplings $g_{1,2}$ and the VEV~$v$ from $\alpha_e(M_Z)$,
$G_F$ and the $W$-boson pole mass $M_W$. As the remaining free parameters we
choose $\tan\beta$, $g'$, $X$ and $v_S$, whereas dependent parameters are
$M_{Z,Z'}$ and $\xi_{ZZ'}$.  Note that the latter depend only on the product
$g' X$, such that there are effectively only three parameters. We will restrict
this parameter space to
\begin{align}
  0.3 \leq & \tan\beta \leq 40, &
  0 \leq & \, g' X \leq 3, &
  0 \mbox{ TeV} \leq & v_S \leq 2 \mbox{ TeV}. &  
\end{align}
The lower bound on $\tan\beta$ guarantees perturbativity of the top-quark Yukawa
coupling~\cite{Branco:2011iw}, whereas $v_S$ is bounded from above by the
requirements \refeq{eq:scale-hierarchy-1} and yields $M_{Z'} \lesssim 1.5$~TeV
within the above limits. Constraints on these parameters arise from the measured
value of $M_Z$, which we impose with an error of $\delta M_Z = 5$~GeV to account for the
use of tree-level relations only. Further constraints come from the partial widths
of $Z\to \ell\bar\ell$ $(\ell = e, \mu, \tau)$, constraining the new physics contributions
of the $Z$-lepton couplings \refeq{eq:Z-lepton-coupl} that depend on the
$\xi_{ZZ'}$ and $g'$ due to gauge mixing. We find a small mixing angle
$\xi_{ZZ'} \lesssim 0.1$ in the above specified parameter space of $\tan\beta$,
$g' X$ and $v_S$ if we impose the bound on new physics contributions to the
partial widths of $Z\to \ell\bar\ell$ from LEP \cite{ALEPH:2005ab}, allowing for
5$\sigma$ deviations from the measured central values, together with the bound
on $M_Z$. This justifies the expansion in the small mixing angle as done in
\reftab{tab:GSMP}.

%
%
%

\section[VLQ decoupling and RG effects]
{\boldmath VLQ decoupling and RG effects
 \label{app:VLQ-decoupl}
}

This appendix contains results of the Wilson coefficients of $\psi^2 \varphi^2 D$
and $\psi^2 \varphi^3$ operators in $\GSMandUpr$-EFTs after the tree-level decoupling
of VLQs at the scale $\mu_M$. We provide further the relations to flavour-changing
$Z$ and $Z'$ couplings \refeq{eq:Zcouplings} and \refeq{eq:Zprimecouplings} after
spontaneous symmetry breaking at the scale $\mu_{\rm EW}$ (neglecting self-mixing).

%
%

\subsection[$\psi^2 \varphi^2 D$ operators]
 {\boldmath $\psi^2 \varphi^2 D$ operators}

The matching in $\GSM$ models at the scale $\mu_M$ of order of the VLQ mass
yields nonvanishing contributions for
\begin{equation}
  \label{eq:GSM:matching:SMEFT:psi2H2D}
\begin{aligned}
  D : &&
  \wc[(1)]{Hq}{ij} & = \wc[(3)]{Hq}{ij} 
  = - \frac{1}{4} \frac{\lambda_i^\ast \lambda_j}{M^2} ,
\\
  T_d : &&
  \wc[(1)]{Hq}{ij} & = - 3\, \wc[(3)]{Hq}{ij} 
  = - \frac{3}{8} \frac{\lambda_i^\ast \lambda_j}{M^2} ,
\\
  T_u : &&
  \wc[(1)]{Hq}{ij} & = 3\, \wc[(3)]{Hq}{ij} 
  = \frac{3}{8} \frac{\lambda_i^\ast \lambda_j}{M^2} ,
\\
  Q_d : &&
  \wc{Hd}{ij} &  
  = - \frac{1}{2} \frac{\lambda_i \lambda_j^\ast}{M^2} , 
\\
  Q_V : &&
  \wc{Hd}{ij} &  
  = \frac{1}{2} \frac{\lambda_i^{V_d} \lambda_j^{V_d\ast}}{M^2} , \quad
  \wc{Hu}{ij} 
  = - \frac{1}{2} \frac{\lambda_i^{V_u} \lambda_j^{V_u\ast}}{M^2} , \quad
  \wc{Hud}{ij} 
  = \frac{\lambda_i^{V_u} \lambda_j^{V_d\ast}}{M^2} ,
\end{aligned}
\end{equation}
in agreement with \cite{delAguila:2000rc}, and analogously for $\GSMUpr(\Phi)$
models with $H \to \Phi$. The matching of $\GSMUpr(S)$ models for VLQs $D$ and
$Q_V$ yields nonvanishing Wilson coefficients 
\begin{equation}
  \label{eq:GSM:matching:SMEFTS:psi2H2D}
\begin{aligned}
  D : &&
  \wc{Sd}{ij} & 
  = - \frac{1}{2} \frac{\lambda_i \lambda_j^\ast}{M^2} , & \qquad
  Q_V : &&
  \wc{Sq}{ij} &  
  = - \frac{1}{2} \frac{\lambda_i^\ast \lambda_j}{M^2} .
\end{aligned}
\end{equation}

The flavour-changing $Z$ and $Z'$ couplings \refeq{eq:Zcouplings} and
\refeq{eq:Zprimecouplings} after spontaneous symmetry breaking  are given in terms
of the Wilson coefficients at the scale $\mu_{\rm EW}$. In the case of $\GSM$-models, 
the tree-level calculation of the process $\bar{f}_i f_j Z_\mu$ from $\GSM$-EFT
\refeq{eq:GSM:EFT} yields
\begin{equation}
\begin{aligned}
  \Delta_L^{u_i u_j}(Z) & 
  = {\cal F}_H \left[ \wc[(1)]{Hq}{,ij} - \wc[(3)]{Hq}{,ij} \right] , & 
  \Delta_L^{d_i d_j}(Z) & 
  = {\cal F}_H \left[ \wc[(1)]{Hq}{,ij} + \wc[(3)]{Hq}{,ij} \right] ,
\\
  \Delta_R^{u_i u_j}(Z) & = {\cal F}_H \, \wc{Hu}{,ij} , &
  \Delta_R^{d_i d_j}(Z) & = {\cal F}_H \, \wc{Hd}{,ij} ,
\end{aligned}
\end{equation}
with ${\cal F}_H \equiv -2 M_Z^2/g_Z$ and generation indices $i,j=1,2,3$.
The variant of $\GSMUpr(S)$-models with the scalar sector of $S$ and $H$ generates
only non-zero couplings to $Z'$. We find  for  $\GSMUpr(S)$-models
\begin{equation}
\begin{aligned}
  \Delta_L^{u_i u_j,\, d_i d_j}(Z') & = {\cal F}_S \, \wc{Sq}{,ij} \,, & 
  \Delta_R^{u_i u_j}(Z') & = {\cal F}_S \, \wc{Su}{,ij} \,, &
  \Delta_R^{d_i d_j}(Z') & = {\cal F}_S \, \wc{Sd}{,ij}\,, 
\end{aligned}
\end{equation}
with the EFT-coefficients ${\cal C}_i$ given in (\ref{eq:GSM:matching:SMEFT:psi2H2D})
and ${\cal F}_S \equiv m_{Z'}^2/(g' X)$. The variant of $\GSMUpr(\Phi)$-models
with the scalar sector of $S$, $H$ and $\Phi$ generates non-zero couplings to $Z'$
and $Z$. The results for $\GSMUpr(\Phi)$ models are similar to $\GSM$ models, with
the difference that they involve $Z-Z'$ mixings:
\begin{equation}
\begin{aligned}
  \Delta_L^{u_i u_j}(V) & = {\cal F}_\Phi(V)
    \left[ \wc[(1)]{\Phi q}{,ij} - \wc[(3)]{\Phi q}{,ij} \right] , &
  \Delta_L^{d_i d_j}(V) & = {\cal F}_\Phi(V)
    \left[ \wc[(1)]{\Phi q}{,ij} + \wc[(3)]{\Phi q}{,ij} \right] ,
\\
  \Delta_R^{u_i u_j}(V) & = {\cal F}_\Phi(V) \, \wc{\Phi u}{,ij} \,, &
  \Delta_R^{d_i d_j}(V) & = {\cal F}_\Phi(V) \, \wc{\Phi d}{,ij} \,,
\end{aligned}
\end{equation}
where $V = Z,\, Z'$ and 
\begin{equation}
\begin{aligned}
  {\cal F}_\Phi(Z) & 
  \equiv -2 \, \frac{M_Z^2}{g_Z} c^2_\beta \left[\cos\xi_{ZZ'} - r' \sin\xi_{ZZ'} \right] ,
\\
  {\cal F}_\Phi(Z') & 
  \equiv +2 \, \frac{M_Z^2}{g_Z} c^2_\beta \left[\sin\xi_{ZZ'} + r' \cos\xi_{ZZ'} \right] .
\end{aligned}
\end{equation}

%
%

\subsection[$\psi^2 \varphi^3$ operators]
 {\boldmath $\psi^2 \varphi^3$ operators
  \label{app:psi2H3-operators}
 }

We define the SM Yukawa couplings of quarks as in \cite{Grzadkowski:2010es}
\begin{align}
  \label{eq:SM-dim-4-Yuk}
  - {\cal L}_{\rm Yuk} & 
  = \bar q_L \, Y_{d} \,H \,d_R  + \bar q_L \, Y_{u}\, \widetilde{H}\, u_R    
    + \mbox{h.c.} .
\end{align}
Nonvanishing Wilson coefficients are generated also for $\psi^2 \varphi^3$ 
operators (see \reftab{tab:psi2H2D-ops} for definitions) as a consequence of the
application of equations of motion (EOM) in the tree-level decoupling of VLQs in 
\refsec{sec:tree-decoupl}. Due to the application of EOMs, these Wilson coefficients
scale with the corresponding Yukawa coupling as
\begin{equation}
\begin{aligned}
  \wc{uH}{ij} & 
  = \left[\Yuk{u} \, \WcD{Hu D} 
       + (\Wc[(1)]{Hq D} - \Wc[(3)]{Hq D}) \, \Yuk{u} \right]_{ij} ,
\\
  \wc{dH}{ij} & 
  = \left[\Yuk{d} \, \WcD{Hd D} 
       + (\Wc[(1)]{Hq D} + \Wc[(3)]{Hq D}) \, \Yuk{d} \right]_{ij} . 
\end{aligned}
\end{equation}
Note the matrix multiplications w.r.t. the generation indices of $Y_{u,d}$
with the respective coefficients $\Wc{H\psi D}$ inside the brackets.

The tree-level matching in $\GSM$-models gives nonvanishing contributions 
at $\mu_M$ to
\begin{equation}
  \label{eq:GSM:matching:SMEFT:psi2phi3}
\begin{aligned}
  D : &&
  [{\cal C}_{HqD}^{(1)} + {\cal C}_{HqD}^{(3)}]_{ij} &
  = \frac{1}{2} \frac{\lambda_i^\ast \lambda_j}{M^2} , &
  [{\cal C}_{HqD}^{(1)} - {\cal C}_{HqD}^{(3)}]_{ij} &
  = 0 , 
\\
  T_d : &&
  [{\cal C}_{HqD}^{(1)} + {\cal C}_{HqD}^{(3)}]_{ij} &
  = \frac{1}{4} \frac{\lambda_i^\ast \lambda_j}{M^2} , &
  [{\cal C}_{HqD}^{(1)} - {\cal C}_{HqD}^{(3)}]_{ij} &
  = \frac{1}{2} \frac{\lambda_i^\ast \lambda_j}{M^2} ,  
\\
  T_u : &&
  [{\cal C}_{HqD}^{(1)} + {\cal C}_{HqD}^{(3)}]_{ij} &
  = \frac{1}{2} \frac{\lambda_i^\ast \lambda_j}{M^2} , &
  [{\cal C}_{HqD}^{(1)} - {\cal C}_{HqD}^{(3)}]_{ij} &
  = \frac{1}{4} \frac{\lambda_i^\ast \lambda_j}{M^2} ,  
\\
  Q_d : &&
  \wc{HdD}{ij} &  
  = \frac{1}{2} \frac{\lambda_i \lambda_j^\ast}{M^2} , &
\\
  Q_V : &&
  \wc{HdD}{ij} &  
  = \frac{1}{2} \frac{\lambda_i^{V_d} \lambda_j^{V_d\ast}}{M^2} , &
  \wc{HuD}{ij} &
  = \frac{1}{2} \frac{\lambda_i^{V_u} \lambda_j^{V_u\ast}}{M^2} , 
\end{aligned}
\end{equation}
in agreement with \cite{delAguila:2000rc}. Analogous Wilson coefficients in
$\GSMUpr(\Phi)$ are found by $H \to \Phi$.

In $\GSMUpr(S)$ models analogous relations
\begin{align}
  \wc{uS}{ij} & 
  = \left[\Yuk{u} \, \WcD{Su D} + \Wc{Sq D} \, \Yuk{u} \right]_{ij} , &
  \wc{dS}{ij} & 
  = \left[\Yuk{d} \, \WcD{Sd D} + \Wc{Sq D} \, \Yuk{d} \right]_{ij}  
\end{align}
hold with nonvanishing
\begin{align}
  \label{eq:GSM:matching:SMEFTS:psi2S2H}
  D : \quad
  \wc{SdD}{ij} & 
  = \frac{\lambda_i \lambda_j^\ast}{M^2} , & \qquad
  Q_V : \quad
  \wc{SqD}{ij} &  
  = \frac{\lambda_i^\ast \lambda_j}{M^2} .
\end{align}

%
%

\subsection[Top-Yukawa RG effects]
 {\boldmath Top-Yukawa RG effects
  \label{app:SMEFT-ADMs}
}

This appendix collects the ADM entries of the $\GSM$-EFT proportional to the
up-type quark Yukawa coupling $Y_u$ from \cite{Jenkins:2013wua}, i.e. neglecting
contributions from $Y_{d,e}$. We list them only for operators that receive leading
logarithmic contributions at the scale $\mu_{\rm EW}$ from the initial Wilson 
coefficients at the scale $\mu_M$ of $\psi^2 H^2 D$ and $\psi^2 H^3$ operators in the
1stLLA via direct mixing, see footnote \ref{footnoteX}. For convenience of the
reader we keep here also $\Wc{Hu}$ and $\Wc{Hud}$, which are absent in the VLQ 
models $D, T_u, T_d, Q_d$, but contribute in $Q_V$ for $\lambda^{V_u} \neq 0$. 

The $H^6$-operator ${\cal O}_H = (H^\dagger H)^3$ receives direct leading 
logarithmic contributions\footnote{Note that if the generation indices
are not given explicitly on Yukawa couplings and Wilson coefficients then a matrix
multiplication is implied.}
\begin{align}
  \dotWc{H} & \equiv (4\pi)^2 \mu \frac{{\rm d} \Wc{H}}{{\rm d}\mu} 
  = -12\, \mbox{Tr} \big[ \Wc{uH} \, \YukD{u} \Yuk{u} \YukD{u} 
                        + \Yuk{u} \YukD{u} \Yuk{u} \, \WcD{uH} \big] ,
\end{align}
via $\Wc{uH} \neq 0$ in models $\mbox{VLQ} = T_u, T_d$. The Wilson coefficent
$\Wc{H}$ changes the Higgs potential and leads to a shift of the VEV
\cite{Alonso:2013hga}.

The $H^4 D^2$-operators ${\cal O}_{H\Box} =(H^\dagger H)\Box(H^\dagger H)$ and
${\cal O}_{HD} = (H^\dagger {\cal D}_\mu H)^\ast (H^\dagger {\cal D}^\mu H)$ 
receive leading logarithmic contributions in LH models $\mbox{VLQ} = D, T_u, T_d$ 
via $\Wc[(1,3)]{Hq}$:
\begin{align}
  \dotWc{H\Box} & = 6\, \mbox{Tr} \left[
      \big(\Wc[(1)]{Hq} - 3 \Wc[(3)]{Hq}\big) \Yuk{u} \YukD{u} 
   - \Wc{Hu} \YukD{u} \Yuk{u} \right] ,
\\
  \dotWc{HD} & = 24\, \mbox{Tr} \left[
     \Wc[(1)]{Hq} \Yuk{u} \YukD{u} - \Wc{Hu} \YukD{u} \Yuk{u} \right] .
\end{align}
Their Wilson coefficients contribute to the Higgs-boson mass and the electroweak
precision observable $T = - 2\pi v^2(g_1^{-2} + g_2^{-2})\, \Wc{HD}$ 
\cite{Alonso:2013hga}.

The $\psi^2 H^3$-operators (see \reftab{tab:psi2H2D-ops})
\begin{align}
  \dotWc{uH} = & 
  - 12\, \mbox{Tr} \big[\Wc[(3)]{Hq} \Yuk{u} \YukD{u}] \Yuk{u} 
  - 2\, \Wc[(1)]{Hq} \Yuk{u} \YukD{u} \Yuk{u}
  + 2\, \Yuk{u} \YukD{u} \Yuk{u} \Wc{Hu}
\\ \nonumber &
  + 6\, \mbox{Tr} \big[\Wc{uH} \YukD{u} \big] \Yuk{u} 
  + 9\, \mbox{Tr} \big[\Yuk{u} Y_u^\dag \big] \Wc{uH} 
  + 5\, \Wc{uH} Y_u^\dag \Yuk{u}
  + \frac{11}{2} \Yuk{u} \YukD{u} \Wc{uH}\,,
\\
  \dotWc{dH} = & 
  - 12\, \mbox{Tr} \big[\Wc[(3)]{Hq} \Yuk{u} \YukD{u}] \Yuk{d}
  + 6\, \Wc[(3)]{Hq} \Yuk{u} \YukD{u} \Yuk{d}
  - 2\, \Yuk{u} \YukD{u} \Yuk{u} \Wc{Hud}
\\ \nonumber &
  + 6\, \mbox{Tr} \big[\Yuk{u} \WcD{uH} \big] \Yuk{d}
  - 2\, \Yuk{u} \WcD{uH} \Yuk{d}
  - \Wc{uH} \YukD{u} \Yuk{d}
  + 9\, \mbox{Tr} \big[\Yuk{u} \YukD{u} \big] \Wc{dH}
  - \frac{3}{2} \Yuk{u} \YukD{u} \Wc{dH}\,, 
\\
  \dotWc{eH} = & 
  - 12\, \mbox{Tr} \big[\Wc[(3)]{Hq} \Yuk{u} \YukD{u}] \Yuk{e}
  + 6\, \mbox{Tr} \big[\Yuk{u} \WcD{uH} \big] \Yuk{e}
\end{align}
have self-mixing for $\Wc{uH,dH}$, and $\Wc{uH}$ mixes also into $\Wc{dH,eH}$. 
They receive also contributions from $\Wc{\psi^2 H^2 D}$. The $\Wc{\psi^2 H^3}$ 
enter fermion-mass matrices \refeq{eq:quark-mass-dim-6} and lead also to 
fermion-Higgs couplings that are in general flavour-off-diagonal.

The $\psi^2  H^2 D$-operators (see \reftab{tab:psi2H2D-ops})
\begin{align}
  \dotWc[(1)]{Hq} = & 
  \; 6\, \mbox{Tr} \big[\Yuk{u} \YukD{u}] \Wc[(1)]{Hq}
  + 2 \left( \Yuk{u} \YukD{u} \Wc[(1)]{Hq} + \Wc[(1)]{Hq} \Yuk{u} \YukD{u} \right) ,
\\ \nonumber &
  - \frac{9}{2} \left( \Yuk{u} \YukD{u} \Wc[(3)]{Hq} 
                     + \Wc[(3)]{Hq} \Yuk{u} \YukD{u} \right)
  - \Yuk{u} \Wc{Hu} \YukD{u} ,
\\
  \dotWc[(3)]{Hq} = &
  \; 6\, \mbox{Tr} \big[\Yuk{u} \YukD{u}] \Wc[(3)]{Hq}
  + \Yuk{u} \YukD{u} \Wc[(3)]{Hq} + \Wc[(3)]{Hq} \Yuk{u} \YukD{u}
  - \frac{3}{2} \left( \Yuk{u} \YukD{u} \Wc[(1)]{Hq} 
                     + \Wc[(1)]{Hq} \Yuk{u} \YukD{u} \right) ,
\\
  \dotWc{Hd} = & \; 6\, \mbox{Tr} \big[\Yuk{u} \YukD{u}] \Wc{Hd} \,,
\\
  \dotWc{Hu} = &
  - 2 \YukD{u} \Wc[(1)]{Hq} \Yuk{u}
  + 6\, \mbox{Tr}[\Yuk{u} \YukD{u}] \Wc{Hu} 
  + 4 \left(\YukD{u} \Yuk{u} \Wc{Hu} + \Wc{Hu} \YukD{u} \Yuk{u} \right) , 
\\
  \dotWc{Hud} = &
  \; 6\; \mbox{Tr}[\Yuk{u} \YukD{u}] \Wc{Hud}
  + 3 \YukD{u} \Yuk{u} \Wc{Hud} 
\end{align}
show a mixing pattern among $\Wc[(1,3)]{Hq}$ as well as $\Wc[(1)]{Hq}$ and
$\Wc{Hu}$. The latter implies that the LH scenarios $D, T_u, T_d$ will 
generate via mixing also a RH coupling $\Wc{Hu}$ via $\Wc[(1)]{Hq}$, which 
is however a one-loop effect compared to the effects of $\Wc[(1)]{Hq}$. 
Both $\Wc{Hd}$ and $\Wc{Hud}$ have only self-mixing.

In the case of $\psi^4$-operators there are $(\overline{L}L)(\overline{L}L)$
operators 
\begin{align}
  \label{eq:ADM:qq1}
  \dotwc[(1)]{qq}{ijkl} & 
  = + \frac{1}{2} \left( [\Yuk{u} \YukD{u}]_{ij} \wc[(1)]{Hq}{kl} 
                     + \wc[(1)]{Hq}{ij} [\Yuk{u} \YukD{u}]_{kl} \right) ,
\\
  \label{eq:ADM:qq3}
  \dotwc[(3)]{qq}{ijkl} &
  = - \frac{1}{2} \left( [\Yuk{u} \YukD{u}]_{ij} \wc[(3)]{Hq}{kl} 
                       + \wc[(3)]{Hq}{ij} [\Yuk{u} \YukD{u}]_{kl} \right) ,
\intertext{the $(\overline{L}L)(\overline{R}R)$ operators}
  \dotwc[(1)]{qu}{ijkl} & 
  = [\Yuk{u} \YukD{u}]_{ij} \wc{Hu}{kl} - 2 \wc[(1)]{Hq}{ij} [\YukD{u} \Yuk{u}]_{kl} ,
\\
  \label{eq:RGE:qd1}
  \dotwc[(1)]{qd}{ijkl} & 
  = [\Yuk{u} \YukD{u}]_{ij} \wc{Hd}{kl} ,
\intertext{and the $(\overline{R}R)(\overline{R}R)$ operators}
  \dotwc{uu}{ijkl} & 
  = - [\YukD{u} \Yuk{u}]_{ij} \wc{Hu}{kl} - \wc{Hu}{ij} [\YukD{u} \Yuk{u}]_{kl} ,
\\
  \label{eq:RGE:ud1}
  \dotwc[(1)]{ud}{ijkl} & 
  = -2 [\YukD{u} \Yuk{u}]_{ij} \wc{Hd}{kl} ,
\end{align}
of which the ones relevant for $|\Delta F|=2$ are given in
\refeq{eq:GSM-EFT:DF2-op-LLLL} and \refeq{eq:GSM-EFT:DF2-op-LLRR}. Hence there
are two additional operators $[{\cal O}^{(1)}_{ud}]_{ijkl} = [\bar{u}_R^i \gamma_\mu u_R^j]
[\bar{d}_R^k \gamma^\mu d_R^l]$ and $[{\cal O}^{(1)}_{qu}]_{ijkl} = [\bar{q}_L^i 
\gamma_\mu q_L^j] [\bar{u}_R^k \gamma^\mu u_R^l]$ under the assumption $\Wc{Hu} = 0$.

%
%
%

\section[Master formulae for $K$ and $B$ decays]
{\boldmath Master formulae for $K$ and $B$ decays}
\label{app:master}

%
%

\subsection[$|\Delta F| = 2$]
 {\boldmath $|\Delta F| = 2$}
\label{app:DF2:EFT}

The effective Lagrangian for neutral meson mixing in the down-type quark
sector ($d_j \bar{d}_i \to \bar{d}_j d_i$ with $i\neq j$) can be written
as~\cite{Buras:2000if}
\begin{align}
  \label{eq:DF2-hamiltonian}
  {\cal H}_{\Delta F = 2}^{ij} &
  = {\cal N}_{ij} \sum_a  C_a^{ij} O_a^{ij} + \mbox{h.c.} ,
\end{align}
where the normalisation factor and the CKM combinations are
\begin{align}
  \label{eq:DF2:norm-factor}
  {\cal N}_{ij} & 
  = \frac{G_F^2}{4 \pi^2} M_W^2 \left(\lambda^{(t)}_{ij}\right)^2 ,
\end{align}
with $ij = sd$ for kaon mixing and $ij = bd,bs$ for $B_d$ and $B_s$
mixing, respectively.
The set of operators consists out of $(5 + 3) = 8$ operators \cite{Buras:2000if},
\begin{equation}
  \label{eq:DF2-operators}
\begin{aligned}
  O_{{\rm VLL}}^{ij} & 
  = [\bar{d}_i \gamma_\mu P_L d_j][\bar{d}_i \gamma^\mu P_L d_j] , &
\\[0.2cm]
  O_{{\rm LR},1}^{ij} & 
  = [\bar{d}_i \gamma_\mu P_L d_j][\bar{d}_i \gamma^\mu P_R d_j] , &
  O_{{\rm LR},2}^{ij} & 
  = [\bar{d}_i P_L d_j][\bar{d}_i P_R d_j] ,
\\[0.2cm]
  O_{{\rm SLL},1}^{ij} & 
  = [\bar{d}_i P_L d_j][\bar{d}_i P_L d_j] , &
  O_{{\rm SLL},2}^{ij} & 
  = -[\bar{d}_i \sigma_{\mu\nu} P_L d_j][\bar{d}_i \sigma^{\mu\nu} P_L d_j] ,
\end{aligned}
\end{equation}
which are built out of colour-singlet currents $[\bar{d}^\alpha_i \ldots d^\alpha_j]
[\bar{d}^\beta_i \ldots d^\beta_j]$, where $\alpha,\, \beta$ denote colour indices.
The chirality-flipped sectors VRR and SRR are obtained from interchanging 
$P_L \leftrightarrow P_R$ in VLL and SLL. Note that the minus sign in
$Q_{{\rm SLL},2}$ arises from different definitions of $\tilde{\sigma}_{\mu\nu} \equiv 
[\gamma_\mu,\, \gamma_\nu]/2$ in Ref.~\cite{Buras:2000if} w.r.t. 
$\sigma_{\mu\nu} = i \tilde{\sigma}_{\mu\nu}$ used here. The ADM's
of the 5 distinct sectors (VLL, SLL, LR, VRR, SRR) have been calculated in 
Refs.~\cite{Buras:2000if, Ciuchini:1997bw} at NLO in QCD, and numerical solutions are given in 
Ref.~\cite{Buras:2001ra}. The NLO ADM's are also available for an alternative basis
\cite{Gorbahn:2009pp} with colour octet operators $Q_{{\rm SLL},2} = 
[\bar{d}^\alpha_i P_L d^\beta_j][\bar{d}^\beta_i P_L d^\alpha_j]$ and analogous
$Q_{{\rm SRR},2}$. 

In the SM only
\begin{align}
  \label{eq:DF2:S0}
  C_{{\rm VLL}}^{ij}(\mu_{\rm EW})|_{\rm SM} & = S_0(x_t), &
  S_0(x) & 
  = \frac{x (4 - 11 x + x^2)}{4\, (x-1)^2} + \frac{3 x^3 \ln x}{2\, (x-1)^3}   
\end{align}
is non-zero at the scale $\mu_{\rm EW}$, depending on the ratio $x_t \equiv 
m_t^2/M_W^2$ of the top-quark and $W$-boson masses. 

The $|\Delta F| = 2$ observables of interest $\Delta M_{K,\,B_d,\,B_s}$,
$\epsilon_K$ and $\sin(2\beta_{d,s})$ derive all from the complex-valued
off-diagonal elements $M_{12}^{ij}$ of the mass-mixing matrices of the neutral
mesons \cite{Buchalla:1995vs, Buras:1998raa}. For the latter we use the full
higher-order SM expressions in combination with the LO new physics
contributions. In particular for $M_{12}^{ds}$, we make use of NLO and in part
NNLO QCD corrections $\eta_{cc,\,tt,\,ct}$ collected in \reftab{tab:input}
and for the hadronic matrix element of $|\Delta S|=2$ operators the value of
$\hat{B}_K$. Concerning $|\Delta B| = 2$, we include the NLO QCD corrections
$\eta_B$ to the SM and use for the hadronic matrix elements the latest results
for $F_{B_{d,s}}\sqrt{\hat{B}_{B_{d,s}}}$ \cite{Bazavov:2016nty}. The hadronic 
matrix elements of $|\Delta S,B|=2$ of left-right operators are given in
\reftab{tab:M12:input}.

%
%

\subsection[$d_j\to d_i \nu\bar\nu$]
 {\boldmath $d_j\to d_i \nu\bar\nu$}
\label{app:d->dvv}

The effective Lagrangian for $d_j\to d_i \nu\bar\nu$ ($i\neq j$) is 
adopted from Ref.~\cite{Altmannshofer:2009ma},
\begin{align}
  \label{eq:EFT:ddvv}
  {\cal L}_{d\to d\nu\bar\nu} &
  = \frac{4 G_F}{\sqrt{2}} \frac{\alpha_e}{4\pi} \, \lambda^{(t)}_{ij} 
    \sum_{a} \sum_{\nu} C_a^{ij,\nu} O_a^{ij,\nu} + \mbox{h.c.} ,
\end{align}
where the sums extend over $a = \{L,R \}$ and neutrino flavour 
$\nu = \{e, \mu, \tau\}$
\begin{align}
  \label{eq:operators:ddvv}
  O_{L\,(R)}^{ij,\nu} & 
  = [\bar{d}_i \gamma_\mu P_{L\,(R)} d_j][\bar\nu \gamma^\mu (1-\gamma_5) \nu] .
\end{align}
In the SM only
\begin{align}
  \label{eq:ddvv-wc-X}
  C_L^{ij,\nu} \big|_{\rm SM} & 
  = \frac{4 B - C}{s_W^2}
  \equiv -\frac{X_0}{s_W^2} 
\end{align}
has non-vanishing contribution at the scale $\mu_{\rm EW}$, whereas
$C_R^{\nu} = 0$.  The functions $B$ and $C$ depend on the ratio
$x_t \equiv m_t^2/M_W^2$ of the top-quark and $W$-boson masses and enter as the
gauge-independent linear combination $X_0(x_t) \equiv C(x_t) - 4 B(x_t)$
\cite{Inami:1980fz, Buchalla:1990qz},
\begin{align}
  \label{eq:X-SM}
  X_0(x) &
  = \frac{x}{8} \left(\frac{x + 2}{x - 1} + {\frac{3 x - 6}{(x - 1)^2}} \ln x \right) .
\end{align}
It is given by
\begin{align}
  \label{eq:X_t,SM}
  X_0 \to X_L^{\rm SM} = 1.481 \pm 0.009,
\end{align}
when including higher order QCD and electroweak corrections \cite{Buchalla:1993bv, 
Misiak:1999yg, Buchalla:1998ba, Brod:2010hi} as extracted in Ref.~\cite{Buras:2015qea} 
from original papers. 

The theoretical predictions for $b\to s\nu\bar\nu$ observables defined in
Eq.~\refeq{eq:b->svv:obs} are based on formulae given in Ref.~\cite{Buras:2014fpa}.
These expressions account for the lepton-non-universal contribution of VLQ's
w.r.t. the neutrino flavour in $\GSMUpr$ models. However, the particular
structure of the gauged $\UonePr$ \refeq{eq:UonePr-lep-charges} leads to a
cancellation of the numerically leading interference contributions of the SM and
new physics \cite{Altmannshofer:2014cfa}.

The $Br(\kpn)$ receives in the SM the numerically leading contribution from the
``top''-sector, when decoupling heavy degrees of freedom at $\mu_{\rm EW}$,
which yields directly the local ${\cal O}_L^{sd,\nu}$ operator
$(\nu = e, \mu, \tau)$.  Further, a non-negligible ``charm''-sector arises from
double-insertions of hadronic and semi-leptonic $|\Delta S| = 1$ operators when
decoupling the charm quark at $\mu_c \sim m_c$, which is enhanced due to the
strong CKM hierarchy
$(\lambda_{sd}^{(t)} \propto \lambda^5) \; \ll \; (\lambda_{sd}^{(c)} \propto
\lambda^2)$,
where $\lambda = |V_{us}|$ is the Cabibbo angle.  This is usually expressed in
the effective Hamiltonian of the SM as \cite{Buras:2006gb}
\begin{align}
  {\cal H}_{\rm eff} & 
  = {\cal N} \sum_\nu \left[\lambda_{sd}^{(c)} X_{c}^\nu 
     + \lambda_{sd}^{(t)} X_L^{\rm SM} \right] {\cal O}_L^{sd,\nu} ,
\end{align}
with ${\cal N} = G_F \alpha_{e}/(2 \sqrt{2}\pi s_W^2)$, where $X_{c}^e = X_{c}^\mu 
\neq X_c^\tau$. 

The NP contributions in VLQ-models cannot compete with the SM contribution to
the tree-level processes entering the ``charm''-sector, since they are
suppressed by an additional factor $(M_W/M_{\rm VLQ})^2$. In consequence, NP
contributes to the ``top''-sector only
\begin{align}
  X_L^{\rm SM} \quad \to \quad  X_{t}^\nu &
  = X_L^{\rm SM} + X_L^{sd,\nu} + X_R^{sd,\nu} 
  \equiv X_L^{\rm SM} + X_{\rm NP}^\nu ,
\end{align}
with $X_{L,R}^{sd,\nu}$ given in Eq.~\refeq{eq:X_LR}, such that the top-sector
becomes neutrino-flavour dependent.

The experimental measurement averages over the three neutrino flavours,
\begin{align}
  \label{eq:Br-kpivv-1}
  Br(\kpn) & =
  \frac{\kappa_+ (1 + \Delta_{\rm EM})}{\lambda^{10}} \frac{1}{3} \sum_{\nu}
  \left[  \mbox{Im}^2 \Big(\lambda_{sd}^{(t)} X_{t}^\nu \Big)
        + \mbox{Re}^2 \Big(\lambda_{sd}^{(c)} X_{c}^\nu
                         + \lambda_{sd}^{(t)} X_{t}^\nu \Big)\right] ,
\end{align}
with the assumption that $\lambda_{sd}^{(c)} X_{c}^\nu$ is real.
The NNLO QCD results of the functions $X_{c}^\nu$ \cite{Buras:2006gb} 
together with long distance contributions \cite{Isidori:2005xm}
are combined into
\begin{align}
  P_c &
  = \frac{1}{\lambda^4} \left( \frac{2}{3} X_{c}^e + \frac{1}{3} X_{c}^\tau \right)
  = \left(\frac{0.2252}{\lambda}\right)^4 (0.404 \pm 0.024),
\end{align}
where $\lambda = 0.2252$ has been used in Ref.~\cite{Buras:2015qea}. The factor 
\begin{align}
  \kappa_+ &
  = r_{K^+} \frac{3 \alpha^2(M_Z) \lambda^8}{2 \pi^2 s_W^4} Br(K\to \pi e\bar{\nu}_e)
  = 0.5173(25) \times 10^{-10} \left[\frac{\lambda}{0.225}\right]^8 
\end{align}
contains the experimental value $Br(K\to \pi e\bar{\nu}_e)$ and the isospin
correction $r_{K^+}$ and has been evaluated in Ref.~\cite{Mescia:2007kn} (table 2)
including various corrections. Further $\Delta_{\rm EM} = -0.003$ for
$E^\gamma_{\rm max} \approx 20$~MeV \cite{Mescia:2007kn}. If one takes into
account the different value of $s_W^2 = 0.231$ taken in Ref.~\cite{Mescia:2007kn}
compared to our value in \reftab{tab:input}, then
$\kappa_+ = 0.5150 \times 10^{-10} \, (\lambda/0.225)^8$.

The sum \refeq{eq:Br-kpivv-1}
contains the SM contribution and further the interference of SM$\times$NP and 
NP$\times$NP. Besides $P_c$ at NNLO in the SM contribution, the NLO numerical
values
\begin{align}
  X_c^e    & = 10.05 \times 10^{-4} , &
  X_c^\tau & = 6.64 \times 10^{-4} , 
\end{align}
for $\mu_c = 1.3$~GeV are used for the interference of SM$\times$NP.
 
The branching fraction of $\klpn$ is obtained again by averaging over the three
neutrino flavours
\begin{align}
  Br(\klpn) & =
  \frac{\kappa_L }{\lambda^{10}} \frac{1}{3} \sum_{\nu}
  \mbox{Im}^2 \Big(\lambda_{sd}^{(t)} X_{t}^\nu \Big) ,
\end{align}
with
\begin{align}
  \kappa_L &
  = \kappa_+ \frac{r_{K_L}}{r_{K_+}} \frac{\tau_{K_L}}{\tau_{K_+}}
  = 2.231(13) \times 10^{-10} \left[\frac{\lambda}{0.225}\right]^8 .
\end{align}
The numerical value is from Ref.~\cite{Mescia:2007kn} (table 2) and it decreases
to $\kappa_L = 2.221 \times 10^{-10} \, (\lambda/0.225)^8$ when rescaling
with our value of $s_W^2$.

%
%

\subsection[$d_j\to d_i\, \ell\bar\ell$]
 {\boldmath $d_j\to d_i\, \ell\bar\ell$}
\label{app:d->dll}

The effective Lagrangian for $d_j\to d_i \ell\bar\ell$ ($i\neq j$) is adopted
from Ref.~\cite{Bobeth:2007dw},
\begin{align}
  \label{eq:EFT:ddll}
  {\cal L}_{d\to d\ell\bar\ell} &
  = \frac{4 G_F}{\sqrt{2}} \frac{\alpha_e}{4\pi} \, \lambda^{(t)}_{ij}
    \sum_{a} \sum_{\ell} C_a^{ij, \ell} O_a^{ij, \ell} + \mbox{h.c.} ,
\end{align}
were the sum over $a$ extends over the $|\Delta F| = 1$ operators
\begin{align}
  \label{eq:operators:ddll}
  O_{9\,(9')}^{ij,\ell} & 
  = [\bar{d}_i \gamma_\mu P_{L\,(R)} d_j][\bar\ell \gamma^\mu \ell] , &
  O_{10\,(10')}^{ij,\ell} &
  = [\bar{d}_i \gamma_\mu P_{L\,(R)} d_j][\bar\ell \gamma^\mu \gamma_5 \ell] ,
\end{align}
whereas scalar $O_{\rm S,P(S',P')}^{\ell}$ and tensorial operators 
$O_{\rm T(T5)}^{\ell}$ are not generated in the context
of VLQ models. In the SM the only non-zero Wilson coefficients,
\begin{align}
  C_9^{ij,\ell} \big|_{\rm SM} & 
  = \frac{1}{s_W^2} \left[(1 - 4 s_W^2) C - B - s_W^2 D \right] 
  \equiv \frac{Y_0}{s_W^2} - 4 Z_0,
\\
  C_{10}^{ij,\ell} \big|_{\rm SM} & 
  = \frac{1}{s_{W}^2} \left(B - C \right) \equiv -\frac{Y_0}{s_W^2},
\end{align}
are lepton-flavour universal and also universal w.r.t down-type quark transitions,
as the CKM elements have been factored out. All other Wilson coefficients vanish
at the scale $\mu_{\rm EW}$. The functions $B,C,D$ depend again on the ratio 
$x_t \equiv m_t^2/M_W^2$ of the top-quark and $W$-boson masses and give
two gauge-independent combinations $Y_0(x_t) \equiv C(x_t) - B(x_t)$ and 
$Z_0(x_t) \equiv C(x_t) + D(x_t)/4$, that are given in the SM as
\begin{align}
\label{eq:Y-SM}
  Y_0(x) & 
  = \frac{x}{8} \left(\frac{x - 4}{x - 1} + \frac{3 x \ln x}{(x - 1)^2} \right) ,
\end{align}
\begin{align}
  \label{eq:Z-SM}
  Z_0(x) & 
  = \frac{18 x^4 - 163 x^3 + 259 x^2 - 108 x}{144 (x-1)^3} 
  + \frac{32 x^4 - 38 x^3 - 15 x^2 + 18 x}{72 (x-1)^4} \ln x
  - \frac{1}{9} \ln x .
\end{align}

In the predictions of $Br(B_{d,s}\to \mu\bar\mu)$ and the mass-eigenstate
rate asymmetry $A_{\Delta \Gamma}(B_{d,s}\to \mu\bar\mu)$ we include for the SM 
contribution the NNLO QCD \cite{Hermann:2013kca} and NLO EW \cite{Bobeth:2013tba} 
corrections, whereas NP contributions are included at LO. The values of the decay 
constants $F_{B_{d,s}}$ are collected in \reftab{tab:input}.

The branching fractions $Br(B^+ \to (\pi^+,\, K^+) \mu\bar\mu)$ at high dilepton invariant mass $q^2$
are predicted within the framework outlined in Refs.~\cite{Grinstein:2004vb, Beylich:2011aq,
Bobeth:2011nj}. We neglect contributions from QCD penguin operators, which have
small Wilson coefficients and the NLO QCD corrections to matrix elements of the
charged-current operators \cite{Seidel:2004jh, Greub:2008cy}, but include the
contributions $\sim V_{ub}^{} V_{ud(s)}^*$. The form factors and their uncertainties
are adapted from lattice calculations \cite{Lattice:2015tia, Bailey:2015nbd} for
$B\to \pi$ and \cite{Bailey:2015dka} for $B\to K$ with a summary given in
\cite{Du:2015tda}. We add additional relative 
uncertainties of 15\% for missing NLO QCD corrections and 10\% for possible
duality violation \cite{Beylich:2011aq} in quadrature.

The predictions for observables of $B\to K^*\mu\bar\mu$ are based on
Refs.~\cite{Bobeth:2008ij} and~\cite{Bobeth:2010wg} for low- and high-$q^2$ regions, 
respectively. The corresponding results for $B\to K^*$ form factors in the two
regions are from the LCSR calculation~\cite{Straub:2015ica} and the lattice
calculations~\cite{Horgan:2013hoa, Horgan:2015vla}.  

The measurement of $Br(K_L \to \mu\bar\mu)$ provides important constraints on its
short-distance (SD) contributions, despite  the dominating long-distance (LD) 
contributions inducing uncertainties that are not entirely under theoretical control.
In particular there is the issue of the sign of the interference of the SD
part $\chi_{\rm SD}$ of the decay amplitude of $K_L \to \mu\bar\mu$ with the LD
parts. Allowing for both signs implies a conservative bound $|\chi_{\rm SD}| \leq 3.1$
\cite{Isidori:2003ts}. Relying on predictions of this sign based on the quite
general assumptions stated in~\cite{Isidori:2003ts, DAmbrosio:1996kjn, GomezDumm:1998gw}
one finds $-3.1 \leq \chi_{\rm SD}\leq 1.7$ which we employ in most of this work. 
Note, however, that a different sign is found\footnote{We thank G.~D'Ambrosio
and J-M.~G{\'e}rard for the discussion on this point.} in \cite{DAmbrosio:1996kjn,
Gerard:2005yk}, implying $-1.7 \leq \chi_{\rm SD}\leq 3.1$. In light of this 
situation, we comment on the impact of the more conservative choice where appropriate,
which includes both sign choices.

%
%

\subsection[$d_j\to d_i\, q\bar{q}$ and $\epe$]
{\boldmath $d_j\to d_i\, q\bar{q}$ and $\epe$}
\label{app:d->dqq}

The effective Lagrangian for $d_j\to d_i q\bar{q}$ ($i\neq j$) is adopted
from Ref.~\cite{Buras:1993dy}, where the definition of the operators can be found and
here we restrict ourselves to $\bar{s}\to \bar{d}$, i.e. $ij=sd$. At the scale
$\mu_{\rm EW}$ ($N_f = 5$) it reads
\begin{equation}
  \label{eq:EFT:ddqq:Nf5}
\begin{aligned}
  {\cal L}_{d\to dq\bar{q}} 
  = - \frac{G_F}{\sqrt{2}} \, \lambda_{sd}^{(u)}
    \Big\{ & (1-\tau) \big[z_1 (O_1 - O_1^c) + z_2 (O_2 - O_2^c)\big]
\\ & 
  + \sum_{a=3}^{10} (\tau v_a + v_a^{\rm NP}) O_a
  + \sum_{a=3}^{10} v'_a O'_a \Big\} + \mbox{h.c.} ,
\end{aligned}
\end{equation}
where $O_{1,2}^{(c)}$ denote current-current operators. The sum over $a$ 
extends over the QCD- and EW-penguin operators and we included their 
chirality-flipped counterparts  $O'_a = O_a [\gamma_5 \to -\gamma_5]$.
Thereby we assume that VLQ contributions to other operators are strongly
suppressed. The Wilson coefficients are denoted as $z_a$, $v_a^{(\rm NP)}$ 
and $v'_a$, taken at the scale $\mu_{\rm EW}$. For the SM-part, CKM 
unitarity was used,
\begin{align}
  \tau & \equiv\lambda_{sd}^{(u)} \big/ \lambda_{sd}^{(t)} ,
\end{align}
and we introduced a new physics contribution $v_a^{\rm NP}$ as shown above,
which is related to the VLQ-contribution \refeq{eq:ddqq-Z-contr} as 
\begin{align}
  v_a^{\rm NP} & = C_a^{sd} , & 
  v'_a & = C_{a'}^{sd} .
\end{align}

The RG evolution at NLO in QCD and QED leads to the effective Hamiltonian at a scale
$\mu\lesssim \mu_c \sim m_c$ ($N_f=3$)
\begin{align}
  \label{eq:EFT:ddqq:Nf3}
  {\cal H}_{d\to dq\bar{q}} &
  = \frac{G_F}{\sqrt{2}} \, \lambda_{sd}^{(u)}
    \left\{ z_1 O_1 + z_2 O_2 + 
    \sum_{a=3}^{10} [z_a + \tau y_a + v_a^{\rm NP}] O_a 
  + \sum_{a=3}^{10} v'_a O'_a \right\} + \mbox{h.c.} ,
\end{align}
after decoupling of $b$- and $c$-quarks at scales $\mu_{b,c}$ \cite{Buras:1993dy},
where $y_a \equiv v_a - z_a$ and all Wilson coefficients are at the scale $\mu$.

The contributions of new physics can then be accounted for in $\epe$ by the
replacement
\begin{align}
  \label{eq:epe-NP-contr}
  y_a(\mu) & \to 
  y_a(\mu) + \frac{v_a^{\rm NP} (\mu)- v'_a(\mu)}{\tau} ,
\end{align}
where the minus sign is due to $\langle (\pi\pi)_I | O_a | K \rangle = 
- \langle (\pi\pi)_I | O'_a | K \rangle$ for the pseudo-scalar pions in the final 
state \cite{Kagan:2004ia}. For the readers convenience we provide a semi-numerical
formula for $\epe$ with initial conditions of Wilson coefficients from 
new physics in QCD- and EW-penguins $a=3^{(\prime)},5^{(\prime)},7^{(\prime)},9^{(\prime)}$ 
at the electroweak scale $\mu_{\rm EW}$:
\begin{equation}
  \label{eq:epe-seminum}
\begin{aligned}
  \frac{\varepsilon'}{\varepsilon} & =
  \left[-2.58 + 24.01 B_6^{(1/2)} - 12.70 B_8^{(3/2)}\right] \times 10^{-4}
  + \sum_a P_a \, \mbox{Im}(v_a^{\rm NP} - v'_a)[\mu_{\rm EW}].
\end{aligned}
\end{equation}
The coefficients are 
\begin{align}
  P_a & 
  = p_a^{(0)} + p_a^{(6)} B_6^{(1/2)} + p_a^{(8)} B_8^{(3/2)}
\end{align}
with $p^{(n)}_a$ given in \reftab{tab:epe-seminum}, where the last column gives $P_a$
for $B_6^{(1/2)}(\mu) = 0.57$ and $B_8^{(3/2)}(\mu) = 0.76$. For this purpose
$\mu_{\rm EW} = M_W$, $\mu_b = m_b(m_b)$, $\mu_c = 1.3$~GeV and $\mu = 1.53$~GeV
have been used. The central value of the SM prediction is
$(\epe)_{\rm SM} = 1.5 \times 10^{-4}$ compared to $1.9 \times 10^{-4}$ in
\cite{Buras:2015yba} due to different numerical inputs.

\begin{table}[!tb]
\renewcommand{\arraystretch}{1.3}
\centering
\begin{tabular}{|c|rrr|r|}
\hline
  $a$
& $p^{(0)}_a$ & $p^{(6)}_a$ & $p^{(8)}_a$ & $P_a$
\\
\hline\hline
  3
& $7.45$ & $-3.40$ & $-3.50$ & $2.85$
\\
  5
& $1.70$ & $30.62$ & $-18.74$ &  $4.91$
\\
  7
& $-102.02$ & $-1.32$ & $2040.38$ & $1447.91$
\\
  9
& $36.72$ & $4.42$ & $-21.28$ & $23.06$
\\
\hline
\end{tabular}  
\renewcommand{\arraystretch}{1.0}
  \caption{
    Values of the coefficients entering the semi-numerical formula of $\epe$
    in Eq.~\refeq{eq:epe-seminum}. 
    The last column gives $P_a$ for $B_6^{(1/2)} = 0.57$ and $B_8^{(3/2)} = 0.76$.
  }
  \label{tab:epe-seminum}
\end{table}

%
%
%

\section{Statistical approach and numerical input
\label{app:num-input}}

The input quantities included in our analysis are collected in \reftab{tab:input}
and \reftab{tab:M12:input}. 
The CKM parameters have to be determined independently of contributions from the VLQs. 
The ``tree-level'' fit carried out by the CKMfitter collaboration achieves such
a determination, taking only measurements into account that are unaffected in our
NP scenarios, \emph{i.e.} (semi-)leptonic tree-level decays, tree-level determinations
of $\gamma$ and $B\to \pi\pi,\pi\rho,\rho\rho$, used as a constraint on $\gamma$.
The results of this fit are again quoted in Table~\ref{tab:input}.

As a statistical procedure, we choose a frequentist approach. The fits include 
as parameters of interest the VLQ couplings and in addition nuisance parameters,
which constitute theoretical uncertainties. The nuisance parameters are listed
in \reftab{tab:input} and consist of
\begin{itemize}
\item CKM parameters from a ``tree-level'' fit\footnote{We thank Sebastien
  Descotes-Genon for providing us an update of a tree-level CKM fit from CKMfitter
  \cite{Charles:2004jd}.};
\item hadronic parameters: decay constants, form factors, $|\Delta F|=2$
  hadronic matrix elements.
\end{itemize}
The 1- and 2-dimensional confidence regions (CL) of parameters are obtained by
profiling over the remaining parameters, i.e. maximisation of the likelihood
function over the subspace of remaining parameters for a fixed value of the
(pair of) parameter(s) of interest. Similarly, correlation plots for
pairs of observables are obtained by profiling over all parameters and 
imposing in addition the specific values for the pair observables. The
2-dimensional 68\% and 95\% confidence regions are determined then for 
two degrees of freedom. The SM predictions of observables are found in the
same way by setting VLQ contributions to zero and profiling only over
the CKM and hadronic nuisance parameters.

\begin{table}[!tb]
\renewcommand{\arraystretch}{1.2}
\center{
\begin{tabular}{|ll|}
\hline
  \multicolumn{2}{|c|}{general \cite{Agashe:2014kda}}  
\\
 $M_W = 80.385(15) \gev$
& $M_Z = 91.1876(21) \gev$ 
\\
  $G_F = 1.16638(1)\times 10^{-5}\gev^{-2}$
& $s_W^2 \equiv \sin^2\!\theta_W = 0.23126(13)$  
\\
  $\alpha(M_Z) = 1/127.9$
& $\alpha_s(M_Z) = 0.1185(6)$
\\
\hline\hline
  \multicolumn{2}{|c|}{quark masses }
\\
  $m_d(2\gev) = 4.68(16)\mev$              \hfill\cite{Aoki:2013ldr} 
&
\\  
  $m_s(2\gev)=93.8(24) \mev$               \hfill\cite{Aoki:2013ldr}
& $m_c(m_c) = 1.275(25) \gev$              \hfill\cite{Agashe:2014kda}
\\
  $m_b(m_b) = 4.18(3)\gev$                 \hfill\cite{Agashe:2014kda}
& $m_t(m_t) = 163(1)\gev$                  \hfill\cite{Allison:2008xk}
\\
\hline\hline
  \multicolumn{2}{|c|}{CKM}
\\
  $\lambda = 0.22544({}^{+33}_{-28})$
& $A = 0.8207(7)(13)$
\\
  $\bar\rho = 0.125({}^{+30}_{-18})$
& $\bar\eta = 0.382({}^{+22}_{-18})$
\\
\hline\hline
  \multicolumn{2}{|c|}{Kaon}
\\
  $m_K = 497.614(24)\mev$                  \hfill\cite{Agashe:2014kda}
& $\kappa_\epsilon = 0.94(2)$              \hfill\cite{Buras:2008nn, Buras:2010pza}
\\
  $F_K/F_\pi = 1.194(5)$                   \hfill\cite{Agashe:2014kda}
& $F_\pi = 130.41(20)\mev$                 \hfill\cite{Agashe:2014kda}
\\
  $\hat B_K = 0.750(15)$                   \hfill\cite{Aoki:2013ldr, Buras:2014maa}
& $\eta_{tt} = 0.5765(65)$                 \hfill\cite{Buras:1990fn}
\\
  $\eta_{ct} = 0.496(47)$                  \hfill\cite{Brod:2010mj}
& $\eta_{cc} =1.87(76)$                    \hfill\cite{Brod:2011ty}
\\
\hline\hline
  \multicolumn{2}{|c|}{$B$-meson}
\\
  $m_{B^\pm} = 5279.29(15)\mev$            \hfill\cite{Agashe:2014kda}
& $\tau_{B^\pm} = 1.638(4)\,\text{ps}$     \hfill\cite{Amhis:2014hma}
\\
  $m_{B_d} = 5279.61(16)\mev$              \hfill\cite{Agashe:2014kda}
& $\tau_{B_d} = 1.520(4) \,\text{ps}$      \hfill\cite{Amhis:2014hma}
\\
  $m_{B_s} = 5366.79(23)\mev$              \hfill\cite{Agashe:2014kda}
& $\tau_{B_s} = 1.505(4)\,\text{ps}$       \hfill\cite{Amhis:2014hma}
\\
  $F_{B_d} = 190.5(42)\mev\quad$ 
& $F_{B_s} = 227.7(45)\mev\quad$           \hfill\cite{Aoki:2013ldr}
\\
  $F_{B_d} (\hat B_{B_d})^{1/2} = 229.4(93)\mev$
& $F_{B_s} (\hat B_{B_s})^{1/2} = 276.0(85)\mev$ \hfill \cite{Bazavov:2016nty}
\\
  $\rho\big(F_{B_s}, F_{B_d}\big) = 61.7\%^* \quad$ 
& $\rho\big(F_{B_s} (\hat B_{B_s})^{1/2}, F_{B_d} (\hat B_{B_d})^{1/2}\big) = 95.1\%^{**}$
\\
  $\eta_B = 0.55(1)$                       \hfill\cite{Buras:1990fn, Urban:1997gw}
& $\Delta\Gamma_s/\Gamma_s = 0.124(9)$     \hfill\cite{Amhis:2014hma}
\\
\hline
\end{tabular}  
}
\renewcommand{\arraystretch}{1.0}
  \caption{\small
    Values of the experimental and theoretical quantities used as input parameters
    as of March 2016. 
    ${}^*:$ Calculated by 
    demanding that the uncertainty of the ratio of the decay constants
    given above should equal the uncertainty given explicitly for the ratio, also
    given in Ref.~\cite{Aoki:2013ldr}.
    ${}^{**}:$ Calculated from information given in Ref.~\cite{Bazavov:2016nty}. 
    Note that their assumption for the $\mathrm{SU(3)}$ breaking from the charm
    sea contribution corresponds to the assumption of a $91.8\%$ correlation
    for this uncertainty between $B_d$ and $B_s$.
  }
  \label{tab:input}
\end{table}

\begin{table}
  \centering
\renewcommand{\arraystretch}{1.5}
\begin{tabular}{|c||cc|c|ccc|}
\hline
  $ij$  & $\muLow$ [GeV] & $N_f$ & $r_\chi$ & $B_1^{ij}$ & $B_4^{ij}$ & $B_5^{ij}$
\\
\hline\hline
  $sd$  & $3.0$          & 3     & $30.8$   & 0.525(16)      & 0.920(20)      & 0.707(45)
\\
\hline\hline
  &  &  &  & $F_{B_j}^2 B_1^{ij}$ & $F_{B_j}^2 B_4^{ij}$ & $F_{B_j}^2 B_5^{ij}$
\\
\hline
  $bd$  & $4.18$         & 5     & $1.6$    & 0.0342(30)      & 0.0390(29)    & 0.0361(36)
\\
  $bs$  & $4.18$         & 5     & $1.6$    & 0.0498(32)      & 0.0534(32)    & 0.0493(37)
\\
\hline
\end{tabular}
\renewcommand{\arraystretch}{1.0}
\caption{\small \label{tab:M12:input}
  Scale settings and number of flavours, $N_f$, as well as numerical
  inputs of bag factors entering $M_{12}^{ij}$, see \cite{Bazavov:2016nty}
  and \cite{Garron:2016mva} for correlations. For the Kaon system threshold
  crossings to $N_f=4$ and $N_f = 3$ have been chosen as $4.18$~GeV and $1.4$~GeV.
  The chirality-factor is given as $r_\chi^{ij} = (M_{M_{ij}} / (m_i(\muLow) + m_j(\muLow))^2$.
  See also \cite{Bobeth:2017xry} for more details on $M_{12}^{ij}$.
}
\end{table}

%
%
%

\renewcommand{\refname}{R\lowercase{eferences}}

\addcontentsline{toc}{section}{References}

\bibliographystyle{JHEP}
\bibliography{AJBeprefs}

\end{document}